\documentclass[review,12pt]{elsarticle}
\usepackage[numbers]{natbib}
\usepackage{pdflscape}

\usepackage{tikz}
\usetikzlibrary{shapes,arrows}

\usepackage{rotating}

\usepackage{framed} 
\usepackage{multicol} 
\usepackage{color}
\usepackage{amssymb}
\usepackage{mathrsfs}
\usepackage{epstopdf}
\usepackage{makeidx}
\usepackage{bm}
\usepackage{psfrag}
\usepackage{amsthm}
\usepackage{algorithm}
\usepackage{algorithmic}
\usepackage{epsf,amsfonts,amsmath}
\usepackage{amssymb,amsmath}
\usepackage{appendix}
\usepackage{graphicx}
\usepackage{multirow}
\usepackage{array}

\newtheorem{remark1}{Remark}

\usepackage{caption}
\usepackage{times}
\usepackage{float}
\usepackage[caption = false]{subfig}
\usepackage[intoc]{nomencl}
\makenomenclature
\renewcommand\nomgroup[1]{%
\ifthenelse{\equal{#1}{}}{%
\item[\textbf{Symbols}] }{
\ifthenelse{\equal{#1}{R}}{%
\item[\textbf{Roman Symbols}]}{
\ifthenelse{\equal{#1}{G}}{%
\item[\textbf{Greek Symbols }]}{
\ifthenelse{\equal{#1}{S}}{%
\item[\textbf{Superscripts  }]}{{
\ifthenelse{\equal{#1}{U}}{%
\item[\textbf{Subscripts }]}{{
\ifthenelse{\equal{#1}{X}}{%
\item[\textbf{Other Symbols }]}
{{}}}}}}}}}}

\journal{Elsevier}

\begin{document}

\begin{frontmatter}

\title{{\color{black}On Uncertainty Quantification of Lithium-ion Batteries: Application to an LiC$_6$/LiCoO$_2$ cell}}

\author[UCB1]{Mohammad Hadigol}
\ead{mohammad.hadigol@colorado.edu}

\author[UCB1]{Kurt Maute}
\ead{maute@colorado.edu}

\author[UCB1]{Alireza Doostan\corref{cor1}}
\ead{alireza.doostan@colorado.edu}

\cortext[cor1]{Corresponding Author: Alireza Doostan}

\address[UCB1]{Aerospace Engineering Sciences Department, University of Colorado, Boulder, CO 80309, USA}

\begin{abstract}
\label{Abstract}

In this work, a stochastic, physics-based model for Lithium-ion batteries (LIBs) is presented in order to study the effects of parametric model uncertainties on the cell capacity, voltage, and concentrations. To this end, the proposed uncertainty quantification (UQ) approach, based on sparse polynomial chaos expansions, relies on a {\it small} number of battery simulations. Within this UQ framework, the identification of most important uncertainty sources is achieved by performing a global sensitivity analysis via computing the so-called Sobol' indices. Such information aids in designing more efficient and targeted quality control procedures, which consequently may result in reducing the LIB production cost. An LiC$_6$/LiCoO$_2$ cell with 19 uncertain parameters discharged at 0.25C, 1C and 4C rates is considered to study the performance and accuracy of the proposed UQ approach. The results suggest that, for the considered cell, the battery discharge rate is a key factor affecting not only the performance variability of the cell, but also the determination of most important random inputs.

\end{abstract}


\end{frontmatter}


%
\section{Introduction}
\label{sec:introduction}
\setcounter{equation}{5}

High energy and power densities of {\color{black} Lithium-ion batteries (LIBs)} alongside their superior safety features have made them the number one energy storage device for a wide range of electric devices from cell phones to hybrid-electric vehicles and aerospace applications \cite{GMD:09, Doyle03, Marsh01, Etacheri11}. Since the launch of {\color{black}the} first commercial LIB in 1991 \cite{Thomas02}, {\color{black} significant work has} been dedicated to modeling \cite{DFN:93, Smith08c, Cai09, Subramanian09a, Forman11, Wenbo14}, {\color{black} design optimization} \cite{Fuller94, Golmon12, Xue13}{\color{black}, and discovering new materials \cite{Chen13, Pang13, Aravindan13} for LIBs}. 

Among the proposed models for simulation of LIBs, the most widely used one is Newman's model {\color{black} \cite{Newman75, DFN:93, Doyle96b},} which is based on the porous electrode and concentrated solution theories. This model involves material and electrochemical properties{\color{black},} such as porosity of electrodes, solid particle size, and diffusion coefficients, which are measured/estimated directly or indirectly via experimental techniques \cite{Shriram07}. Measurements of most physical quantities are accompanied by uncertainty due to accuracy limitations or natural, cell-to-cell variabilities. {\color{black}For example, measurements of the solid particle size in the LIB electrodes {\color{black}typically} result in distributed values for the particle size \cite{Shriram07, malvern} rather than a single deterministic value which is mostly used in the LIB simulations. Performing additional measurements may lead to a better characterization of the solid particle size, but cannot {\color{black}completely} eliminate the irreducible uncertainties due to natural variations. Hence, a deterministic treatment of the solid particle size may result in predictions which do not agree with the experiments well. Consequently, LIB models which incorporate discrete or continuous particle size distributions have been developed \citep{Darling97, Nagarajan98, Wang02c, Stephenson07, Shriram07}.} Quantifying the impact of such uncertainties is essential for {\color{black}reliable} model-based predictions, and {\color{black}is the focus of} the emerging {\color{black}field} of Uncertainty Quantification (UQ) in computational engineering and science. {\color{black}In general, UQ provides tools for assessing the credibility of model predictions and facilitating decision making under uncertainty \cite{Chakraborty13, Bickel08, Sankararaman2013b}. UQ is also utilized for quantitative validation of simulation models \cite{Xiu10a, LeMaitre07, Mcfarland08} and robust design optimization under uncertainty \cite{Bertsimas2014, Duvigneau13, Banaszuk201177}.  }


One may represent the uncertain model parameters by random variables/processes. This is the subject of a major class of UQ approaches known as probabilistic techniques. Among these methods, stochastic spectral methods \cite{Ghanem91a,Xiu10a} based on polynomial chaos (PC) expansions \cite{Wiener38,Cameron47} have received special attention due to their advantages over traditional UQ techniques such as perturbation-based and Monte Carlo sampling (MCS) methods. In particular, under certain regularity conditions, these schemes converge faster than MCS methods \citep{LeMaitre10} and, unlike perturbation methods, are not restricted to problems with small uncertainty levels \citep{Ghanem91a}. Stochastic spectral methods are based on expanding the solution of interest in PC bases. The coefficients of these expansions are then computed, for instance, intrusively via Galerkin projection \citep{Ghanem91a}, or non-intrusively {\color{black} via regression \citep{Hosder06, Doostan11a, Hampton2015, Hampton15b} or quadrature integration \citep{Constantine12a}}. For complex systems such as LIBs, non-intrusive methods {\color{black}are more attractive than} intrusive ones since they allow the use of simulations as black boxes. In other words, there is no need to modify the available deterministic solvers when one uses non-intrusive PC expansions. In addition, {\color{black}since {\color{black}only} independent solution realizations are needed}, parallel implementation is straightforward.

Up to date, {\color{black}the} majority of the LIB {\color{black}physics-based} simulations have treated {\color{black}the underlying} model parameters {\color{black}deterministically} and ignored the effects of uncertainties in the model parameters on the performance of LIBs. There are a few works in the literature which have addressed the variability of LIB {\color{black}physics-based} model parameters. {\color{black}In \cite{Shriram07},} PC expansion is employed within the Newman's model to demonstrate the reduction in the cell potentials as a result of the uncertainty in the particle size of the anode electrode. Effects of variability in cycling rate, particle size, diffusivity, and electrical conductivity of the cathode electrode on the LIB performance was examined in \citep{Dua10}, where surrogate models are developed using the Newman's model together with techniques such as Kriging, polynomial response, and radial-basis neural networks. It was found that the randomness in electrical conductivity has the minimal effects on the cell performance, while the impact of the remaining parameters depend on the the cycling rate. In \cite{Shriram12}, impedance spectroscopy {\color{black}\cite{Sikha01122008}} is utilized to compare the relative importance of randomness in porosity, particle size, length and tortuosity of the cathode and separator {\color{black}using Nyquist stability criterion}. {\color{black}It was shown that} the particle size has the largest impact on the fluctuations in the impedance. In \citep{Ramadesigan11}, a reduced order LIB model presented in {\color{black}\citep{Subramanian09a}} and a Bayesian inference technique are used to estimate the effective kinetic and transport parameters from experimental data. {\color{black}These works either employ a reduced order model of the LIB or only consider uncertainties in a few number of model parameters to avoid high computational costs. }

This work presents a sampling-based PC approach to study the effects of uncertainty in various model parameters on the cell capacity, voltage, and concentrations of an LIB described by the Newman's model \cite{Newman75, DFN:93, Doyle96b}. The proposed PC approach, first introduced in \cite{Doostan11a}, relies on the sparsity of expansion coefficients to accurately compute the statistics of quantities of interest with a small number of battery simulations. Through its application to an LiC$_6$/LiCoO$_2$ LIB model, we demonstrate that, unlike the previously mentioned works on UQ of LIBs, the proposed PC approach is capable of taking into account a large number of uncertain parameters. While the proposed PC-based UQ framework is general, the results we present are specific to the particular LiC$_6$/LiCoO$_2$ cell we consider and the the choices of uncertainty models for its parameters. The latter are, as much as possible, identified from the reported experiments on identical or similar cells/materials.

Additionally, this UQ framework enables performing a global sensitivity analysis (SA) to identify the most important uncertain parameters affecting the variability of the output quantities. Such an analysis may be used toward reducing cell-to-cell variations and designing more efficient and targeted quality control procedures to reduce the manufacturing cost of LIBs \cite{Shriram12, Shin13}. We also review the standard experimental techniques used for measuring the model parameters and discuss their impacts on the cell capacity and/or power.

{\color{black}
It is worth highlighting that the present work is focused on modeling and propagation of parametric uncertainties, as opposed to model form uncertainties which may consider multiple competing models. 
}

The remainder of this paper is organized as follows. Section \ref{sec:decoupled} {\color{black}reviews the LIB model used in this study}. In Section \ref{sec:PCE}, {\color{black}we present our stochastic LIB modeling approach which is based on non-intrusive PC expansions.} We then continue with reviewing a global sensitivity analysis approach via the Sobol' indices. Standard experimental techniques for measuring the model parameters and their underlying distributions are discussed in Section \ref{sec:rv_LIB}. Finally, an LiC$_6$/LiCoO$_2$ cell is considered as our numerical example in Section \ref{sec:Numerical_Examples}. 
\section{LIB governing equations}
\label{sec:decoupled}

\begin{figure}[htb]
  \begin{center}
	\includegraphics [width=9cm]{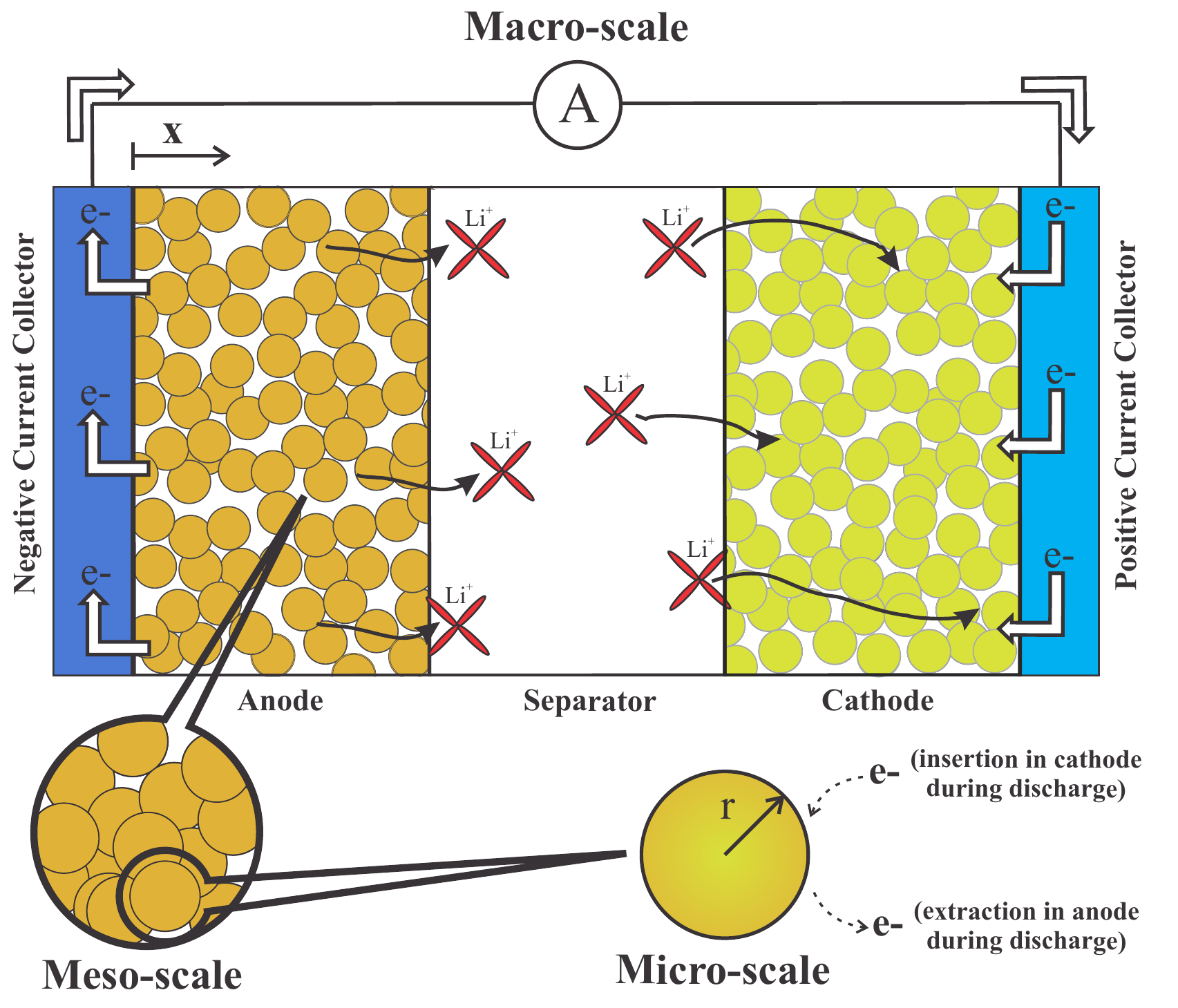}
    \caption{Schematic of a full cell LIB.}
    \label{fig:schematic_LIB}
   \end{center}
\end{figure}

{\color{black}An} LIB {\color{black}schematic} is presented in Fig. \ref{fig:schematic_LIB}. During the discharge process, lithium ions in the solid particles of the anode diffuse to the particle surface where they oxidize into Li$^+$ ions and electrons and transfer to the electrolyte liquid (deintercalation). Electrons flow through the external circuit {\color{black}to} the positive electrode. Meanwhile, Li$^+$ ions travel via diffusion and migration through the electrolyte and separator to the cathode where they are reduced and diffused into the solid particles (intercalation). When the LIB is charged, this process is reversed. 

The Newman's LIB model is developed based on porous electrode and concentrated solution theories {\color{black}\cite{Newman75, DFN:93, Doyle96b}}. The governing equations of this model for the salt concentration in liquid phase $c$, lithium concentration in solid phase $c_s$, liquid phase potential $\phi_{e}$ and solid phase potential $\phi_{s}$ are presented in Table \ref{table:coupled_GE}. {\color{black}Symbols {\color{black}used} in this table are defined in the nomenclature.} {\color{black}The assumptions made in this model are summarized as: (i) transport properties are independent of temperature and simulations are isothermal; 
(ii) $D$, $D_s$ and $t_+^0$ do not depend on the concentrations; 
(iii) volume changes within the cell are ignored; (iv) zero surface electrolyte inter-phase (SEI) resistance is {\color{black}considered}; (v) no double layer capacitive effects are {\color{black}considered}; and (vi) solid particles are spherical.}

\begin{table}[h]
\caption{Coupled non-linear governing equations of LIB.} 
\centering
\resizebox{\textwidth}{!}{%
\begin{tabular}[t]{ >{\normalsize} p{2cm}  >{\normalsize} c  c }   
\hline \hline
  & \textbf{Governing equation} & \textbf{Boundary conditions} \\  
\hline \hline
 Electrolyte phase diffusion & 
\begin{minipage}{9cm} \begin{equation} \label{eqn:c} \begin{aligned} \frac{\partial ( \epsilon c )}{\partial t} = \nabla (\epsilon D^{\mathrm{eff}} \nabla c) + \frac{1-t_+^0}{F} j_{vol} \end{aligned} \end{equation} \end{minipage}
&  \begin{minipage}{6cm} \begin{equation} \begin{aligned} \nabla c|_{x=0} = \nabla c|_{x=L} = 0 \end{aligned}  \nonumber \end{equation} \end{minipage}  \\ 
\hline
Solid phase diffusion & 
\begin{minipage}{9cm} \begin{equation} \label{eqn:c_s} \begin{aligned} \frac{\partial c_s }{\partial t} = \frac{1}{r^2} \frac{\partial}{ \partial r} \Big( D_s r^2 \frac{\partial}{ \partial r} c_s \Big) \end{aligned} \end{equation} \end{minipage}
&  \begin{minipage}{6cm} \begin{eqnarray} && \nabla c_s|_{r=0} = 0 \nonumber \\ &&  \nabla c_s|_{r=r_s} = -\frac{j_{vol}}{a F D_s}  \nonumber \end{eqnarray} \end{minipage} \\ 
\hline
Liquid phase potential & 
\begin{minipage}{9cm} \begin{equation} \label{eqn:phi_e} \begin{aligned} \nabla ( \kappa^{\mathrm{eff}} \nabla \phi_{e}) - \nabla (\kappa_D^{\mathrm{eff}} \nabla \ln c) + j_{vol} = 0 \end{aligned} \end{equation} \end{minipage}
& \begin{minipage}{6cm} \begin{eqnarray} && \nabla \phi_{e}|_{x=0} = \nabla \phi_{e}|_{x=L} = 0 \nonumber \\ && \phi_{e}|_{x=L} = 0  \nonumber \end{eqnarray} \end{minipage} \\ 
\hline
Solid phase potential & 
\begin{minipage}{9cm} \begin{equation} \label{eqn:phi_s} \begin{aligned} \nabla ( \sigma^{\mathrm{eff}} \nabla \phi_{s}) - j_{vol} = 0 \end{aligned} \end{equation} \end{minipage} 
&  \begin{minipage}{6cm} \begin{eqnarray}  && \nabla \phi_{s}|_{x=0} = \nabla \phi_{s}|_{x=L} = \frac{-I}{\sigma^{\mathrm{eff}}} \nonumber \\
&&  \nabla \phi_{s}|_{x=L_a} = \nabla \phi_{s}|_{x=L_a+L_s} = 0 \nonumber \end{eqnarray} \end{minipage} \\ 
\hline
Reaction kinetics & \multicolumn{2}{l }{ \begin{minipage}{10cm} \begin{eqnarray} \label{eqn:j_vol} \begin{aligned} && j_{vol} = a i_{ex} \Big[ \exp \Big( \frac{0.5 F \eta}{RT} \Big) - \exp \Big( - \frac{0.5 F \eta}{RT} \Big)\Big] \\ && i_{ex} = Fk(c_s^{\mathrm{surf}})^{0.5}(c_{s,max} - c_s^{\mathrm{surf}})^{0.5}(c)^{0.5} \end{aligned} \end{eqnarray} \end{minipage} }   \\
\hline
\end{tabular}}
\label{table:coupled_GE} 
\end{table}

In the literature, numerical techniques such as finite difference \citep{DFN:93}, finite volume \citep{Popov11}, and finite elements \citep{Wang98} have been used to solve this system of coupled non-linear equations simultaneously. {\color{black}In order to reduce the computational cost of the LIB simulations, using the concepts of particular and homogeneous solutions to Ordinary Differential Equations (ODEs), Reimers \citep{Reimers13} suggested a decoupled formulation {\color{black}of Newman's model}, which we use in this study.} {\color{black}For the sake of completeness, this decoupling technique is presented in \ref{sec:LIB_decoupled}.}

\section{UQ via polynomial chaos expansion}
\label{sec:PCE}

Polynomial chaos expansion was first introduced by Wiener in 1938 \citep{Wiener38}. It was reintroduced to the engineering field in 1991 by Ghanem and Spanos \citep{Ghanem91a} for the problems with Gaussian input uncertainties and later extended {\color{black}to} non-Gaussian random inputs by using the orthogonal {\color{black}polynomials} of the Askey scheme (generalized PC expansion) \citep{Xiu02}. PC expansion provides a framework to approximate the solution of a stochastic system by projecting it onto a basis of polynomials of the random inputs, which we review in the following. 

Let $\left(\Omega,\mathcal{T},\mathcal{P} \right)$ be a complete probability space, where $\Omega$ is the sample set and $\mathcal{P}$ is a probability measure on the $\sigma-$field $\mathcal{T}$. Also assume that the system input uncertainty has been discretized and approximated by random variables, such that the vector $\bm{\xi} = \left(\xi_1,\cdots, \xi_d\right):\Omega \rightarrow \mathbb{R}^{d}$, $d \in \mathbb{N}$, represents the set of independent random inputs. {\color{black}We also assume that probability density function (PDF) of {\color{black}the} random variable $\xi_k$ is denoted by $\rho(\xi_k)$, while $\rho(\bm{\xi})$ represents the joint PDF of $\bm{\xi}$}. Let us assume that the finite variance output quantity of interest (QoI) defined on $\left(\Omega,\mathcal{T},\mathcal{P} \right)$ is denoted by $u(\bm{\xi})$. The truncated PC representation of $u(\bm{\xi})$, denoted by $\hat{u}(\bm{\xi})$, is 
\begin{equation} 
\label{eqn:RV_PCE}
\hat{u}(\bm{\xi}) = \sum_{\bm{i} \in \mathscr{I}_{d,p}} \alpha_{\bm{i}} \psi_{\bm{i}}(\bm{\xi}),
\end{equation}
where $\alpha_{\bm{i}}$ are the deterministic coefficients and $\psi_{\bm{i}}(\bm{\xi})$ are the multivariate PC basis functions. The basis functions $\psi_{\bm i}(\bm\xi)$ in (\ref{eqn:RV_PCE}) are generated from

\begin{equation} 
\label{eqn:multivariate_polynomials}
\psi_{\bm{i}}(\bm{\xi}) = \prod_{k=1}^{d} \psi_{{i}_k}(\xi_k), \ \ \ \ \bm{i} \in \mathscr{I}_{d,p},
\end{equation}
where $\psi_{{i}_k}(\xi_k)$, $k=1,\dots,d$, are univariate polynomials of degree ${i}_k  \in \mathbb{N}_0 := \mathbb{N}\cup \lbrace 0\rbrace$ orthogonal with respect to $\rho(\xi_k)$ {\color{black}(see, e.g., Table \ref{table:Askey})}, i.e.,
\begin{equation}
\label{eqn:1d_PCE}
\mathbb{E} [\psi_{i_k} \psi_{j_k}]  = \int \psi_{i_k}(\xi_k) \psi_{j_k}(\xi_k) \rho(\xi_k) \mathrm{d}\xi_k= \delta_{i_kj_k} \mathbb{E} [ \psi_{i_k}^2 ],
\end{equation}
where $\delta_{i_kj_k}$ is the Kronecker delta and $\mathbb{E}[\cdot]$ denotes the mathematical expectation operator. The multi-index $\bm{i}$ in (\ref{eqn:RV_PCE}) is ${\bm{i}} = (i_1, \cdots , i_d) \in \mathscr{I}_{d,p}$ and the set of multi-indices $\mathscr{I}_{d,p}$ is defined by
\begin{equation} 
\mathscr{I}_{d,p} = \lbrace \bm i = (i_1, \cdots , i_d)\in \mathbb{N}_0^d: \Vert\bm{i}\Vert_1 \leqslant p \rbrace,
\end{equation}
where $\Vert \cdot \Vert_1$ is the $l_1$ norm and the {\color{black}size} of  $\mathscr{I}_{d,p}$, hence the number $P$ of PC basis functions of total order not larger than $p$ in dimension $d$, is given by 

\begin{equation} 
\label{eqn:P+1}
P=\vert \mathscr{I}_{d,p}\vert= \frac{(p+d)!}{p!d!}.
\end{equation}

{\color{black}Due to} the orthogonality of the polynomials $\psi_{{i}_k}(\xi_k)$ and given that the $\xi_k$ are independent, the PC basis functions $\psi_{\bm i}(\bm\xi)$ are also orthogonal, i.e., $\mathbb{E}[\psi_{\bm i}\psi_{\bm j}]=\delta_{\bm i,\bm j}\mathbb{E}[\psi_{\bm i}^2]$. The truncated PC expansion in (\ref{eqn:RV_PCE}) converges in the mean-square sense as $ p \rightarrow \infty$ when $u(\bm\xi)$ has finite variance and the coefficients $\alpha_{\bm{i}}$ are computed from the projection equation $\alpha_{\bm{i}} = \mathbb{E}[u(\cdot)\psi_{\bm i}(\cdot)]/\mathbb{E}[\psi_{\bm i}^2]$
 \citep{Xiu02}. For convenience, we also index the PC basis functions by $\{1,\dots,P\}$ so that there is a one-to-one correspondence between $\{\psi_j(\bm\xi)\}_{j=1}^{P}$ and $\{ \psi_{\bm i}(\bm\xi) \}_{\bm i\in\mathscr{I}_{p,d}}$.

\begin{table}[h]
\caption{Correspondence of Wiener-Askey polynomial chaos and {\color{black}probability} distribution of the random variables \citep{Xiu02}.} 
\centering
\begin{tabular}[t]{ c   c  c }   
\hline \hline
{\color{black}$\rho(\xi_k)$} & {\color{black}Polynomial type} & Support \\  
\hline \hline
Gaussian & Hermite & (-$\infty$,+$\infty$) \\  
\hline
Gamma & Laguerre & (0,+$\infty$) \\  
\hline
Beta & Jacobi & [a,b] \\  
\hline
Uniform & Legendre & [a,b] \\  
\hline
\end{tabular}
\label{table:Askey} 
\end{table}

\subsection{Non-intrusive polynomial chaos expansion}
\label{sec:Non-intrusive PCE}

The main task in {\color{black}PC-based methods} is to compute the coefficients of the {\color{black}solution expansion} either intrusively \citep{Ghanem91a} or non-intrusively \citep{Eldred09}. In an intrusive approach, {\color{black}the governing equations are} projected onto {\color{black} the subspace spanned by the PC basis via the Galerkin formulation}. The final system of equations to be solved in an intrusive PC expansion method is $P$ times larger than the size of the deterministic  {\color{black}counterpart}. This approach {\color{black}may} require some modifications of the existing deterministic solvers, which for complex problems such as LIBs, {\color{black}may} be difficult and time-consuming to implement. On the other hand, non-intrusive methods facilitate the use of existing deterministic solvers and treat them as a black box. {\color{black} The first task is to generate a set of $N$ deterministic or random samples of $\bm\xi$, denoted by $\{{\bm{\xi}}^{(i)}\}_{i=1}^{N}$. Next, corresponding to these samples, $N$ realizations of the output QoI, $\{u({\bm{\xi}}^{(i)})\}_{i=1}^{N}$, are computed using an available deterministic solver.} The last step is solving for the PC coefficients using these realizations. Several methods such as least squares regression \citep{Hosder06}, pseudo-spectral collocation \citep{Xiu10a}, Monte Carlo sampling \citep{LeMaitre10}, and compressive sampling (CS) \citep{Doostan11a} have been employed in the literature {\color{black}for this purpose}. Once the PC coefficients are computed, the mean, $\mathbb{E}[\cdot]$, and variance, $\mathrm{var}[\cdot]$, of $u(\bm{\xi})$ can be directly approximated by

\begin{equation}
\label{eqn:mean_PCE}
\mathbb{E}[\hat{u}] =  \alpha_{\bm{0}},
\end{equation}
and                 
\begin{equation}
\label{eqn:var_PCE}
\mathrm{var}[\hat{u}] =  \sum_{\substack{\bm{i} \in \mathscr{I}_{d,p}\\ \bm{i} \neq {\color{black}\bm{0}}}} \alpha_{\bm{i}}^2.
\end{equation}

In the following, we will review the least squares regression and the compressive sampling methods. 

\subsubsection{Least squares regression}
\label{sec:least_squares_regression}
The least squares regression technique is basically the regression of the exact solution $u(\bm{\xi})$ in the PC bases \citep{Sudret08d}. Given the set of samples $\{{\bm{\xi}}^{(i)}\}_{i=1}^{N}$, generated randomly, for instance, according to $\rho(\bm\xi)$, and the corresponding solution realizations $\{u({\bm{\xi}}^{(i)})\}_{i=1}^{N}$, the discrete representation of (\ref{eqn:RV_PCE}) can be written as
\begin{equation}
\label{eqn:regression3}
{\color{black}\bm{\Psi} \bm{\alpha}  \approx \bm{u}},
\end{equation}
where $\bm{u} = ( u({\bm{\xi}}^{(1)}), \cdots,  u({\bm{\xi}}^{(N)}) )^T \in \mathbb{R}^N$ contains the realizations of the QoI, {\color{black}$\bm{\Psi}(i,j)=\psi_j({\bm{\xi}}^{(i)}) \in \mathbb{R}^{N \times P}$} is the measurement matrix containing samples of the PC basis, and $\bm{\alpha} = ( \alpha_1, \cdots,  \alpha_P )^T \in \mathbb{R}^P$ is the vector of PC coefficients. 

PC coefficients may be approximated by solving the least squares problem
\begin{equation}
\label{eqn:regression1}
\min_{\color{black}{\bm{\alpha}}} \Vert \bm{u} - \bm{\Psi} \bm{\alpha} \Vert_2,
\end{equation}
where $\Vert \cdot \Vert_2$ is the $l_2$ norm. {\color{black}When $\bm{\Psi}$ is full rank}, the solution to (\ref{eqn:regression1}) {\color{black}is} computed from the normal equation 
\begin{equation}
\label{eqn:regression2}
(\bm{\Psi}^T \bm{\Psi}) \bm{\alpha} = \bm{\Psi}^T \bm{u}.
\end{equation}

In general, a stable solution $\bm\alpha$ to (\ref{eqn:regression2}) requires $N > P$ realizations of $u(\bm{\xi})$. {\color{black}In \cite[Theorems 2.2 and 3.1]{Hampton15b}, it is shown that for $d$-dimensional Legendre polynomials of total order $p$, for the case of $d>p$, a stable solution recovery from (\ref{eqn:regression1}) can be guaranteed with a number of samples given by 
\begin{equation}
\label{eqn:samples_regression}
{\color{black}N \geq 3^p \ C \ P \ \text{log}(P),}
\end{equation}
where $C$ is an absolute constant. This suggests that the number of samples $N$ depends linearly on $P$ (up to a logarithmic factor) for the least squares regression method. }

For high dimensional complex problems, such as LIBs, generating $N > P$ realizations {\color{black}may} be computationally expensive. {\color{black}In such cases, when the solution $u(\bm{\xi})$ depends {\it smoothly} on $\bm{\xi}$, the PC coefficients are often sparse, i.e., many of them are negligible. In these cases, CS may be employed to compute the coefficients with $N < P$ realizations \citep{Doostan11a, Hampton15b, Peng201492, Hampton2015, Schiavazzi_2014, Yan_2012, Jones2015, Yang201387}. Specifically, the importance sampling approach of \citep{Hampton2015} ensures an accurate computation of $\bm\alpha$ with a number of solution realizations that depends linearly (up to a logarithmic factor in $P$) on the number of dominant coefficients. We next review the CS approach, which we use later for the UQ of LIBs}.

\subsubsection{Compressive sampling}
\label{sec:compressive_sampling}

Compressive sampling/sensing is an emerging {\color{black}direction} in signal processing which enables (up to) exact reconstruction of signals admitting sparse representations with a small number of signal measurements \citep{Donoho06b, Bruckstein09, Candes06c}. It was first introduced to the UQ field in 2011 by Doostan and Owhadi \citep{Doostan11a} where they used CS to approximate sparse PC solutions to stochastic PDEs. The main requirement in CS methods is to have a sparse solution at the stochastic level, that is a small fraction of PC coefficients in (\ref{eqn:RV_PCE}) are dominant and contribute to the solution statistics. The ultimate goal of CS is to approximate the sparse PC coefficients $\bm{\alpha}$ accurately and robustly with $N < P$ {\color{black}realizations of $u(\bm{\xi})$}.

With $N < P$, the underdetermined linear system
in {\color{black}(\ref{eqn:regression3})} is ill-posed and generally has infinitely many solutions. Sparsity of the PC coefficients $\bm{\alpha}$ allows a regularization of {\color{black}(\ref{eqn:regression3})} to ensure a well-posed solution \citep{Doostan11a}. This can be achieved by solving the Basis Pursuit Denoising (BPDN) problem
\begin{equation}
\label{eqn:BPDN}
{\color{black}\min_{\bm{\alpha}}} \Vert \bm{\alpha} \Vert_1 \quad  \mathrm{subject} \ \  \mathrm{to} \quad \Vert \bm{\Psi} \bm{\alpha} - \bm{u} \Vert_2 \leqslant \gamma.
\end{equation}

{\color{black}Minimization of} the ${l}_1$ norm in (\ref{eqn:BPDN}) promotes sparsity in $\bm\alpha$, while the ${l}_2$ {\color{black}residual} norm {\color{black}controls} the accuracy of the truncated PC expansion {\color{black}with the tolerance} $\gamma$. Several numerical techniques are available in the literature to solve the BPDN problem \citep{Doostan11a}. Among those, we adopt the Spectral Projected Gradient algorithm (SPGL1) of \citep{Berg08} 
implemented in the \texttt{SPGL1 package} for MATLAB \citep{spgl1:2007}.

The accuracy of {\color{black}$\bm{\alpha}$ computed from (\ref{eqn:BPDN})} depends on the sample size $N$ and the truncation error $\Vert \bm{\Psi} \bm{\alpha} - \bm{u} \Vert_2 $ in (\ref{eqn:BPDN}). In general, the truncation error may be decreased by increasing the order $p$ of the PC basis, which leads to a larger number of coefficients, {\color{black}$P$}. This, in turn, requires a larger number of samples $N$ to maintain the stability of the BPDN problem. The minimum sampling rate depends on the type of the PC basis, the sparsity of $\bm\alpha$, and the sampling distribution according to which $\{\bm{\xi}^{(i)}\}_{i=1}^N$ are generated, and is shown to linearly depend on the number of dominant PC coefficients {\color{black}\citep[Theorems 3.1 and 4.2]{Hampton2015}. More precisely, for  $d$-dimensional Legendre polynomials of total order $p$, where $d>p$, the number of samples $N$ required to guarantee a stable solution recovery from (\ref{eqn:BPDN}) is given by
\begin{equation}
\label{eqn:samples_BPDN}
{\color{black}N \geq 3^p \ C \ S \ \text{log}(P),}
\end{equation}
where $S$ is the number of dominant coefficients. In (\ref{eqn:samples_BPDN}), $N$ depends primarily on $S$ and weakly on $P$ through the $\text{log}(P)$ term. For high dimensional complex problems, we usually have $P \gg S$, hence, a comparison of (\ref{eqn:samples_regression}) and (\ref{eqn:samples_BPDN}) suggests that the CS approach requires considerably smaller number of samples in comparison to the least squares regression.}

{

In practice, one may start by approximating a lower order PC expansion when $N$ is small and increase $p$ when a larger number of samples become available. Another important factor in the sparse reconstruction is the selection of the truncation error tolerance $\gamma$. The ideal value for the tolerance is $\gamma \approx \Vert \bm{\Psi} \bm{\alpha}_{\mathrm{exact}} - \bm{u} \Vert_2 $. Since the exact PC coefficients $\bm{\alpha}_{\mathrm{exact}}$ are not known, selecting larger values than $\Vert \bm{\Psi} \bm{\alpha}_{\mathrm{exact}} - \bm{u} \Vert_2 $ for $\gamma$ deteriorates the accuracy of the approximation, while smaller choices may result in over-fitting the solution samples and, thus, {\color{black}less accurate results}. In the numerical results of Section \ref{sec:Numerical_Examples}, we employ the cross-validation approach in \cite[Section 3.5]{Doostan11a} to optimally choose $\gamma$. {\color{black}For the sake of {\color{black}completeness}, this cross-validation approach is summarized in Algorithm \ref{alg:CV}}. For more details about the CS method, {\color{black}the} interested reader is referred to \citep{Doostan11a, Hampton2015}. 

{\color{black}
\begin{algorithm}[H]
\caption{{\color{black}Algorithm for cross-validation estimation of $\gamma$.}} 
\label{alg:CV}
\begin{algorithmic}[1]
{\color{black}
\STATE Divide the $N$ solution samples to $N_r$ reconstruction and $N_v$ validation samples.
\STATE Choose multiple values for $\gamma_r$ such that the exact truncation error $\Vert \bm{\Psi} \bm{\alpha}_{\mathrm{exact}} - \bm{u} \Vert_2 $ of the reconstruction samples is within the range of $\gamma_r$ values.
\STATE For each value of $\gamma_r$ solve the BPDN problem (\ref{eqn:BPDN}) using the $N_r$ reconstruction samples to compute $\bm{\alpha}_r$. 
\STATE For each value of $\gamma_r$, compute the truncation error $\gamma_v := \Vert \bm{\Psi}_v \bm{\alpha}_r -\bm{u}_v \Vert_2$ of the $N_v$ validation samples.  
\STATE Find the minimum value of $\gamma_v$ and and its corresponding $\hat{\gamma}_r := \gamma_r$.
\STATE Set $\gamma = \sqrt{\frac{N}{N_r}}\hat{\gamma}_r $.
}
\end{algorithmic}
\end{algorithm}
}

\subsection{Global sensitivity analysis}
\label{sec:GSA}

{\color{black}Identification of the most important random inputs affecting the variations in the cell voltage, capacity, and concentrations is one of the objectives of this study.} This {\color{black}is} achieved by performing a global sensitivity analysis (SA) to quantify the specific effects of random inputs {\color{black}on} the variance of the QoI. 

Among the available techniques to perform global SA, we use the Sobol' indices \citep{Sobol01} which {\color{black}are} widely used due to their generality and accuracy. Sudret \citep{Sudret08d} introduced an analytic approach to compute the Sobol' indices as a post-processing of the PC coefficients. {\color{black}Let us assume the PC coefficients in (\ref{eqn:RV_PCE}) are computed. The
first order PC-based Sobol' index $S_k$, which represents the sole effects of the random input $\xi_k$ on the variability of $u(\bm{\xi})$, is given by}
\begin{equation}
\label{eqn:1st_order_sobol}
S_k = \sum_{\bm{i} \in \mathscr{I}_{k}} {\color{black}\alpha_{\bm{i}}^2}/\mathrm{var}[u], \quad \mathscr{I}_{k} = \lbrace \bm i \in \mathbb{N}_0^d: {i}_k > 0,  {i}_{m \neq k} = 0 \rbrace,
\end{equation}
where $\mathrm{var}[u]$ is given in (\ref{eqn:var_PCE}). {\color{black} In computing $S_k$, it is assumed that all random inputs except $\xi_k$ are fixed, therefore, $S_k$ does not represent the effects {\color{black}of} the interactions between $\xi_k$ and other random inputs. In order to quantify the total effects of {\color{black}the} random input $\xi_k$, including the interactions between random inputs} on the variability of $u(\bm{\xi})$, one needs to compute the total PC-based Sobol' indices defined as
\begin{equation}
\label{eqn:total_order_sobol}
S_k^T = \sum_{\bm{i} \in \mathscr{I}_{k}^T} {\color{black}\alpha_{\bm{i}}^2}/\mathrm{var}[u], \quad \mathscr{I}_{k}^T = \lbrace \bm i \in \mathbb{N}_0^d: {i}_k > 0 \rbrace. 
\end{equation}

The smaller $S_k^T$, the less important random input $\xi_k$. For the cases when $S_k^T \ll 1$, $\xi_k$ is considered as insignificant and may be replaced by {\color{black}its mean} value without {\color{black}considerable} effects on the variability of $u(\bm{\xi})$. In this study, we employ $S_k^T$ as a measure to identify the most important random inputs of {\color{black}the LIB model considered}. 

\section{{\color{black}Uncertainty in LIB model parameters}}
\label{sec:rv_LIB}

The model parameters of {\color{black}LIBs} are measured experimentally {\color{black}and are accompanied by uncertainty due to natural or experimental variability}. Some of these parameters are measured using {\color{black}complex} electrochemical techniques{\color{black},} while {\color{black}others are} obtained via simple experiments. In the following, we will discuss {\color{black}a number of such techniques}. It should be noted that because of the limited data available in the literature, we could not avoid making assumptions {\color{black}on} the {\color{black}probability distribution for} some of {\color{black}the parameters. Additionally, the reported uncertainty models are specific to} the LiC$_6$/LiCoO$_2$ cell we consider here. 

\subsection{Porosity, $\epsilon$}
\label{sec:porosity}
Porosity is defined as the ratio of the pore volume to the bulk volume. Manufacturers choose the porosity as a trade off between power and energy, i.e., {\color{black}the} higher the porosity, {\color{black}the} higher the power and {\color{black}the} lower the capacity \cite{Shearing10}. 
There are several methods to measure the porosity, {\color{black}such as} the Method of Standard Porosimetry (MSP) \cite{DuBeshter14}, porosity measurement using liquid or gas absorption methods according to the American Society for Testing and Materials (ASTM) D-2873 \cite{Zhang07}, and X-ray tomography \cite{Chung14}.

{\color{black}In \cite{Shriram12}, a uniform distribution [0.28, 032] is considered for the porosity of LiCoO$_2$ {\color{black}based on experimental data}, which has $\pm 6.7\%$ (of the mean) variation around the mean. DuBeshter et al. \cite{DuBeshter14} reported $\pm 4.5\%$ variability for porosity of the graphite anode electrode, while a higher value of $\pm12.6\%$ is {\color{black}reported} for the separator. Hence, in the present study, we assume a uniform distribution for the porosity with $\pm6.7\%$, $\pm4.5\%$, and $\pm12.6\%$ variation around the mean in LiCoO$_2$ cathode, LiC$_6$ anode, and separator, respectively. Defining the coefficient of variation (COV) as the ratio of the standard deviation to the mean, the assumed variations translate to COVs of 0.026, 0.073, and 0.038 in anode, separator, and cathode, respectively.}

\subsection{Solid particle size, $r_s$}
\label{sec:porosity}

{\color{black}The flux of Li$^+$ at the electrode-electrolyte interface} is affected by {\color{black}the solid particle size in electrodes}, as it defines the available surface area for the reaction. The maximum battery power may be increased by decreasing the particle size of the electrode material and increasing the
surface area {\color{black}per volume}. 

{\color{black}The particle size distribution} may be measured by a laser diffraction and scattering device \cite{Chou08}, X-ray computed tomography (XCT) \cite{Izzo08}, or focused ion beam tomography \cite{Shearing09}.

Santhanagopalan and White \cite{Shriram07} considered a normal distribution for {\color{black}a} graphite anode with the mean and standard deviation of 6.2 $\mu$m and 0.42 $\mu$m, respectively, to quantify the effects of random particle size. For the particle size distribution of LiCoO$_2$, a normal distribution with mean of 7.7 $\mu$m and standard deviation of approximately 1.5 $\mu$m is reported in \cite{malvern}. The corresponding COV values are 0.0677 and 0.1948 for the graphite anode and LiCoO$_2$ cathode, respectively. Nominal values, i.e., mean, of the particle sizes in both electrodes are equal to 2.0 $\mu$m in our cell. {\color{black}Because of the limited data available in the literature on the particle size distribution of our specific electrodes}, we use these COVs to find the corresponding standard deviations. Based on these COVs, we assume a normal distribution with a mean of 2.0 $\mu$m and standard deviation of 0.1354 $\mu$m for the graphite anode. For the cathode electrode, {\color{black}the} same mean value {\color{black}but} a standard deviation of 0.3896 $\mu$m is considered.   

\begin{remark1}
When one draws samples from a Gaussian distribution for the solid particle size $r_s$, {\color{black}one} should assure that all the $r_s$ samples {\color{black}are} strictly positive. {\color{black}For this purpose}, we {\color{black}here} employed a truncated Gaussian distribution bounded between {\color{black}mean} $ \pm 3$ standard deviation {\color{black}of} $r_s$. 
\end{remark1}
\subsection{Bruggeman coefficient, $\mathrm{brugg}$}
\label{sec:brugg}

In porous electrode theory, instead of specifying the exact position and shapes of all pores and particles, a volume-averaged formulation is used \cite{Thomas02}. In this model, the effective transport properties of the liquid phase, $D^{\mathrm{eff}}$ and $\kappa^{\mathrm{eff}}$, are obtained using the porosity $\epsilon$ and tortuosity $\tau$ via

\begin{equation} 
\label{eqn:eff_D_kappa}
\kappa^{\mathrm{eff}} = \frac{\epsilon \kappa}{\tau}, \qquad D^{\mathrm{eff}} = \frac{\epsilon D}{\tau}.
\end{equation}
Tortuosity $\tau$ is a {\color{black}geometric} parameter and depends on the porous electrode structure \cite{Thorat09}. Although in recent years researchers have attempted to measure {\color{black}$\tau$} via experimental techniques \cite{Thorat09, Ebner14, DuBeshter14, Shearing10}, because of the experimental complexities, {\color{black}$\tau$ has been commonly used} as {\color{black}a model parameter that is calibrated by experiments} \cite{Doyle96b}. More precisely, {\color{black}$\tau$} is computed via the well-known Bruggeman relation 
\begin{equation} 
\label{eqn:brugg_tor}
\tau = \epsilon^{(1-\mathrm{brugg})},
\end{equation}
where the Bruggeman exponent $\mathrm{brugg}$ is chosen to {\color{black}match} numerical results with experimental data, and is usually assumed to be 1.5 \cite{Thorat09}.

Since the Bruggeman relation is widely used in LIB simulations, instead of taking {\color{black}$\tau$} as a random input parameter, we choose the Bruggeman exponent $\mathrm{brugg}$ to be an uncertain parameter in determining the effective properties. In \cite{DuBeshter14}, a uniform distribution with $\pm4.9\%$ of the mean variation around the mean is reported for the Bruggeman exponent of the LiC$_6$ anode, while $\pm33.3\%$ is considered for the {\color{black}Bruggeman exponent in separator} $\mathrm{brugg}_s$ \cite{Shriram12}. For this study, {\color{black}we assume a uniform distribution for the Bruggeman exponent with $\pm5.0\%$ variations around the mean in both electrodes and $\pm20.0\%$ in separator, which corresponds to COVs of 0.029 and 0.115, respectively.}

\subsection{Li$^+$ transference number, $t_+^0$}
\label{sec:tn}
When {\color{black}an LIB} is discharged, electrolyte salt dissociates into Li$^+$ and PF$_6^-$ ions. The {\color{black}portion} of the current that is carried by Li$^+$ ions is called the Li$^+$ transference number. {\color{black}The} higher the Li$^+$ transport number, {\color{black}the} higher the battery power {\color{black}{\cite{Zugmann20113926}}}. 

{\color{black}Hittorf's} method \cite{Bard80}, ac impedance spectroscopy \cite{MACDONALD87}, or pulsed field gradient NMR (pfg-NMR) technique \cite{Ferry98} {\color{black}are} used to measure the transference number in electrolytes. The experimental error of pfg-NMR technique for binary solutions is estimated to be around 5.0\% \cite{Ole05}. Hence, we let the transference number uniformly change between 0.345 and 0.381 in this study. 

\subsection{Salt diffusion coefficient in the liquid phase, $D$}
\label{sec:dce}
The salt diffusion coefficient {\color{black}$D$} is a measure of the friction forces between the ions and the solvents \cite{Nyman11}. In order to restrict the performance-limiting salt concentration gradients{\color{black},} which form in the electrolyte during polarization, it is critical to have {\color{black}a} high salt diffusion coefficient \cite{Nazri03}. High {\color{black}values of $D$} {\color{black}lead} to higher battery power. Experimental methods such as cyclic voltammetry (CV) \cite{Yang14} and electrochemical impedance spectroscopy (EIS) \cite{Qing01} have been used to measure {\color{black}$D$}. 

Salt diffusion coefficient is usually reported as a constant, but in \cite{Ole05} it is assumed {\color{black}to be} a function {\color{black}of temperature and salt concentration. In this study, we assume a uniform distribution for $D$ with $\pm10.0\%$ variation around the mean, which corresponds to COV of 0.0577.}

\subsection{Diffusion coefficient of the solid, $D_s$}
\label{sec:dcs}

Diffusion coefficient of the solid phase plays an important role in the performance of LIBs since {\color{black}it affects the intercalation flux}. Electrochemical techniques such as CV, GITT and EIS have been used to determine $D_s$ \cite{Xie08}. $D_s$ {\color{black}has} been mostly treated as a constant value in {\color{black}LIB} simulations \citep{Fuller94}, {\color{black}while} it has been shown that $D_s$ depends on the intercalation level, {\color{black}i.e., the ratio of $c_s^{\mathrm{surf}}/c_{s,max}$} \cite{Levi97}. The values of $D_s$ for the same materials, reported by different research groups, may differ by several orders of magnitude.  {\color{black}These large differences may suggest that a Fickian diffusion model which is used to calibrate $D_s$ does not correctly describe the transport of Lithium within the active particles and other transport models need to be considered. }

{\color{black}In this study, we assume that $D_s$ does not depend on the intercalation level. Hence, for $D_s$, we assume a uniform distribution with the same COV we assumed for $D$.} 

\subsection{Electronic conductivity of the solid, $\sigma$}
\label{sec:csp}
Capacity of LIBs may be improved by increasing the solid phase electronic conductivity {\color{black}by using} conductive additives {\color{black}in} the electrode materials \cite{Chen10}. Two-point and four-point probe techniques have been used to measure the electronic conductivity of electrode materials \cite{Thorat11}. Although $\sigma$ depends on temperature and state of the charge \cite{Sauvage07}, it is mostly treated as a constant in LIB simulations. 

Reported values of $\sigma$ in the literature for LiC$_6$ anode are mostly around 100 $\mathrm{S} \cdot \mathrm{m^{-1}}$ \citep{Subramanian09a, Doyle03, Fuller94}, while the electronic conductivity of LiCoO$_2$ cathode is reported to be 100 $\mathrm{S} \cdot \mathrm{m^{-1}}$ \citep{Subramanian09a} and 10 $\mathrm{S} \cdot \mathrm{m^{-1}}$ \citep{Doyle03}. In our LIB model, the nominal values of $\sigma$ is equal to 100 $\mathrm{S} \cdot \mathrm{m^{-1}}$ for both electrodes. {\color{black}Because of the limited available data in the literature on the measurement uncertainties in $\sigma$, we assume that in both electrodes it changes uniformly with $\pm10.0\%$ variation.}

\subsection{Reaction rate constant, $k$}
\label{sec:rrc}

Exchange current density $i_{ex}$ is measured at the initial state of the battery \cite{Piao99}. Using the relation $i_{ex} = Fk(c_s^{\mathrm{surf}})^{0.5}(c_{s,max} - c_s^{\mathrm{surf}})^{0.5}(c)^{0.5}$, one can calculate the reaction rate constant $k$ for each electrode at the initial values of $c_{s}^{\mathrm{surf}}$ and $c$. Although $k$ shows an Arrhenius {\color{black}type} dependence on temperature and depends on the nature of the electrode surface \cite{Newman04}, it is usually treated as a constant in LIB simulations. High{\color{black}er} reaction rate constants are favored for Li-ion batteries since the reaction is more reversible{\color{black},} and polarization effects are {\color{black}lower}.

Because of the lack of experimental data on the reaction rate constant, we assumed that $k$ has a uniform distribution with $\pm10.0\%$ variation around the mean in both electrodes. 

A list of random parameters considered in this study is presented in Table \ref{table:random_parameters_battery}. In addition to the discussed input parameters, we also took the lengths {\color{black}$L$} of electrodes and separator to be random in order to study the effect of uncertainties in geometrical parameters of the battery. It also should be noted that in our battery model, the effective ionic conductivity of the liquid phase $ \kappa^{\mathrm{eff}}$ is considered to be a function of liquid concentration. Hence, $ \kappa^{\mathrm{eff}}$ will be automatically a random parameter since the porosity is considered to be random. We note that the anodic and cathodic transfer coefficients (considered to be 0.5 in this study) as well as the open circuit potentials are also subjected to uncertainties; however, we treat them as deterministic parameters in this study.

\begin{table}[H]
\caption{List of random and deterministic LIB inputs used in this study.} 
\centering
\begin{tabular}{| c | c | c |}   
\hline
Random Input & Nominal Value & Distribution \\ 
\hline
$r_{s,a}$ [$\mu$m] & 2 & Gaussian, $\mu=2$, $\sigma=0.1354$ \\ 
\hline
$r_{s,c}$ [$\mu$m] & 2 & Gaussian, $\mu=2$, $\sigma=0.3896$ \\ 
\hline
$\epsilon_a$ & 0.485 & Uniform, [0.46, 0.51] \\ 
\hline
$\epsilon_s$ & 0.724 & Uniform, [0.63, 0.81] \\ 
\hline
$\epsilon_c$ & 0.385 & Uniform, [0.36, 0.41] \\ 
\hline
$\mathrm{brugg}_a$ & 4 & Uniform, [3.8, 4.2] \\ 
\hline
$\mathrm{brugg}_s$ & 4 & Uniform, [3.2, 4.8] \\ 
\hline
$\mathrm{brugg}_c$ & 4 & Uniform, [3.8, 4.2] \\ 
\hline
$t_+^0$ & 0.363 & Uniform, [0.345, 0.381] \\ 
\hline
$D$ [$\mathrm{m^{2}} \cdot \mathrm{s^{-1}}$] & $7.5 \times 10^{-10}$  & Uniform, [6.75, 8.25] $\times 10^{-10}$ \\ 
\hline
$D_{s,a}$ [$\mathrm{m^{2}} \cdot \mathrm{s^{-1}}$] & $3.9 \times 10^{-14}$  & Uniform, [3.51, 4.29] $\times 10^{-14}$ \\ 
\hline
$D_{s,c}$ [$\mathrm{m^{2}} \cdot \mathrm{s^{-1}}$] & $1 \times 10^{-14}$  & Uniform, [0.9, 1.1] $\times 10^{-14}$ \\  
\hline
$\sigma_a$ [$\mathrm{S} \cdot \mathrm{m^{-1}}$] & 100  & Uniform, [90, 110] \\ 
\hline
$\sigma_c$ [$\mathrm{S} \cdot \mathrm{m^{-1}}$] & 100 & Uniform, [90, 110] \\ 
\hline
$k_a$ [$\mathrm{m}^4 \cdot \mathrm{mol} \cdot \mathrm{s}$]& $5.03 \times 10^{-11}$ & Uniform, [4.52, 5.53] $\times 10^{-11}$ \\ 
\hline
$k_c$ [$\mathrm{m}^4 \cdot \mathrm{mol} \cdot \mathrm{s}$] & $2.334 \times 10^{-11}$  & Uniform, [ 2.10, 2.56] $\times 10^{-11}$ \\ 
\hline 
$L_a$ [$\mu$m]&  80 & Uniform, [77, 83]  \\ 
\hline 
$L_s$ [$\mu$m]&  25 & Uniform, [22, 28]  \\ 
\hline 
$L_c$ [$\mu$m]&  88 & Uniform, [85, 91]  \\ 
\hline 
\end{tabular} 
\label{table:random_parameters_battery} 
\end{table}

\section{Numerical Example}
\label{sec:Numerical_Examples}

In this section, we present a numerical example to demonstrate the application of our proposed UQ approach to LIBs. The one-dimensional LiC$_6$/LiCoO$_2$ cell we consider here is studied in \citep{Subramanian09a}. We present our results for three different {\color{black}discharge} rates {\color{black}of} 0.25C, 1C, and 4C to study {\color{black}its effects on the} propagation of uncertainty. {\color{black}We note that assuming a constant discharge rate is an idealized scenario. {\color{black}Typically, the battery loading varies over the course of the discharge process, depending on an application-dependent usage of the battery. Hence, in addition to the previously mentioned uncertain parameters, the discharge rate should also be considered as an uncertain input for the LIB model.} For the sake of comparison, we treat the discharge rate as a deterministic input in this study, while our proposed UQ framework can incorporate random discharge rates with no difficulties, perhaps, with a cost of running additional battery simulations.}

{\color{black}For the spatial discretization of the LIB governing equations, we used the Finite Difference Method on the non-uniform grids described in \citep{Reimers13}. We also performed a mesh convergence analysis to ensure that spatial discretization errors are inconsequential.

\begin{remark1}
Working with a fine mesh and a small time step is advised when one employs sampling-based UQ techniques since a coarse discretization may not return a converged/accurate solution for all samples. In our calculations, we found that having 120 non-uniform grid points in each region and 26 non-uniform grid points in the spherical solid particles are sufficient to obtain accurate realizations for all of the samples. Moreover, constant time steps of 1.0, 0.1, and 0.05 seconds were used for 0.25C, 1C, and 4C rates of discharge, respectively. 
\end{remark1}}

The cell model incorporating the nominal values in Table \ref{table:random_parameters_battery} is referred to as the \textit{deterministic} model, while the \textit{stochastic} model uses the distributed input parameters of Table \ref{table:random_parameters_battery}. The number of random inputs of our stochastic LIB model is $d=19$, which may be considered {\it high}. We use the CS technique in this study to approximate the sparse PC coefficients since{\color{black},} in comparison to least squares regression, CS requires smaller number of battery simulations. 

Having at hand a deterministic LIB solver, the next step in sampling-based PC expansion is to generate $N$ realizations of model parameters listed in Table 3. To this end, we assign an independent random variable $\xi_k$, $k=1,\dots,d$, to each parameter, and consider an appropriate linear transformation of each $\xi_k$ to match the PDF of the corresponding parameter in Table \ref{table:random_parameters_battery}. For Gaussian and uniform PDFs, we use $\xi_k$'s that are standard Gaussians and uniformly distributed between $[-1,1]$, respectively. The independent samples of $\bm\xi$, i.e., $\{\bm\xi^{(i)}\}_{i=1}^{N}$, are used to generate $N$ independent samples of the model parameters, for which the LIB model is simulated to obtain $N$ realizations of the output QoIs, $\{u(\bm\xi^{(i)})\}_{i=1}^{N}$. In our numerical experiments, we simulate each battery realization until a cut-off potential of 2.8 V is reached. Then using  $\{{\bm{\xi}}^{(i)}\}_{i=1}^{N}$ and $\{u({\bm{\xi}}^{(i)})\}_{i=1}^{N}$, we solve the BPDN problem in (\ref{eqn:BPDN}) to approximate the vector of PC coefficients $\bm{\alpha}$. These in turn will be used to compute the statistics of $u({\bm{\xi}})${\color{black},} such as the mean and variance given in (\ref{eqn:mean_PCE}) and (\ref{eqn:var_PCE}), respectively, as well as the total Sobol' indices defined in (\ref{eqn:total_order_sobol}). 

We find that a PC expansion of order $p=3$ and $N=1000$ battery simulations are needed to achieve a {\it validation error} smaller than 1.0\% as specified next. Following the cross-validation procedure described in \citep[Section 3.5]{Doostan11a}, we divide the $N=1000$ samples {\color{black}into} $N_r=900$ reconstruction and $N_v=100$ validation samples to estimate the optimum value of $\gamma$ in (\ref{eqn:BPDN}), using which we compute the solution $\bm \alpha$ from (\ref{eqn:BPDN}). To verify the accuracy of the resulting PC expansion, we compute the validation error 
\begin{equation}
\label{eqn:rel_error}
\text{relative error}= \frac{\Vert \bm{u}_v - \bm{\Psi}_v \bm{\alpha}_r \Vert_2}{\Vert \bm{u}_v \Vert_2},
\end{equation}
where $\bm{u}_v$ is the vector of $N_v=1000$ additional realizations of QoI (not used in computing $\bm\alpha$) and $\bm{\Psi}_v$ is the measurement matrix corresponding to $\bm{u}_v$. Stated differently, the error in (\ref{eqn:rel_error}) determines the accuracy of constructed PC model in predicting independent QoI realizations. In practice, there is no need to generate additional samples for validation and we only used a {\color{black}larger} number of validation samples to demonstrate the accuracy and robustness of our proposed method.

{\color{black}
\begin{remark1}
\label{remark3}
In practice, we rely on the validation error to estimate the solution accuracy and decide to possibly increase the total order of PC expansion $p$. When the validation error does not reduce by increasing the number of {\color{black}numerically generated LIB} samples for a given expansion order $p$, one may achieve smaller validations errors by increasing $p$ with a cost of increased number of LIB simulations. 
\end{remark1}}
}

{\color{black}Algorithm \ref{alg:CS_main} summarizes the main steps in our proposed UQ framework for LIBs for a fixed PC expansion order $p$.  }

{\color{black}
\begin{algorithm}[H]
\caption{{\color{black}Summary of the main steps in the proposed UQ framework for LIBs.}} 
\label{alg:CS_main}
\begin{algorithmic}[1]
{\color{black}
\STATE Identify uncertain parameters and their probability distributions (Section \ref{sec:rv_LIB}).
\WHILE{the validation error {\color{black}in (\ref{eqn:rel_error})} is larger than a user-defined threshold}
\STATE Generate $N$ realizations of the uncertain parameters {\color{black}based on their PDFs. }
\STATE Perform deterministic LIB simulations to obtain the output QoI (Section \ref{sec:decoupled}). 
\STATE Evaluate the corresponding realization of the PC basis (Section \ref{sec:Non-intrusive PCE}). 
\STATE Perform the cross-validation approach in Algorithm \ref{alg:CV} to optimally estimate $\delta$. 
\STATE Solve the BPDN problem in (\ref{eqn:BPDN}) to approximate the PC coefficients.
\STATE Increase $N$. 
\ENDWHILE
\STATE Compute the statistics of the output QoI via (\ref{eqn:mean_PCE}) and (\ref{eqn:var_PCE}) and the total Sobol' indices via (\ref{eqn:total_order_sobol}).
}
\end{algorithmic}
\end{algorithm}
}

\subsection{Results}
\label{sec:Numerical_Examples_results}

\subsubsection{Effects of input uncertainties on cell capacity}
\label{sec:Numerical_Examples_capacity}

Capacity is defined as the available energy stored in a fully charged LIB and is one of the most important factors affecting the battery performance. However, effects of {\color{black}LIB model} uncertainties on capacity estimation {\color{black}are not yet fully explored}.  

\begin{figure}
\hspace{-0.6in}
\subfloat[]{\includegraphics[width = 3.2in]{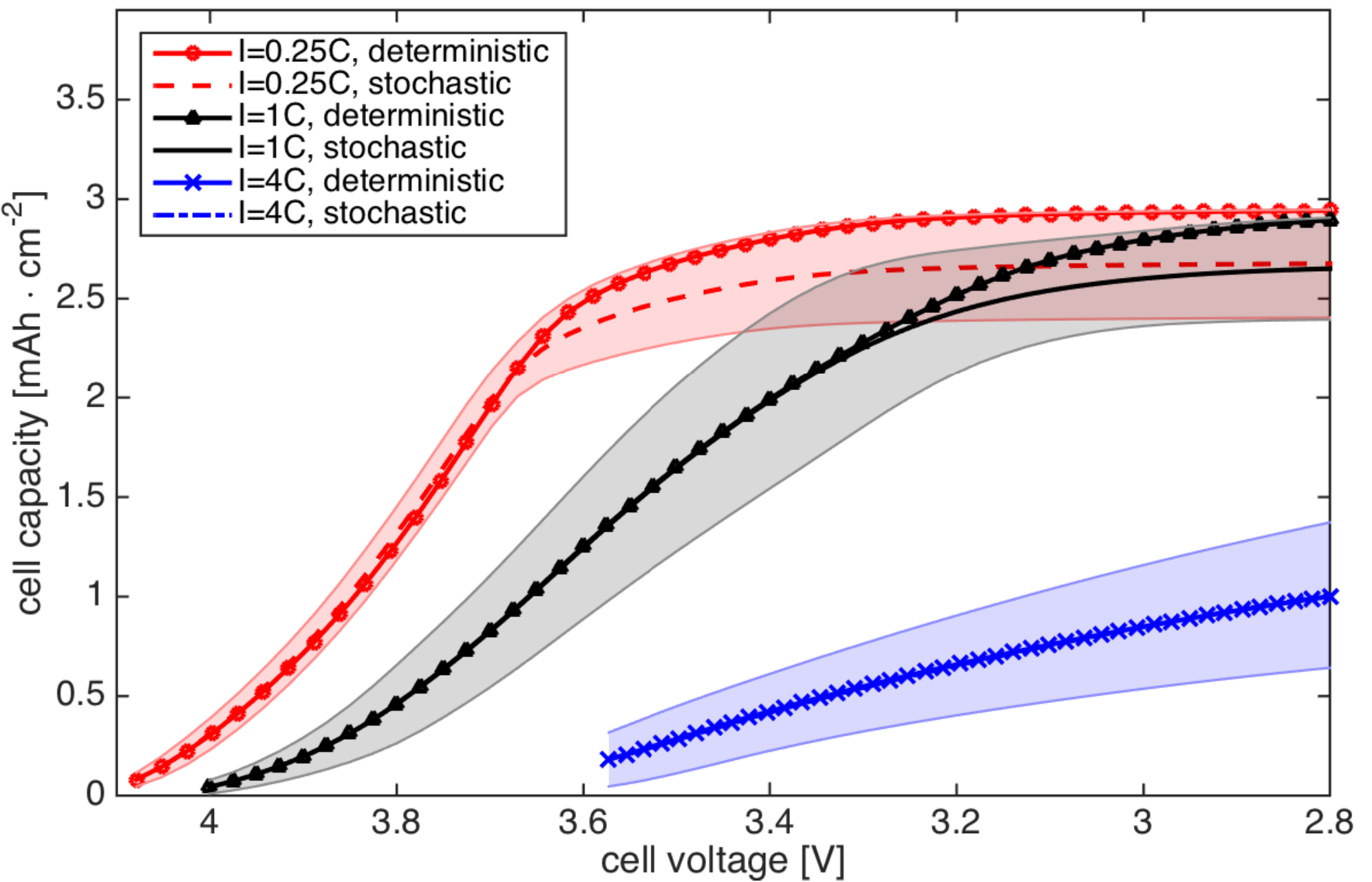}} 
\hspace{0.1in} 
\subfloat[]{\includegraphics[width = 3.2in]{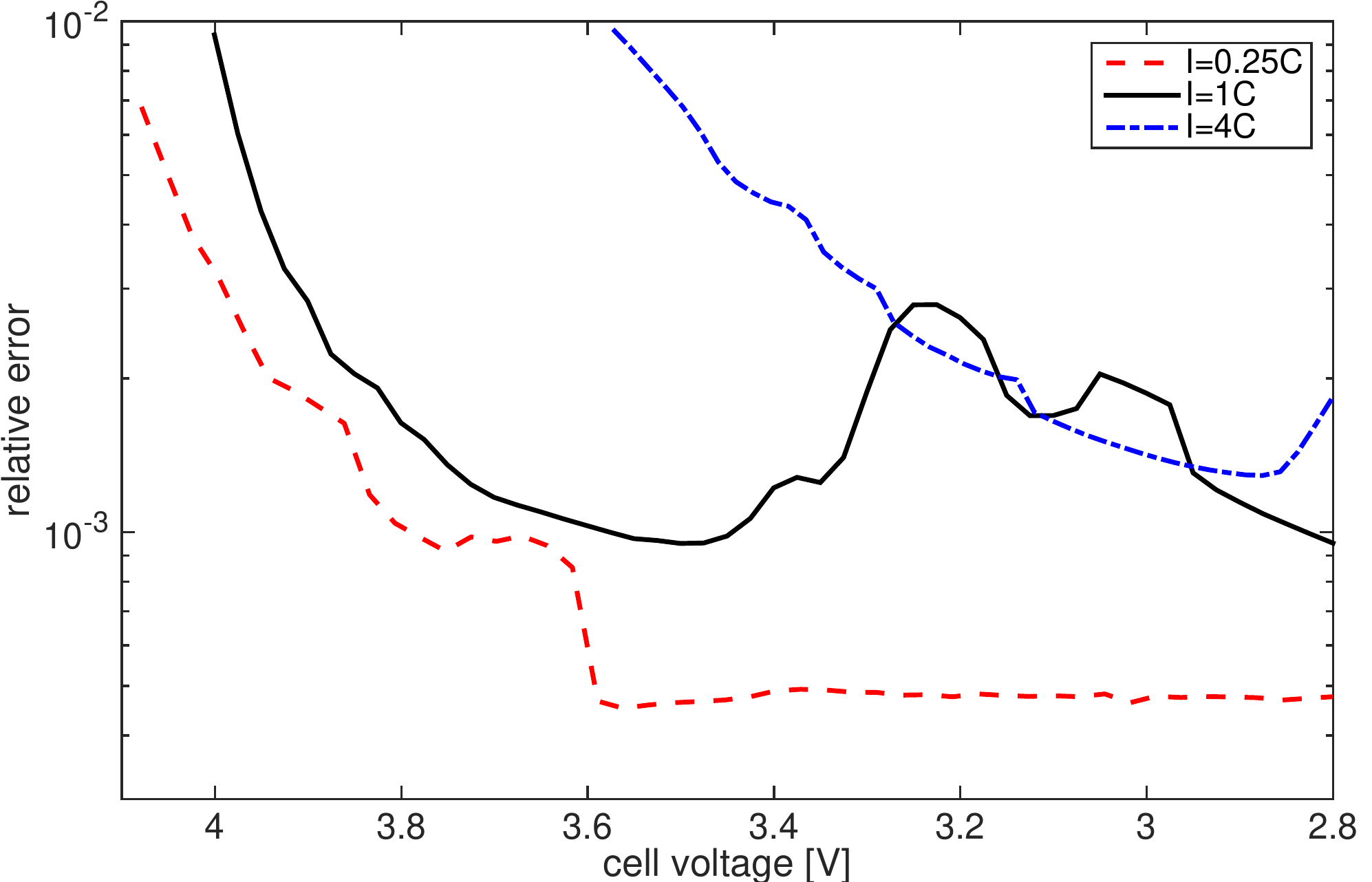}} \par 
\hspace{-0.6in} 
\subfloat[]{\includegraphics[width = 3.2in]{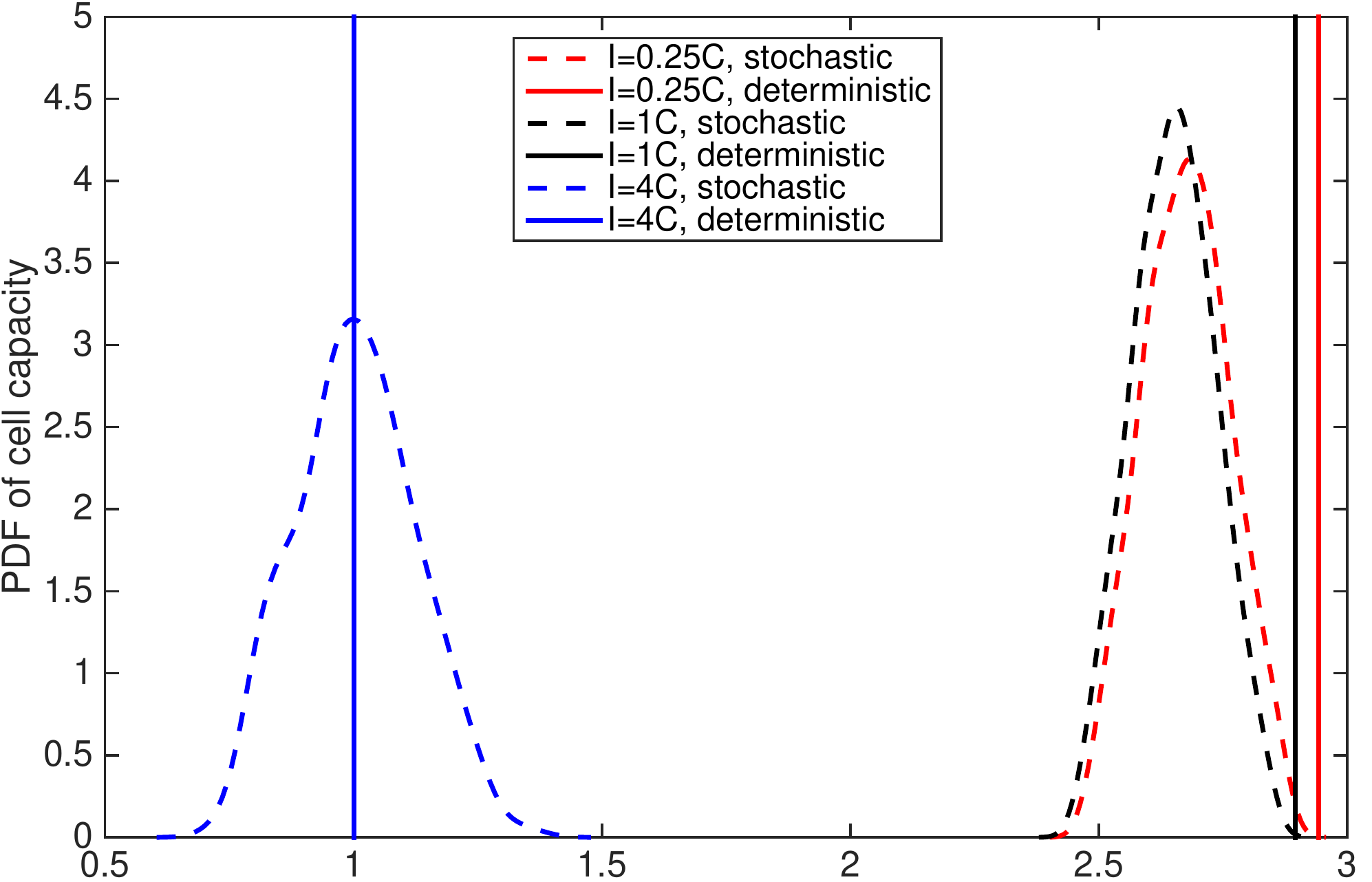}} 
\hspace{0.1in} 
\subfloat[]{\includegraphics[width = 3.2in]{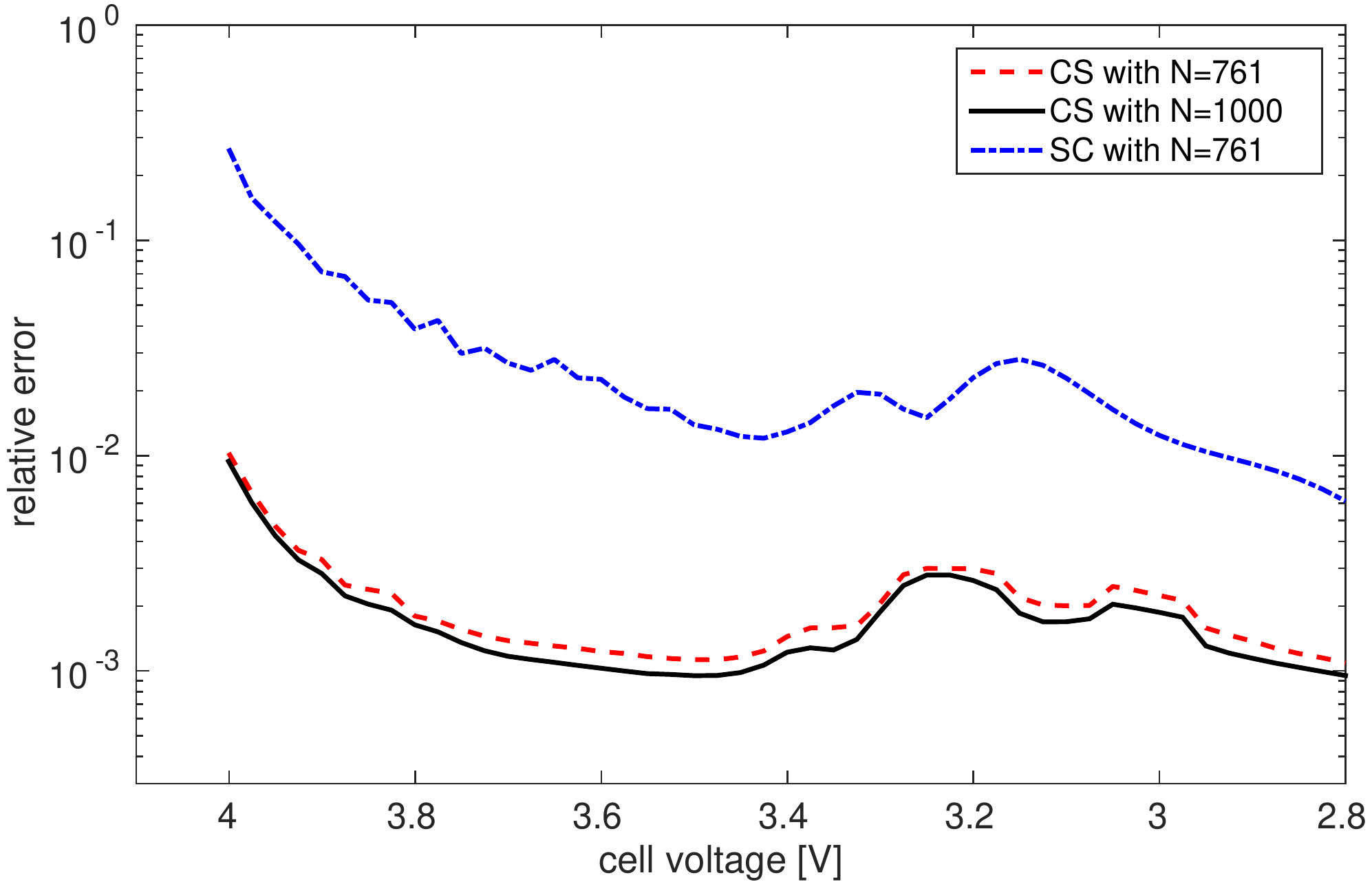}} 

\caption{Comparison of stochastic and deterministic battery models for $I=$ 0.25C, 1C and 4C rates of galvanostatic discharge. (a) Stochastic and deterministic cell capacities as a function of the cell voltage. The shaded areas are probability bounds of three standard deviations around the mean; (b) Relative error as a function of the cell voltage; (c) PDFs of the cell capacity; (d) Comparison of the relative error obtained by the CS method and the level two SC technique on a Clenshaw-Curtis grid \cite{Nobile2008} for 1C rate of discharge. Level three SC method requires 9976 LIB simulations.}
\label{fig:mean_sd_er_cap_V}
\end{figure}

Fig. \ref{fig:mean_sd_er_cap_V} demonstrates the effects of LIB model uncertainties in estimating the cell capacity. In Fig. \ref{fig:mean_sd_er_cap_V}(a), cell capacities as a function of the cell voltage obtained from deterministic and stochastic models are presented for 0.25C, 1C and 4C rates of galvanostatic discharge. Shaded areas around the mean of stochastic cell capacity represent three standard deviation {\color{black}intervals}. As it can be seen, {\color{black}noticeable} deviation from the deterministic capacity starts when the cell voltage experiences a rapid drop-off. {\color{black}As the discharge rate is decreased}, the onset of this {\color{black}deviation} happens at higher cell voltages, i.e., at $\phi_{\mathrm{cell}} \approx 3.7$ V for $I=$ 0.25C and $\phi_{\mathrm{cell}} \approx 3.3$ V for $I=$ 1C. {This plot shows that for low to medium rates of discharge, a deterministic simulation of LIB overestimates the cell capacity. For} 4C discharge rate, the mean of stochastic cell capacity overlaps with the cell capacity predicted by the deterministic model. One may suspect that at high discharge rates, input uncertainties have no significant effect on the cell capacity, but the probability bounds around the mean of cell voltage for $I=$ 4C suggest otherwise. As it can be seen, the largest standard deviation of the cell capacity at the end of discharge {\color{black}is for the 4C discharge rate.} 

Fig. \ref{fig:mean_sd_er_cap_V}(b) shows the validation error in approximating the PC coefficients of the cell capacity. With $p=3$ and $d=19$, the number of PC coefficients $P=\vert \mathscr{I}_{19,3}\vert=1540$, which is larger than $N=1000$. The accurate solution approximation in this under-sampled setting is achieved by minimizing the $l_1$ norm of $\bm\alpha$ in the BPDN problem (\ref{eqn:BPDN}). 

A comparison of deterministic (solid line) and distributed (dashed line) cell capacities is also presented in Fig. \ref{fig:mean_sd_er_cap_V}(c). This plot suggests that the probability of achieving deterministic cell capacity when the cell is subjected to assumed input uncertainties is very small for low to medium discharge rates, while this probability is considerably larger for higher discharge rates.

{\color{black}
An alternative sampling-based approach to compute the PC coefficients is the stochastic collocation (SC) method \cite{Mathelin03, Xiu05a, Babuska05R}. The main idea behind the SC technique is to sample the output QoI at particular points in the stochastic space and then approximate the solution via interpolation or the solution statistics by numerical integration. For high-dimensional problems, sparse tensor products first introduced by Smolyak \cite{Smolyak63}, which may be built upon the Clenshaw-Curtis abscissas, has been proposed to relax the \textit{curse-of-dimensionality} associated with full tensor products \cite{Nobile2008}. The accuracy of this method is controlled by the so-called level parameter. {\color{black}The number of required LIB simulations increases with the magnitude of the level parameter.} SC may also be used to compute the PC coefficients via the projection equation $\alpha_{\bm{i}} = \mathbb{E}[u(\cdot)\psi_{\bm i}(\cdot)]/\mathbb{E}[\psi_{\bm i}^2]$. In particular, the numerator in this equation is a $d$-dimensional integral over $\Omega$ which may be computed using tensor product quadrature or cubature rules. Detailed description of these techniques is beyond the scope of this paper and the interested reader is referred to the provided references. 

In this study, we employ the SC technique with sparse tensor products based on Clenshaw-Curtis abscissas to compare with the proposed CS approach. For the level parameter of two, one needs 761 LIB simulations in the SC method. Fig. \ref{fig:mean_sd_er_cap_V}(d) compares the validation error in approximating the PC coefficients of the cell capacity obtained by SC and CS methods for the cell discharged at 1C rate. As it can be seen, CS with 761 samples results in the validation errors that are smaller (about one order of magnitude) than those reported by the SC technique. Increasing the level parameter of the SC method to three, in order to increase the accuracy, requires 9976 LIB simulations. {\color{black}A comparison between the level two SC and CS approaches confirms the advantage of our proposed UQ framework, hence, we did not perform level three SC due to the high computational costs}. Since we use random samples in CS, unlike in the SC technique, the number of samples are not dictated by the method and one may freely choose a minimum number of additional LIB simulations to achieve the desired accuracy. 
}

\begin{figure}
\hspace{-1.0in}
\subfloat[]{\includegraphics[width = 3.7in]{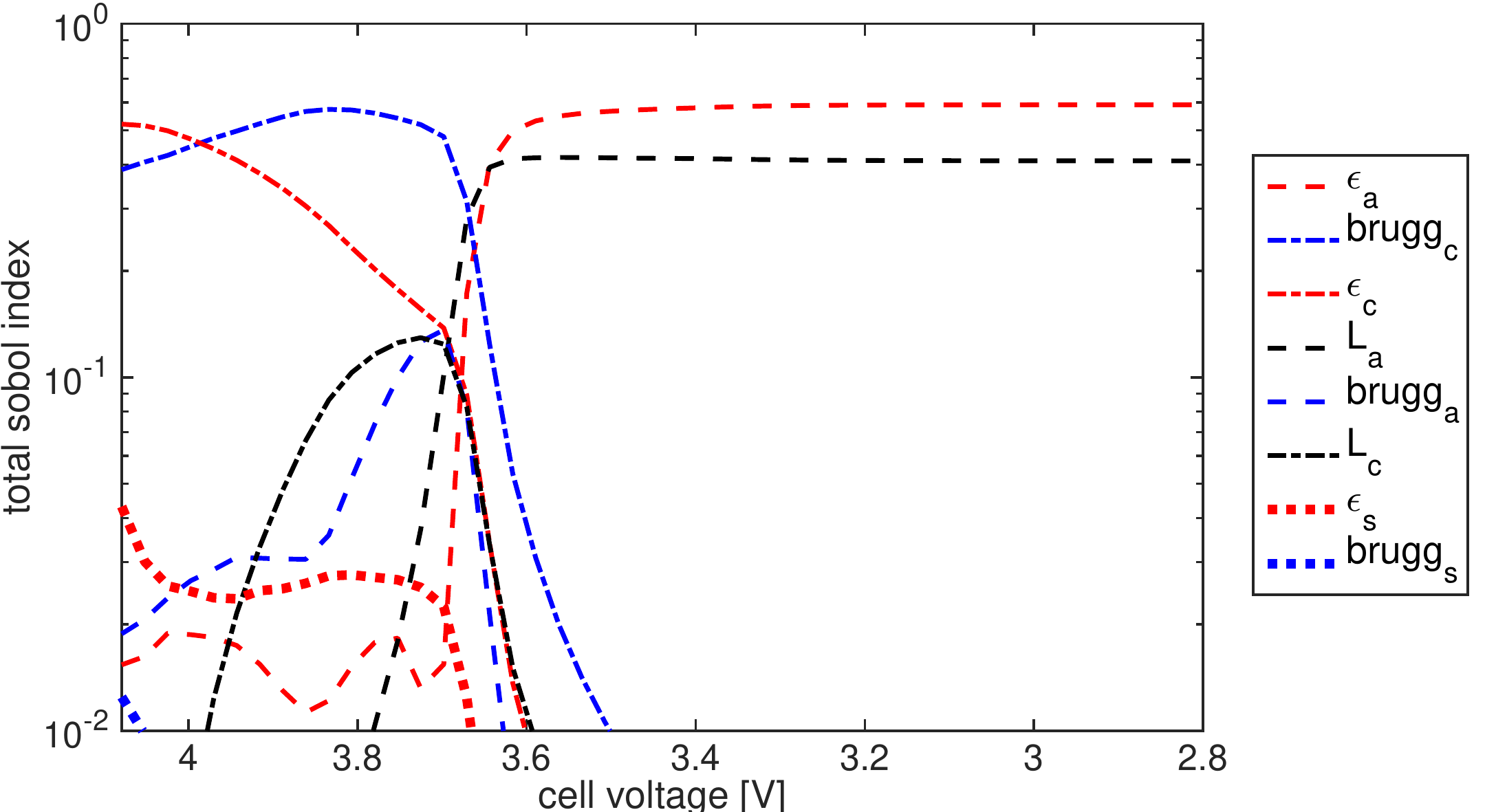}} 
\subfloat[]{\includegraphics[width = 3.7in]{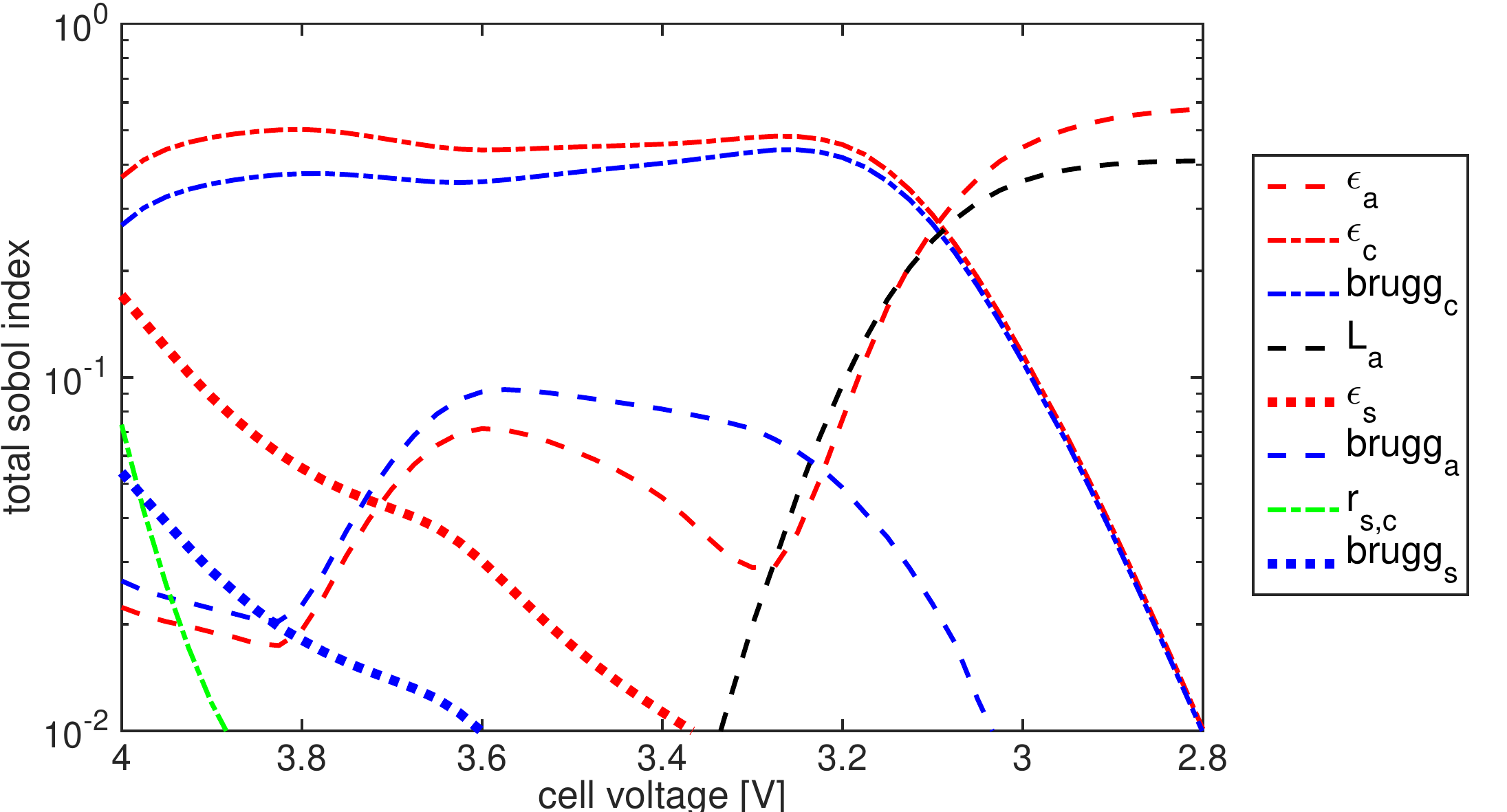}} \\
\hbox to 23.0mm{}
\subfloat[]{\includegraphics[width = 3.7in]{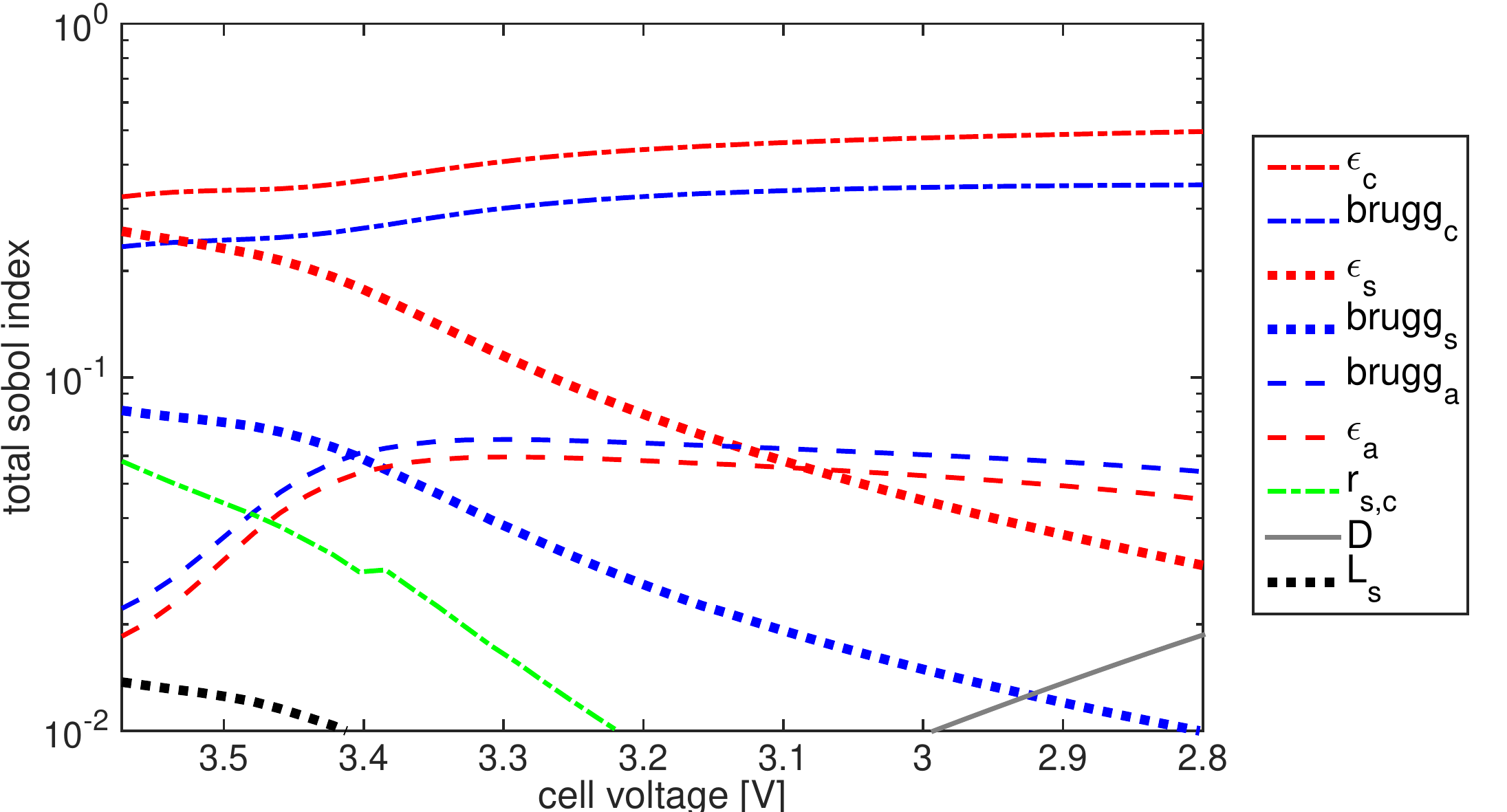}}
\caption{Global sensitivity analysis of the cell capacity for various discharge rates: (a) $I=$ 0.25C; (b)  $I=$ 1C; (c) $I=$ 4C.}
\label{fig:sobol_cap_V}
\end{figure}

As discussed in Section \ref{sec:GSA}, we compute the total Sobol' indices given in (\ref{eqn:total_order_sobol}) to identify the most important random inputs. We assume that {\color{black}a} random input {\color{black}$\xi_k$} with {\color{black}a maximum total Sobol' index $S_k^T$ smaller than 0.01} has no significant effects on the variability of the output QoI and may be treated as a deterministic input. This selection criteria is problem dependent and may be changed depending on the {\color{black}application requirements}. Fig. \ref{fig:sobol_cap_V} shows the total Sobol' indices associated with the cell capacity over a course of discharge for $I=$ 0.25C, 1C and 4C. In this paper, we only present significant Sobol' indices, i.e., those with $\text{max}(S_k^T) > 0.01$, in the global SA plots. Moreover, {\color{black}the} legends of the Sobol' index plots are sorted such that the first legend corresponds to the largest $S_k^T$, while the last legend represents the smallest Sobol' index. The following observations from  Fig. \ref{fig:sobol_cap_V} {\color{black}are worthwhile highlighting}:

\begin{itemize}
  \item {\color{black}Independent of the discharge rate}, porosity $\epsilon$ and Bruggeman coefficient $\mathrm{brugg}$ in anode, separator, and cathode are among the most important random inputs. In other words, variation in the the cell capacity is highly affected by uncertainty in tortuosity $\tau$. 
  \item For all three discharge rates,  Fig. \ref{fig:sobol_cap_V} shows that $\sigma_a$, $\sigma_c$, $t_+^0$, $D_{s,a}$, $D_{s,c}$, $k_a$, $k_c$, and $r_{s,a}$ are insignificant random inputs and their uncertainties have no important effects on {\color{black}the} variability of the cell capacity. Consequently, expensive and accurate quality control measures for these parameters are not required when one aims at reducing the variations in cell capacity. 
  \item As the discharge rate increases, uncertainties in the length of electrodes, i.e., $L_a$ and $L_c$, become less important{\color{black},} while the effects of variability in the length of separator $L_s$ are more pronounced for $I=$ 4C. 
  \item For $I=$ 4C, {\color{black}the} diffusion coefficient of the liquid phase $D$ is an important random input{\color{black},} while for low to medium rates, its corresponding Sobol' index $S^T_{D}$ is smaller than 0.01. 
  \item By increasing the discharge rate, {\color{black}the} number of important random inputs at the end of discharge increases; two for $I=$ 0.25C, four for $I=$ 1C, and seven for $I=$ 4C. In other words, for high {\color{black}discharge} rate LIB applications, accurate quality control measures are needed for a larger number of LIB parameters. 
  \item For low discharge rates, variability in the solid particle size $r_s$ has no significant effects on the variations of the cell capacity.    
\end{itemize}

In general, Fig. \ref{fig:sobol_cap_V} suggests that in determining the most significant random inputs on the variations of the LIB capacity, considering the discharge rate is crucial. Again, we emphasize {\color{black}that} these {\color{black}observations may change when more accurate LIB models that account for cell degradations and other physical phenomena are employed.}

\begin{figure}
\hspace{-0.6in}
\subfloat[]{\includegraphics[width = 3.2in]{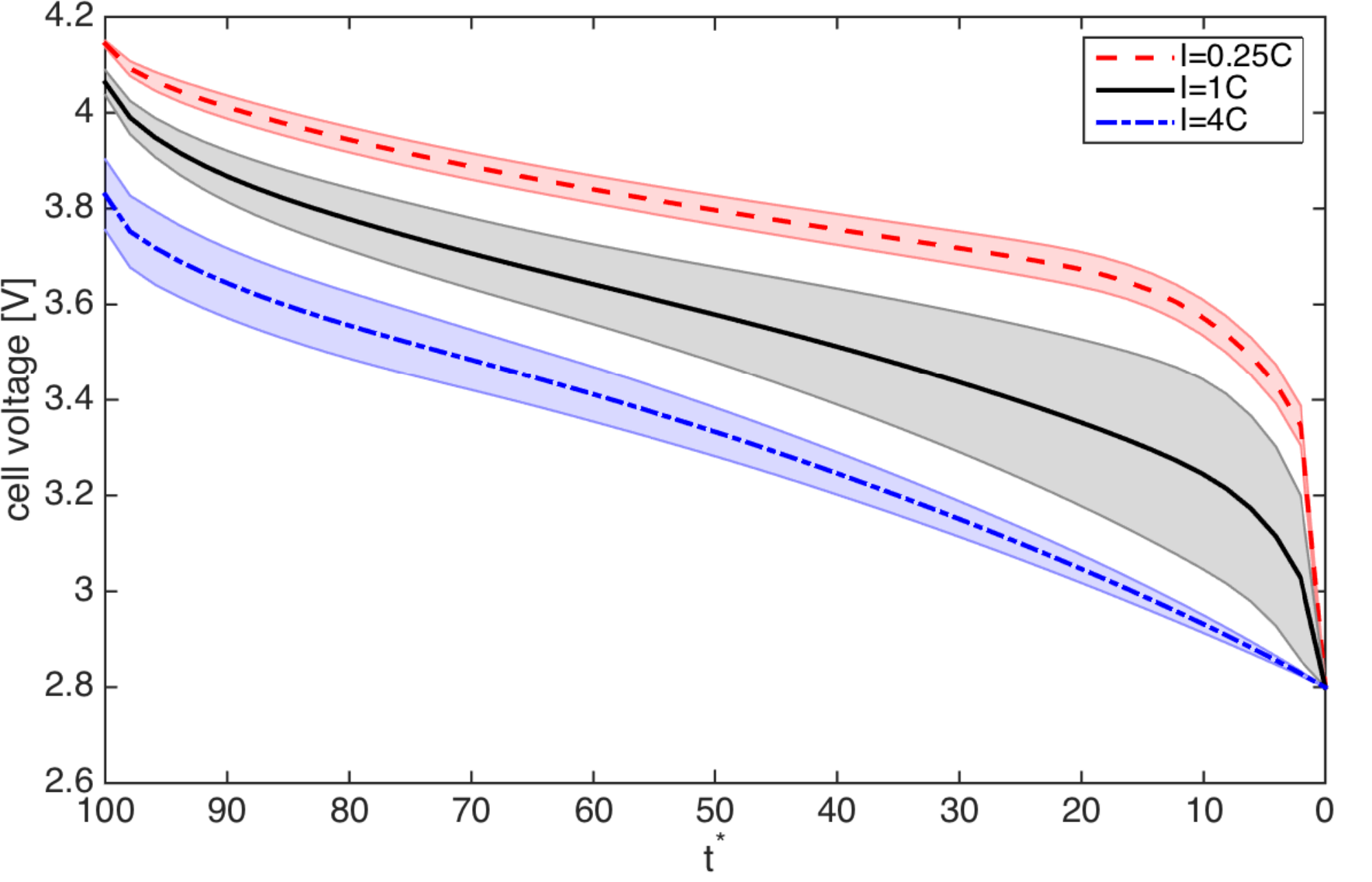}}
\hspace{0.1in} 
\subfloat[]{\includegraphics[width = 3.2in]{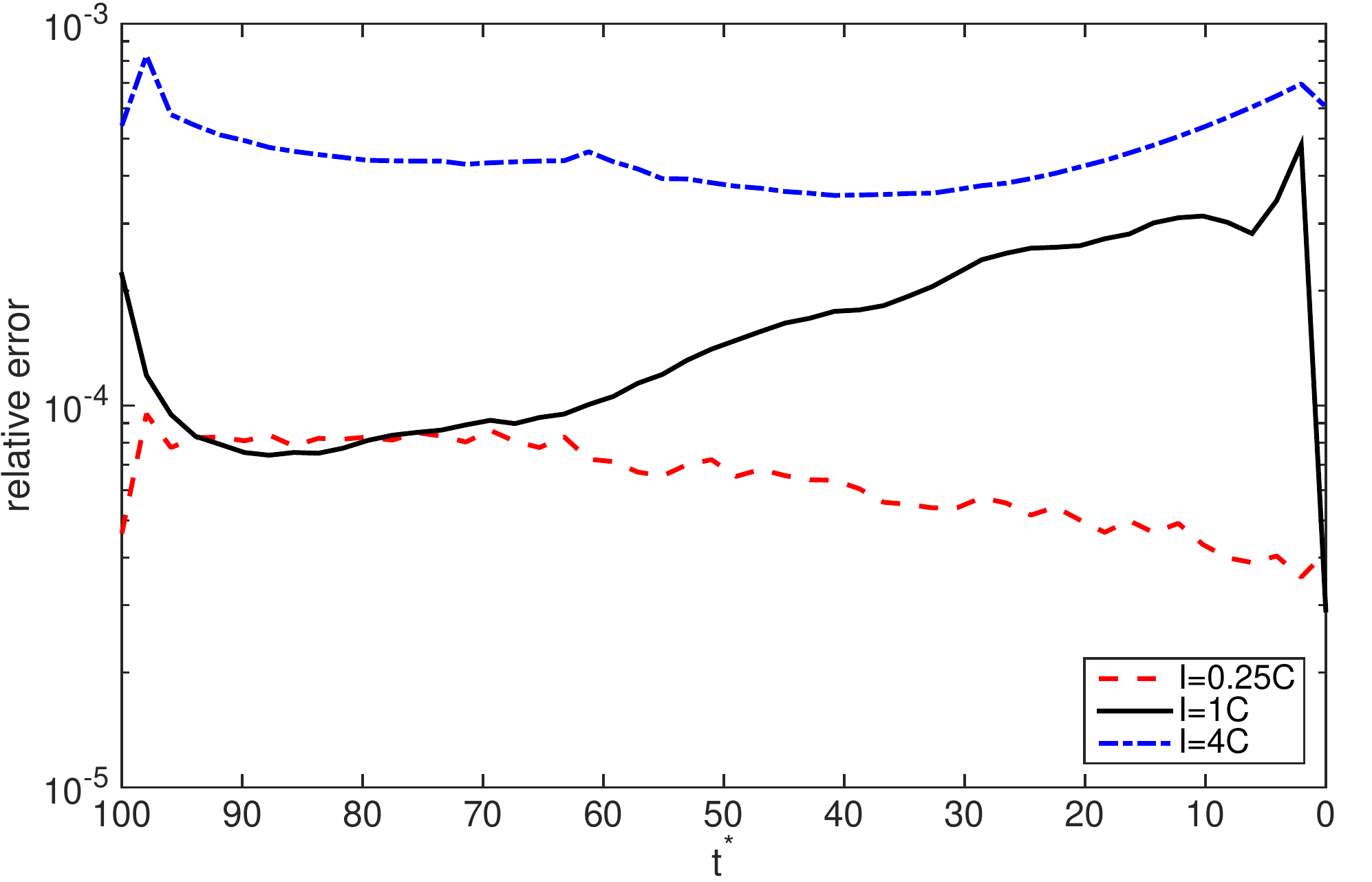}} 
\caption{Effects of input uncertainties on the cell voltage for $I=$ 0.25C, 1C and 4C rates of discharge. (a) Mean and standard deviation of the cell voltage. The shaded areas are probability bounds of three standard deviations around the mean; (b) Relative error.}
\label{fig:V_SOC_mean_sd_er}
\end{figure}

{\color{black}
\subsubsection{{\color{black}Statistics in normalized time}}
\label{sec:soc}
Since the input samples are different, the battery {\color{black}realizations} reach the cut-off potential of 2.8 V at different times. As an example, for 1C rate of discharge, some realizations of the battery reach the cut-off potential before $t=$ 3200 s while other realizations need more time, {\color{black}e.g., $t_{max}=$ 3426 s}. {\color{black}This asynchronous behavior leads to non-smooth dependence of the cell voltage to random inputs and deteriorates the accuracy of PC approximation when the battery approaches the end of discharge.}

To overcome this issue, we introduce an uncertain time {\color{black}scaling} $t^*$. In order to rescale the deterministic realizations, we assume at $t=0$, $t^*$ = 100, and {\color{black}we} set $t^*=0$ when the battery reaches the cut-off potential of 2.8 V. Hence, at $t^*$ = 100 the battery is fully charged and at $t^*$ = 0 it is fully discharged. This rescaling approach enables us to maintain the accuracy of approximation over the entire discharge process without increasing the computational cost or complexity of the problem. {\color{black}We note that our definition of $t^*$ involves a constraint on the cell voltage{\color{black},} such that at $t^*$ = 0 the cell voltage for all realizations is equal to 2.8 V, which translates to zero variability for the cell voltage at $t^*$ = 0. Our $t^*$ definition imposes no constraints on other QoI such as solid and liquid phase concentrations. }

\subsubsection{Effects of input uncertainties on cell voltage and concentrations}
\label{sec:Numerical_Examples_SOC}

{\color{black}Quantification of the effects of input variations on the cell voltage and concentrations over the charge/discharge processes could provide a better understanding of the cell behavior.} In the following, we present the effects of LIB input uncertainties on the stochastic behavior of the cell voltage $\phi_{\mathrm{cell}}$, liquid phase concentration $c$, and solid phase concentration at the surface of the solid particle $c_s^{\mathrm{surf}}$ {\color{black}as functions of the normalized time $t^*$}. In order to study the variations of $c$ in all three main regions of the LIB, we {\color{black}compute} $c$ at three different locations in the cell; middle of anode, separator and cathode. Similarly, $c_s^{\mathrm{surf}}$ is {\color{black}computed} in the middle of electrodes. 

Fig. \ref{fig:V_SOC_mean_sd_er}(a) shows the mean and probability bounds of three standard deviations around the mean of cell voltage as the functions of $t^*$ for $I=$ 0.25C, 1C and 4C. Although at  $t^*$ = 100, larger variabilities in $\phi_{\mathrm{cell}}$ correspond to higher discharge rates, {\color{black}$\phi_{\mathrm{cell}}$} experiences its largest standard deviation at 1C rate of discharge, highlighting {\color{black}again} the importance of discharge rate on UQ analysis of the LIBs. {\color{black}The validation error of the PC solution constructed from $N=1000$ samples is displayed in Fig. \ref{fig:V_SOC_mean_sd_er}(b).} {\color{black}In fact, $N=1000$ samples were enough to accurately approximate $c$ and $c_s^{\mathrm{surf}}$ as well.}
 
\begin{figure}
\vspace{-1.0in}
\hspace{-1.0in}
\subfloat[]{\includegraphics[width = 3.7in]{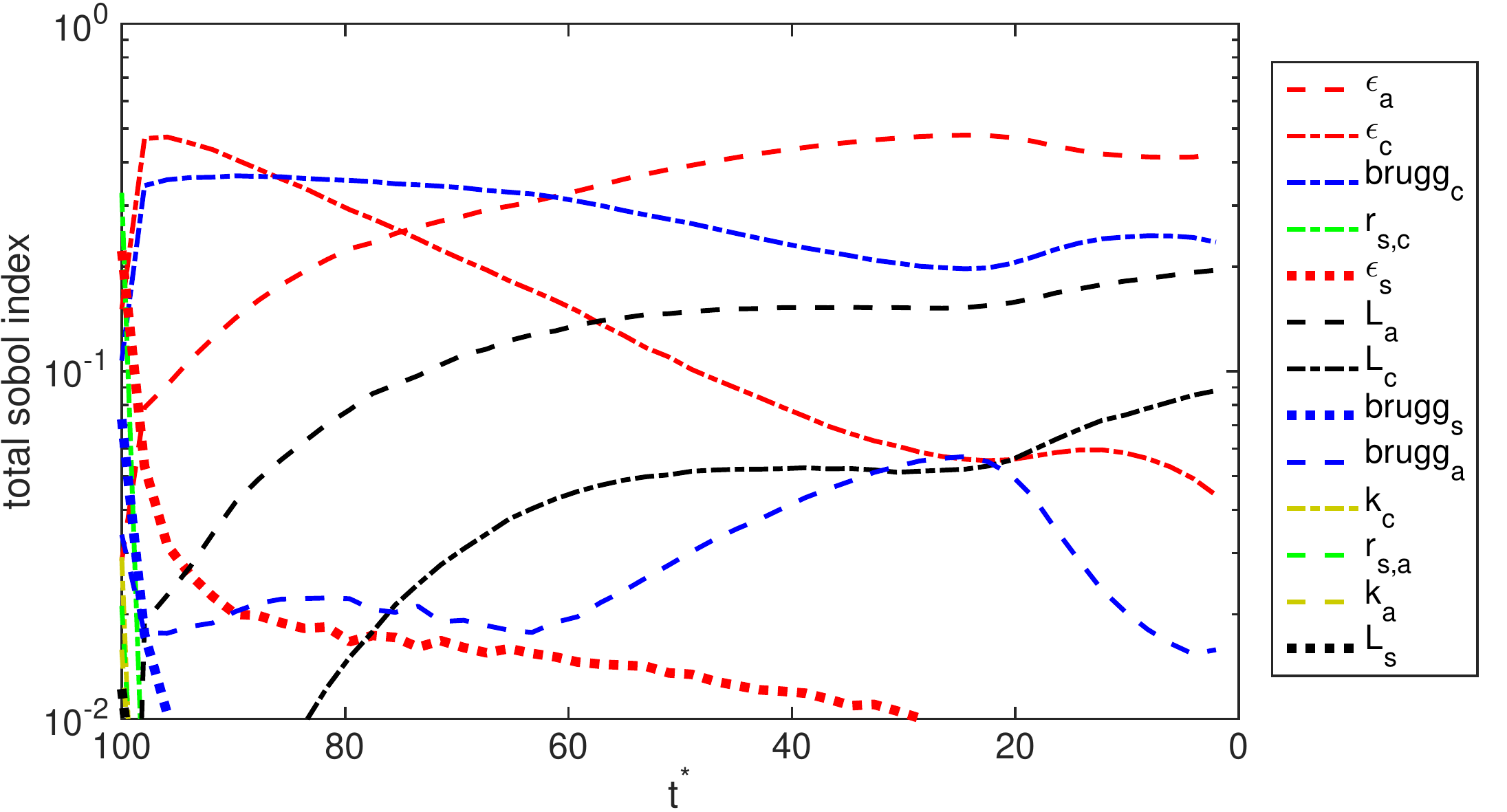}}
\subfloat[]{\includegraphics[width = 3.7in]{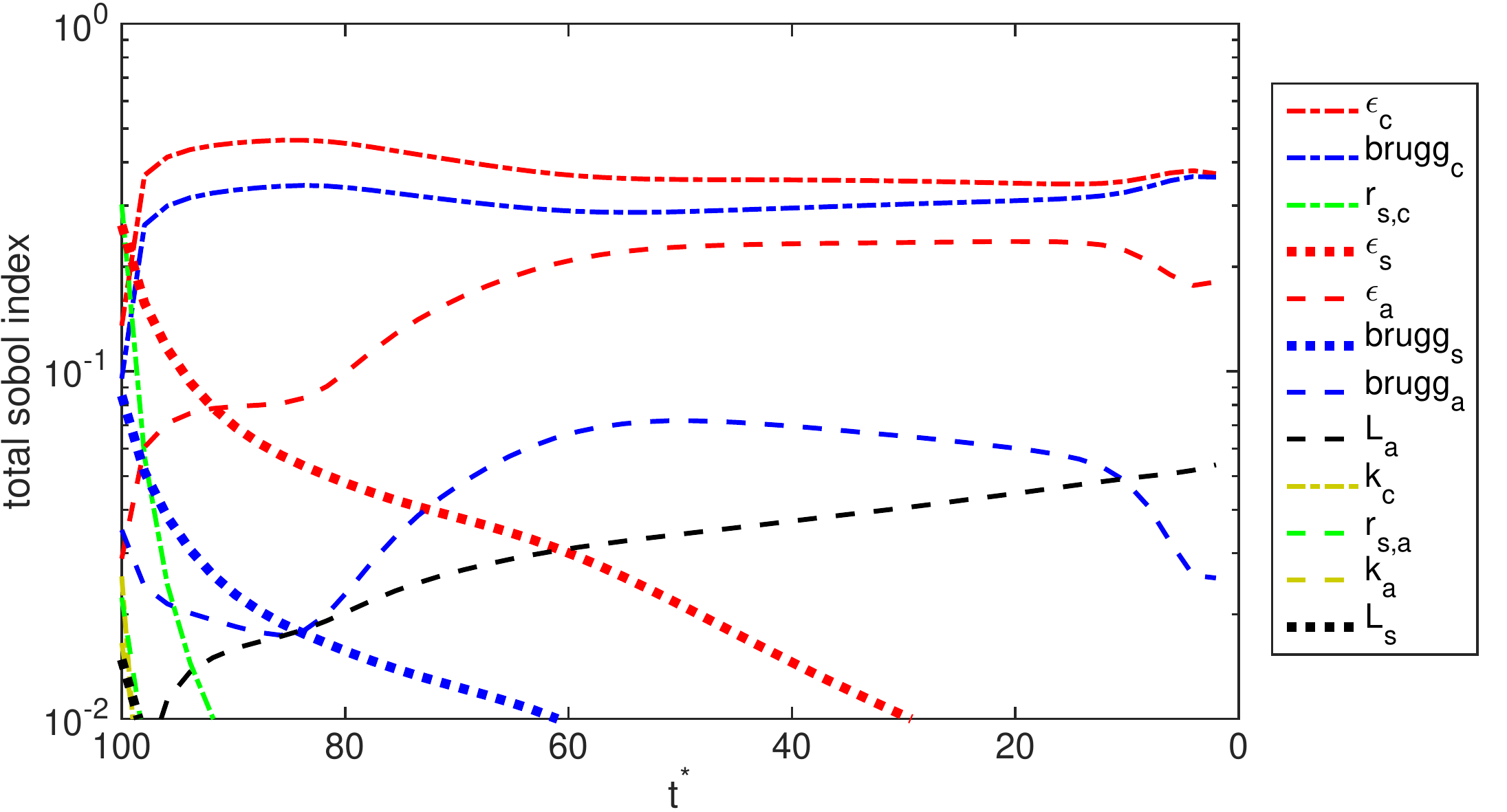}} \\
\hbox to 23.0mm{}
\subfloat[]{\includegraphics[width = 3.7in]{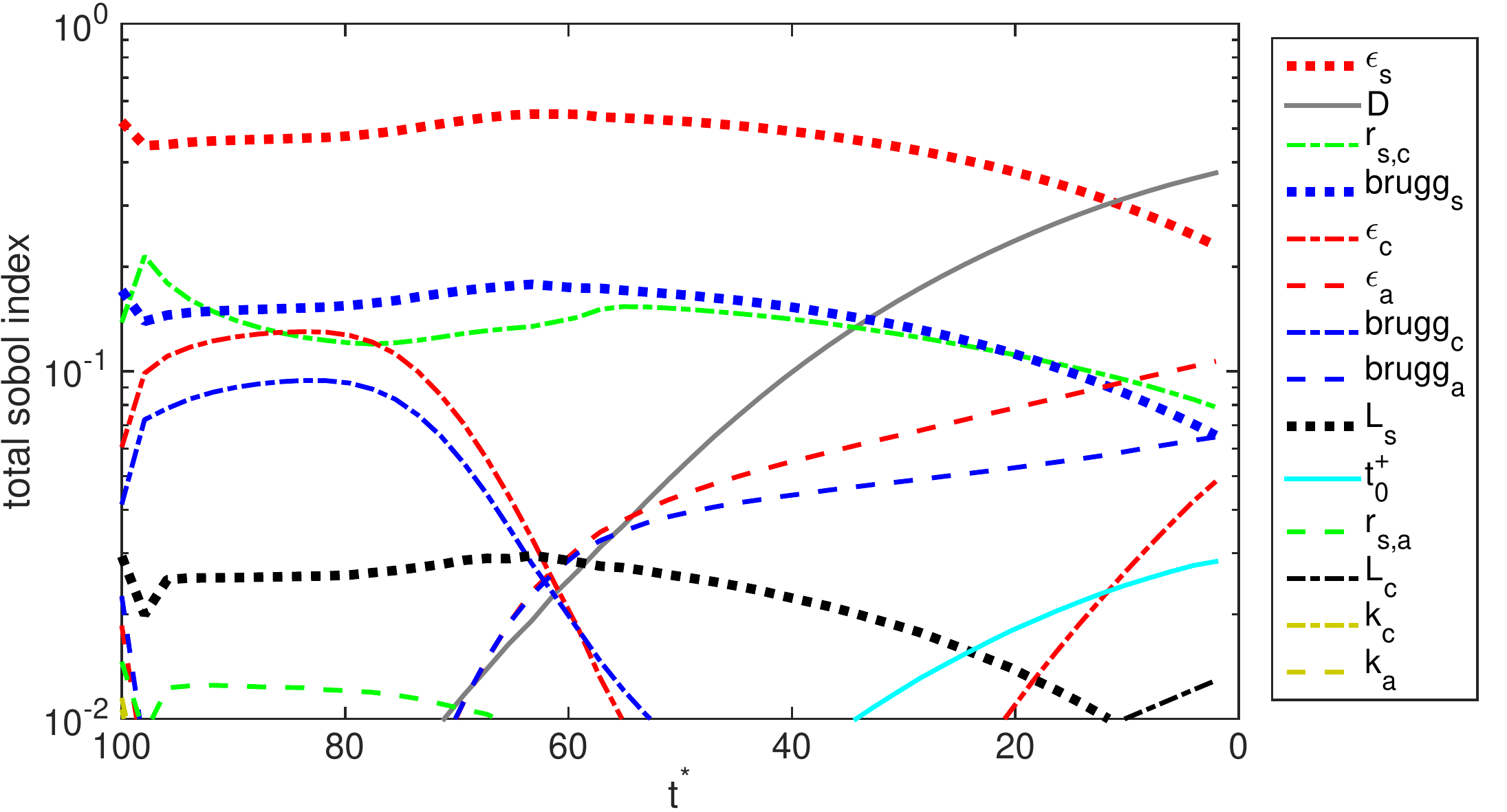}} 
\caption{Global sensitivity analysis of the cell voltage for: (a) $I=$ 0.25C; (b) $I=$ 1C; (c) $I=$ 4C.}
\label{fig:V_SOC_sobol}
\end{figure}

The corresponding total Sobol' indices of the discharge curves in Fig. \ref{fig:V_SOC_mean_sd_er} are presented in Fig. \ref{fig:V_SOC_sobol}{\color{black}, from which we highlight that:}

\begin{itemize}
  \item Similar to the cell capacity, porosity $\epsilon$ and Bruggeman coefficient $\mathrm{brugg}$ in all three regions are among the most important random inputs which contribute to the variability of the cell voltage for all three discharge rates. 
  \item For all three discharge rates, unlike the cell capacity, uncertainties in $k_a$, $k_c$, and $r_{s,a}$ contribute to the cell voltage variations, although, their impacts are limited to the onset of discharge for low to medium {\color{black}discharge rates}. 
  \item The {\color{black}uncertainty} in $\sigma_a$, $\sigma_c$, $D_{s,a}$, and $D_{s,c}$ has no significant effects on the variability of the cell voltage. 
  \item As the discharge rate increases, random parameters in separator, i.e., $\epsilon_s$, $\mathrm{brugg}_s$, and $L_s$, contribute more and more to the cell voltage variations{\color{black},} such that for 4C rate of discharge, $\epsilon_s$ becomes the most important random input affecting the {\color{black}cell voltage variability.} A similar behavior was observed in Fig. \ref{fig:sobol_cap_V}. 
    \item For $I=$ 4C, variations in the cell voltage are highly affected by the uncertainties in {\color{black}the} diffusion coefficient of the liquid phase $D$. Although, for low to medium discharge rates, $D$ is an insignificant random input, for high discharge rates, its Sobol' index $S^T_{D}$ grows {\color{black}considerably}. Moreover, unlike the cell capacity, Li$^+$ transference number $t_+^0$ {\color{black}has a high contribution in the cell voltage variability,} when the battery is discharged at 4C rate. 
  \item Similar to the cell capacity, effects of uncertainties in the length of electrodes, $L_a$ and $L_c$, on the variations of $\phi_{\mathrm{cell}}$ decreases as the rate of discharge increases.   
\end{itemize}

\begin{figure}
\vspace{-0.5in}
\hspace{-0.5in}
\subfloat[]{\includegraphics[width = 3.20in]{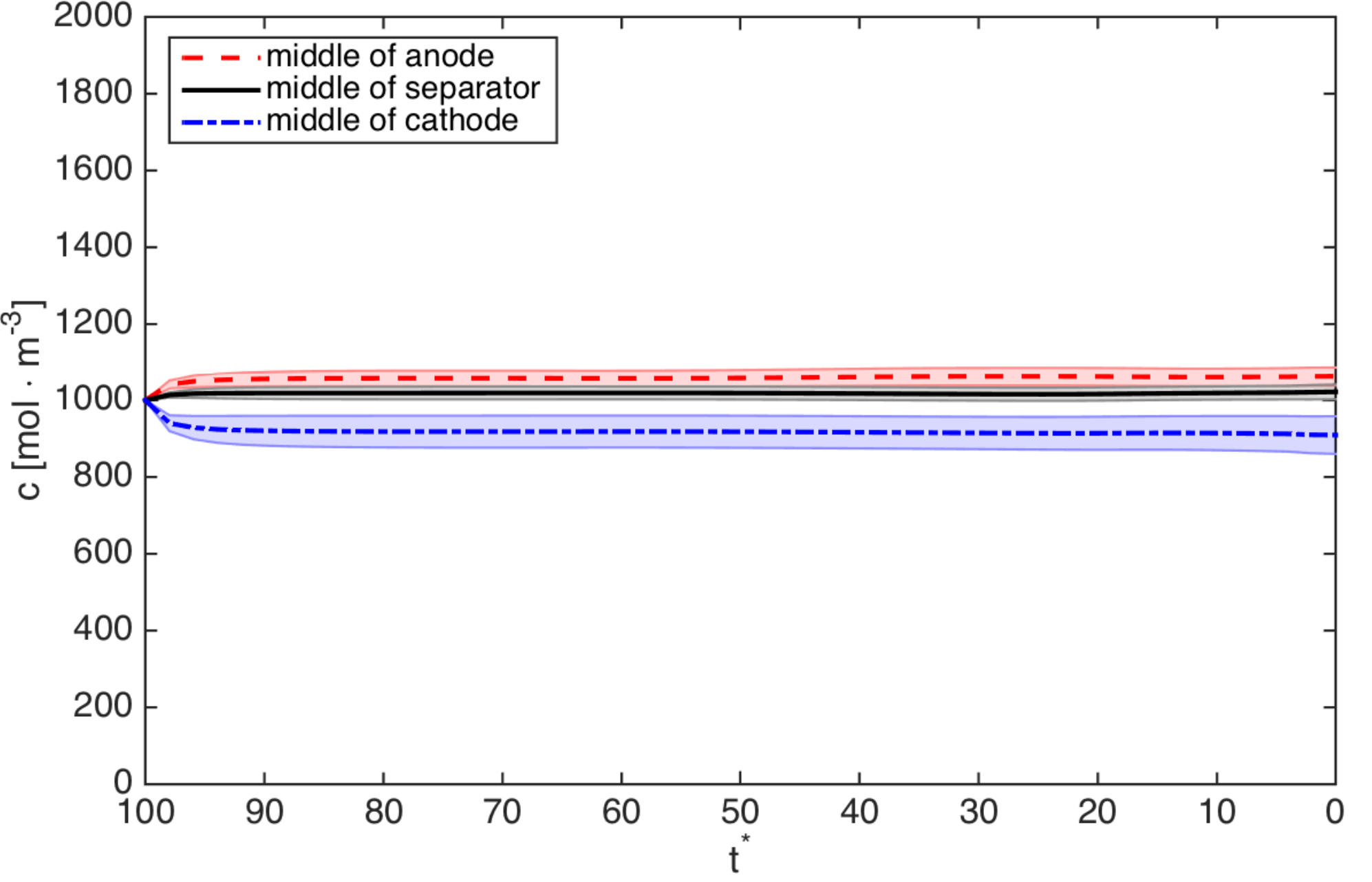}}  
\subfloat[]{\includegraphics[width = 3.20in]{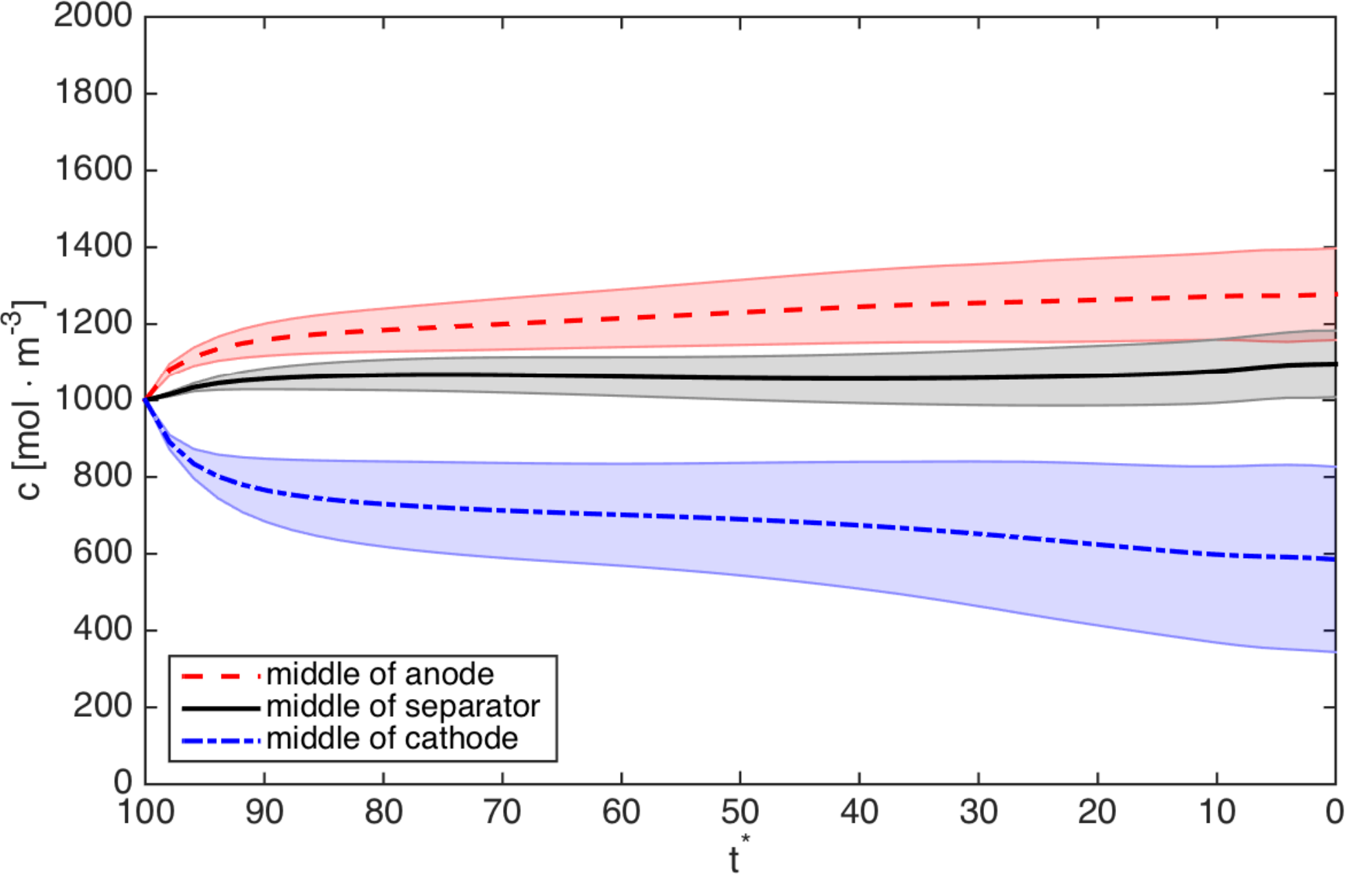}}  \par 
\hspace{-0.5in}
\subfloat[]{\includegraphics[width = 3.20in]{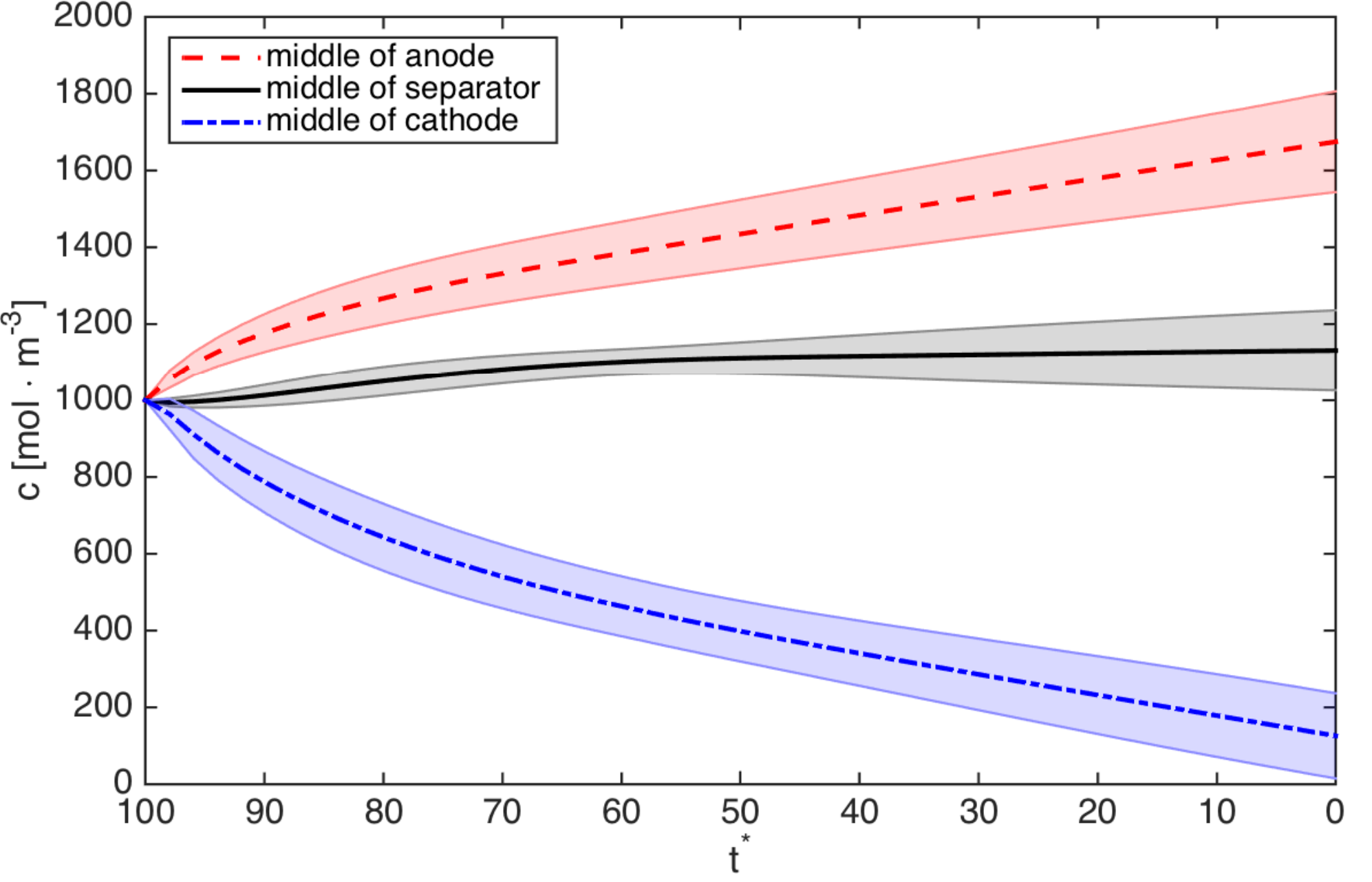}}
\subfloat[]{\includegraphics[width = 3.2in]{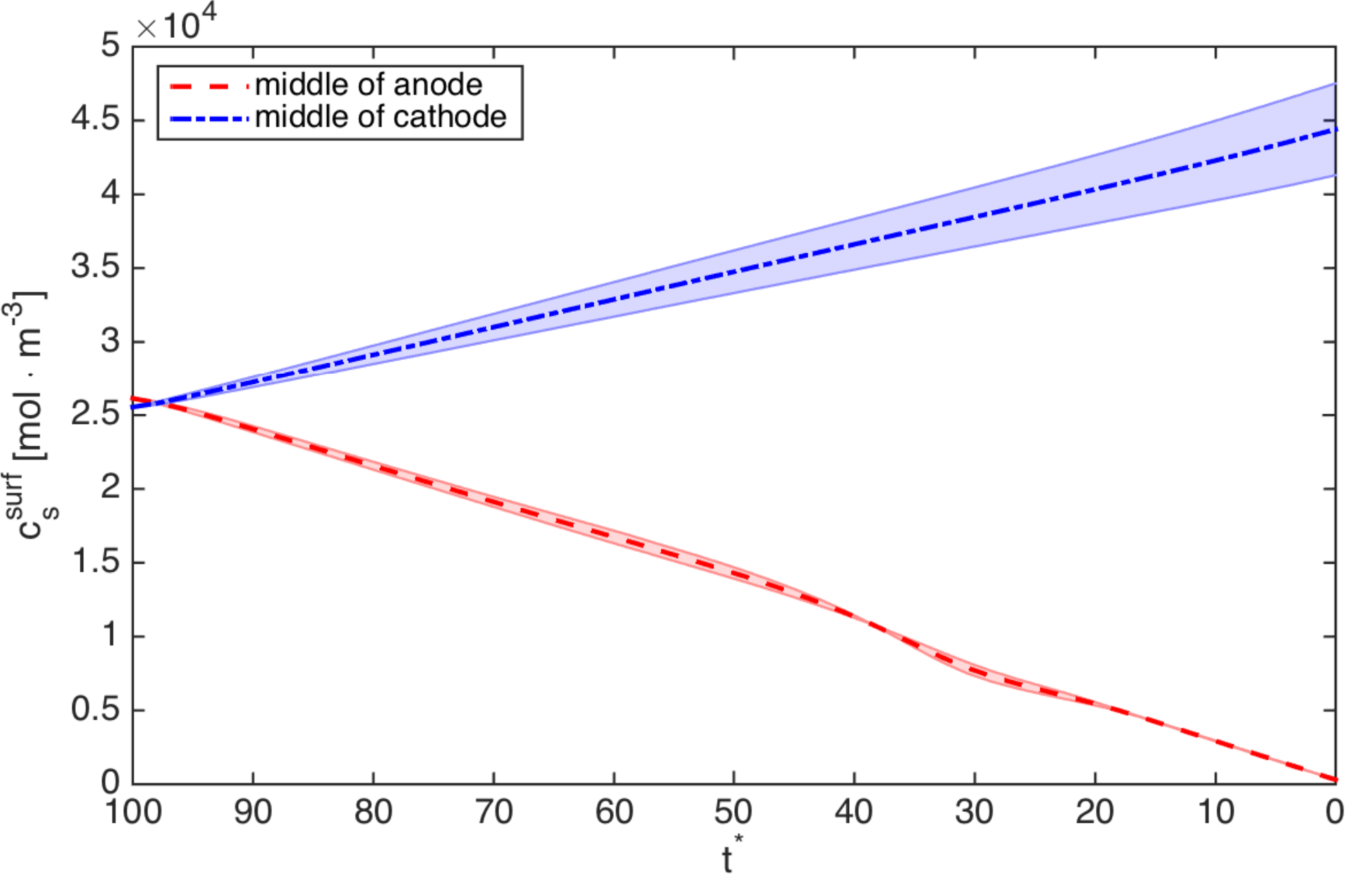}} \par 
\hspace{-0.5in}
\subfloat[]{\includegraphics[width = 3.2in]{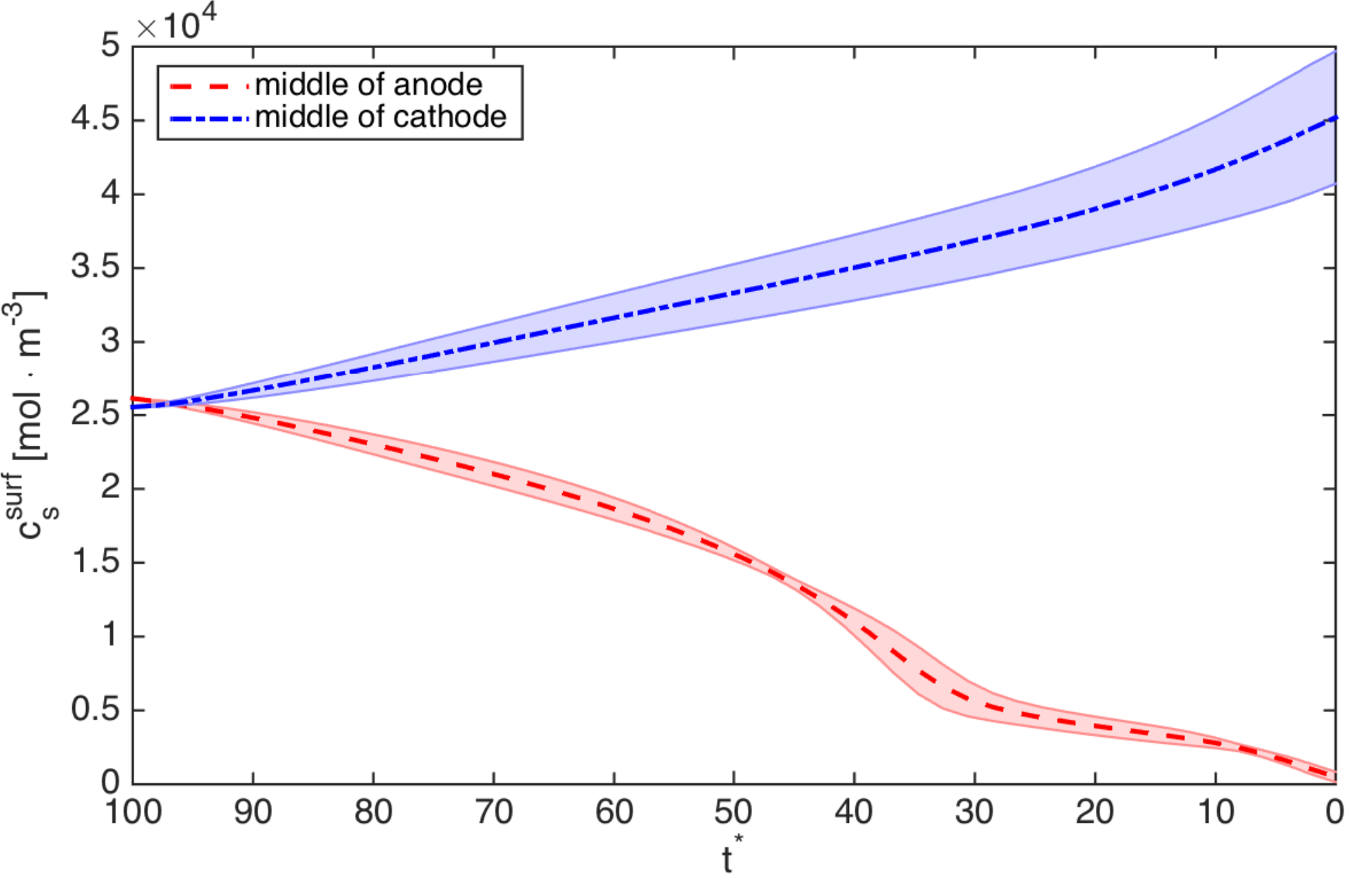}}  
\subfloat[]{\includegraphics[width = 3.2in]{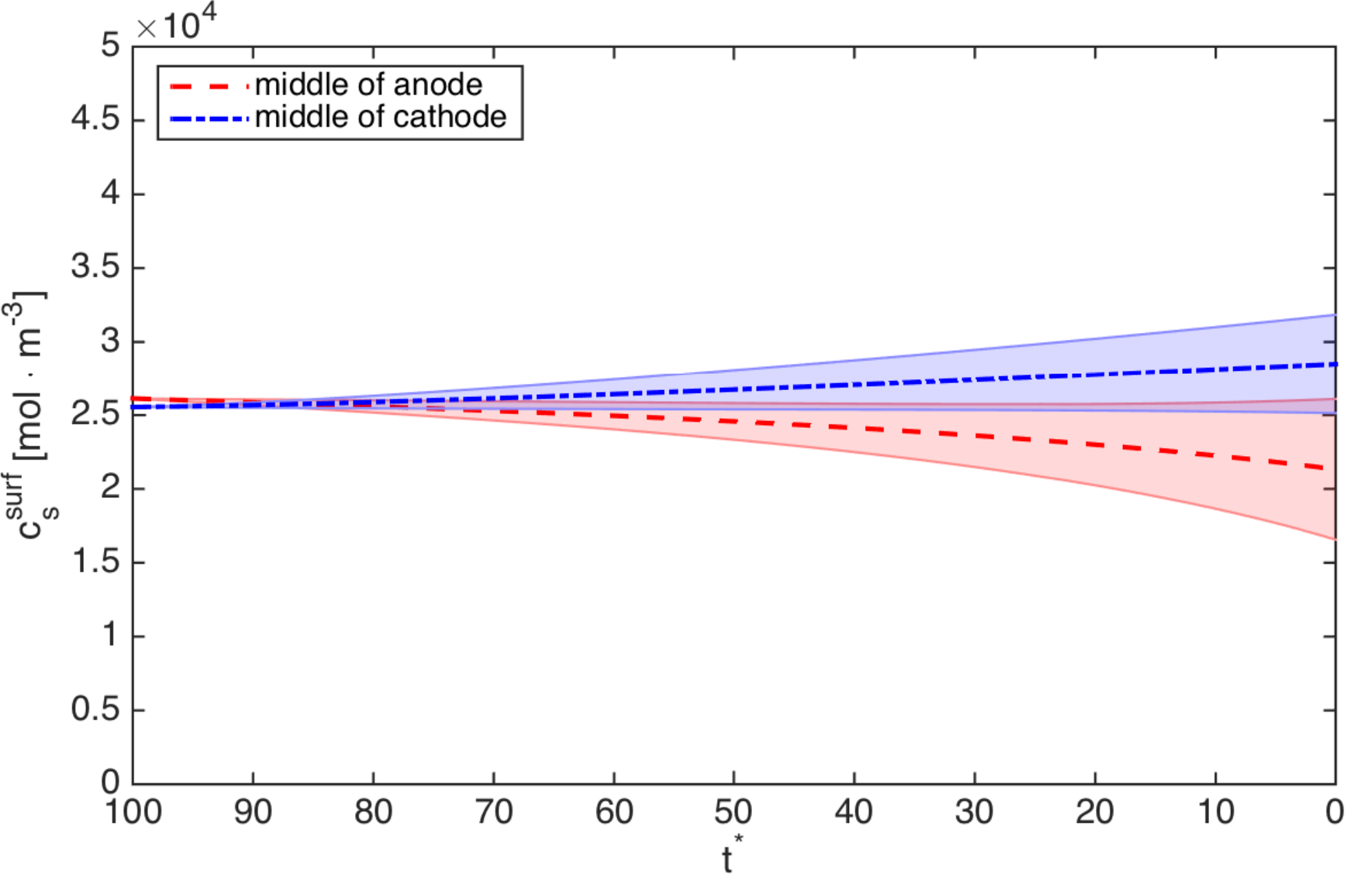}}
\caption{Mean and standard deviation of the liquid phase concentration $c$ in the middle of anode, separator and cathode for: (a) $I=$ 0.25C; (b) $I=$ 1C; (c) 4C rates of discharge. Mean and standard deviation of the solid phase concentration $c_{s}^{\mathrm{surf}}$ in the middle of anode and cathode for: (d) $I=$ 0.25C; (e) $I=$ 1C; (f) 4C rates of discharge. The shaded areas are probability bounds of three standard deviations around the mean.}
\label{fig:c_SOC_mean_sd_er}
\end{figure}
{\color{black}

\begin{sidewaysfigure}
\thispagestyle{empty}
\vspace{-0.75in}
\hspace{-1.0in}
\subfloat[]{\includegraphics[width = 3.2in]{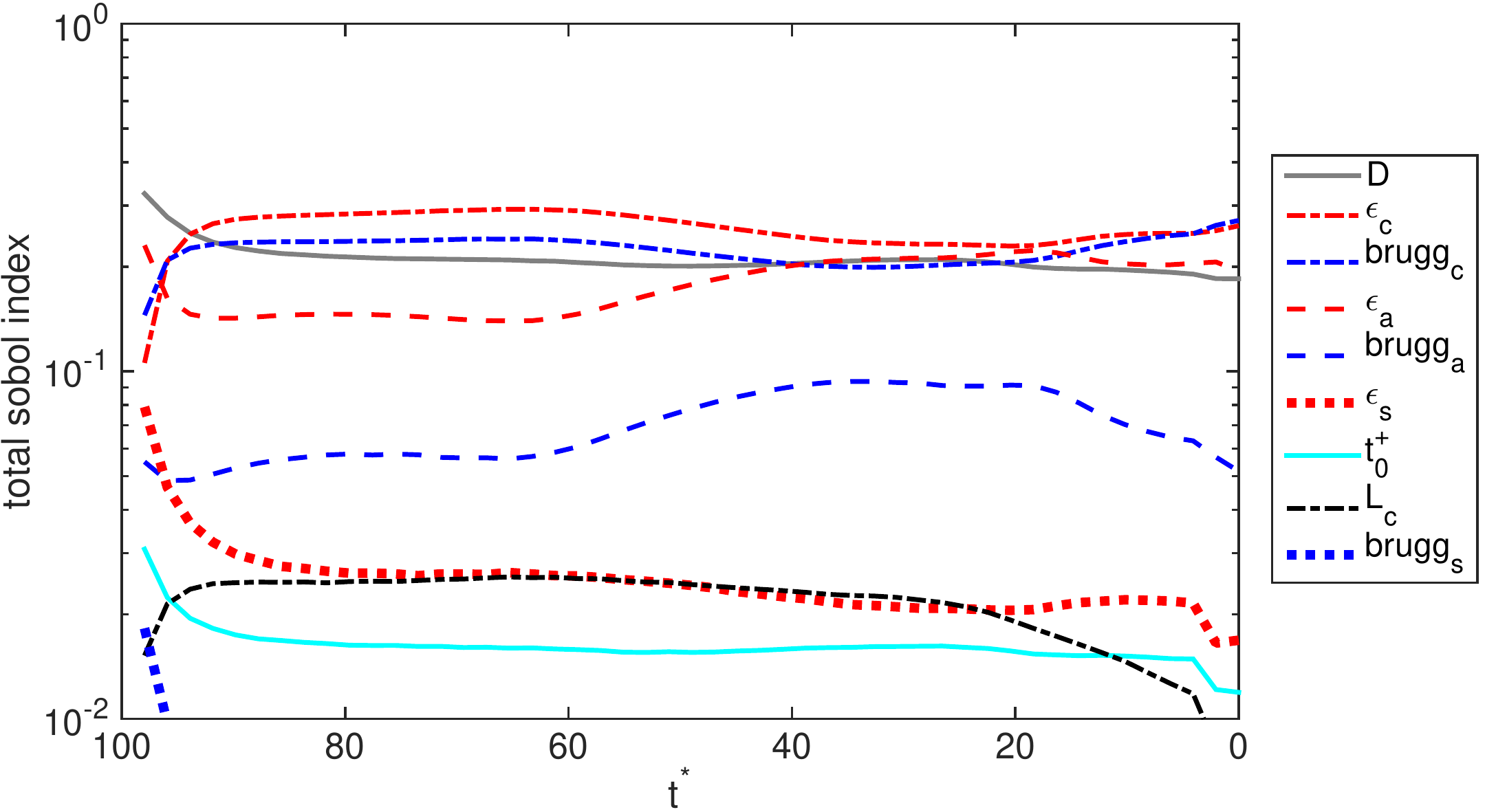}} 
\subfloat[]{\includegraphics[width = 3.2in]{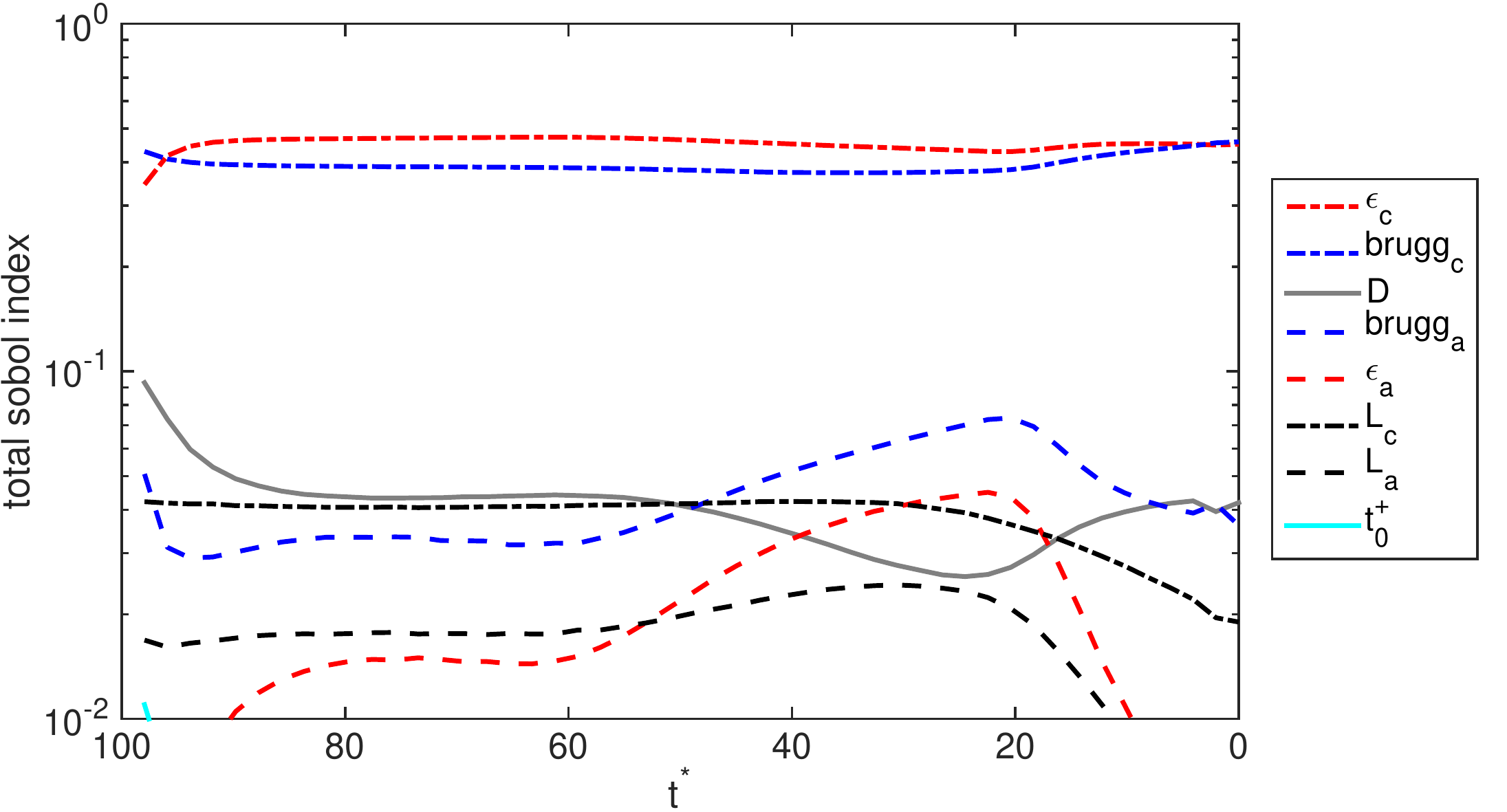}} 
\subfloat[]{\includegraphics[width = 3.2in]{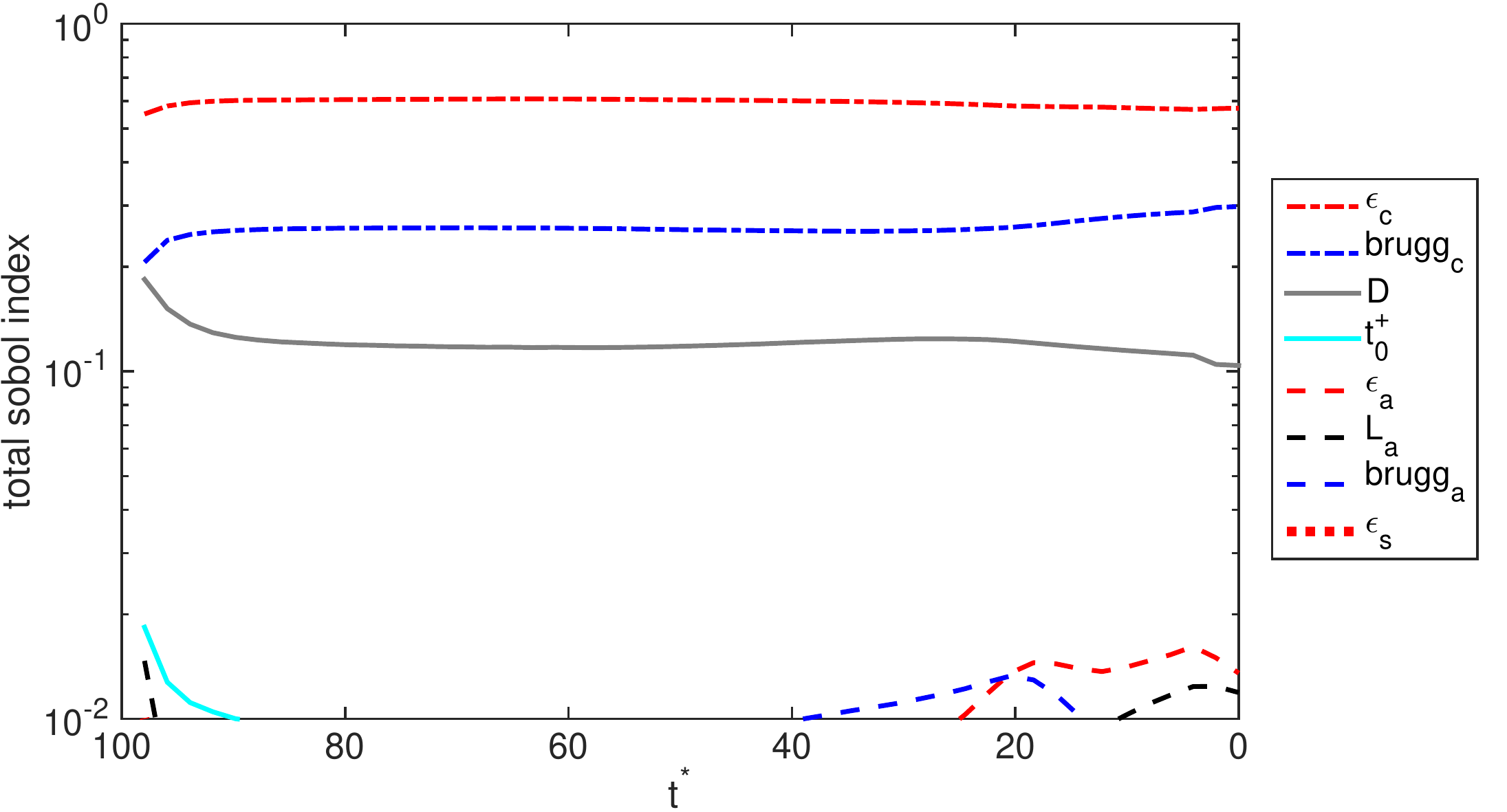}} \par 
\hspace{-1.0in}
\subfloat[]{\includegraphics[width = 3.2in]{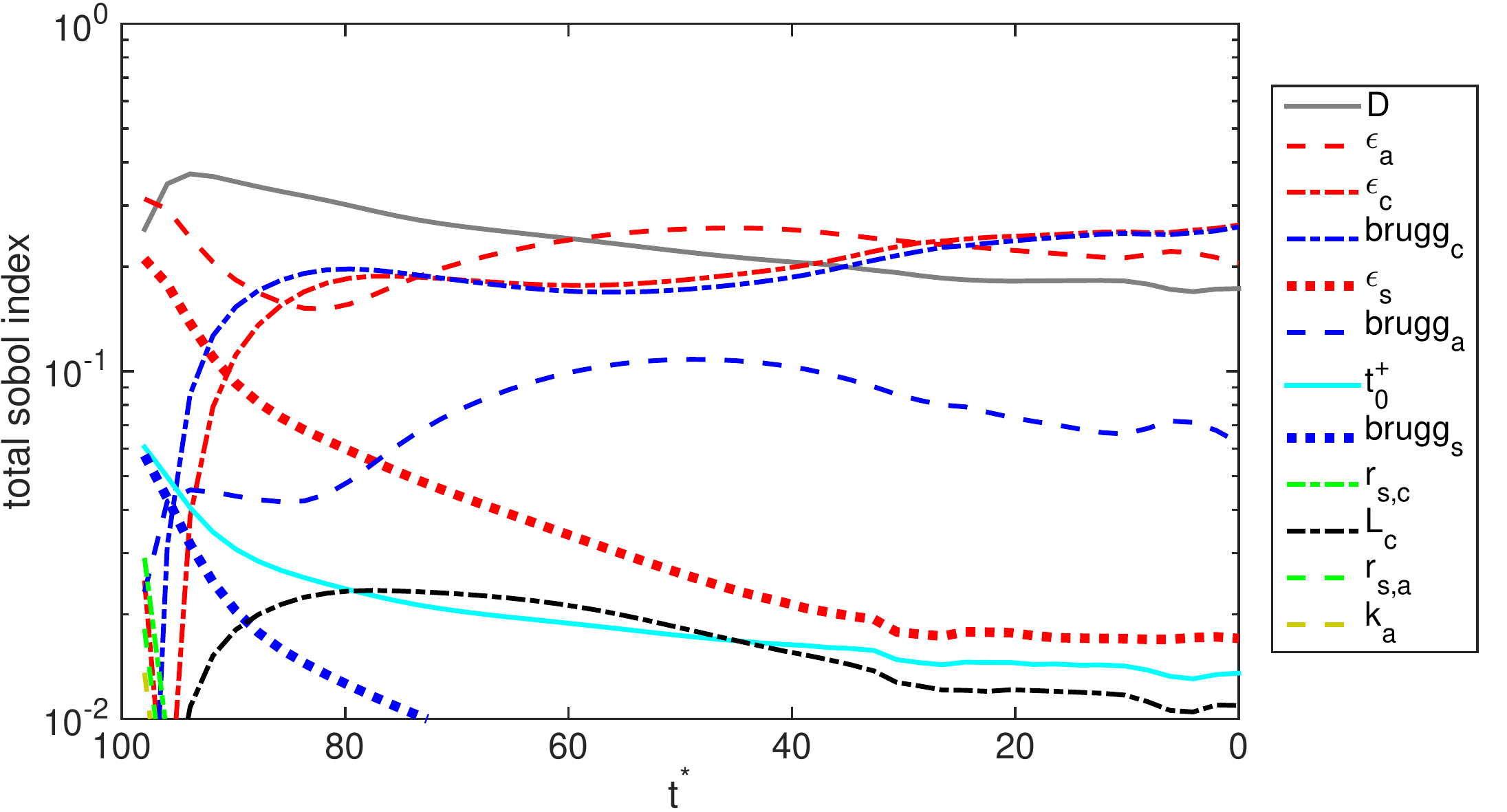}} 
\subfloat[]{\includegraphics[width = 3.2in]{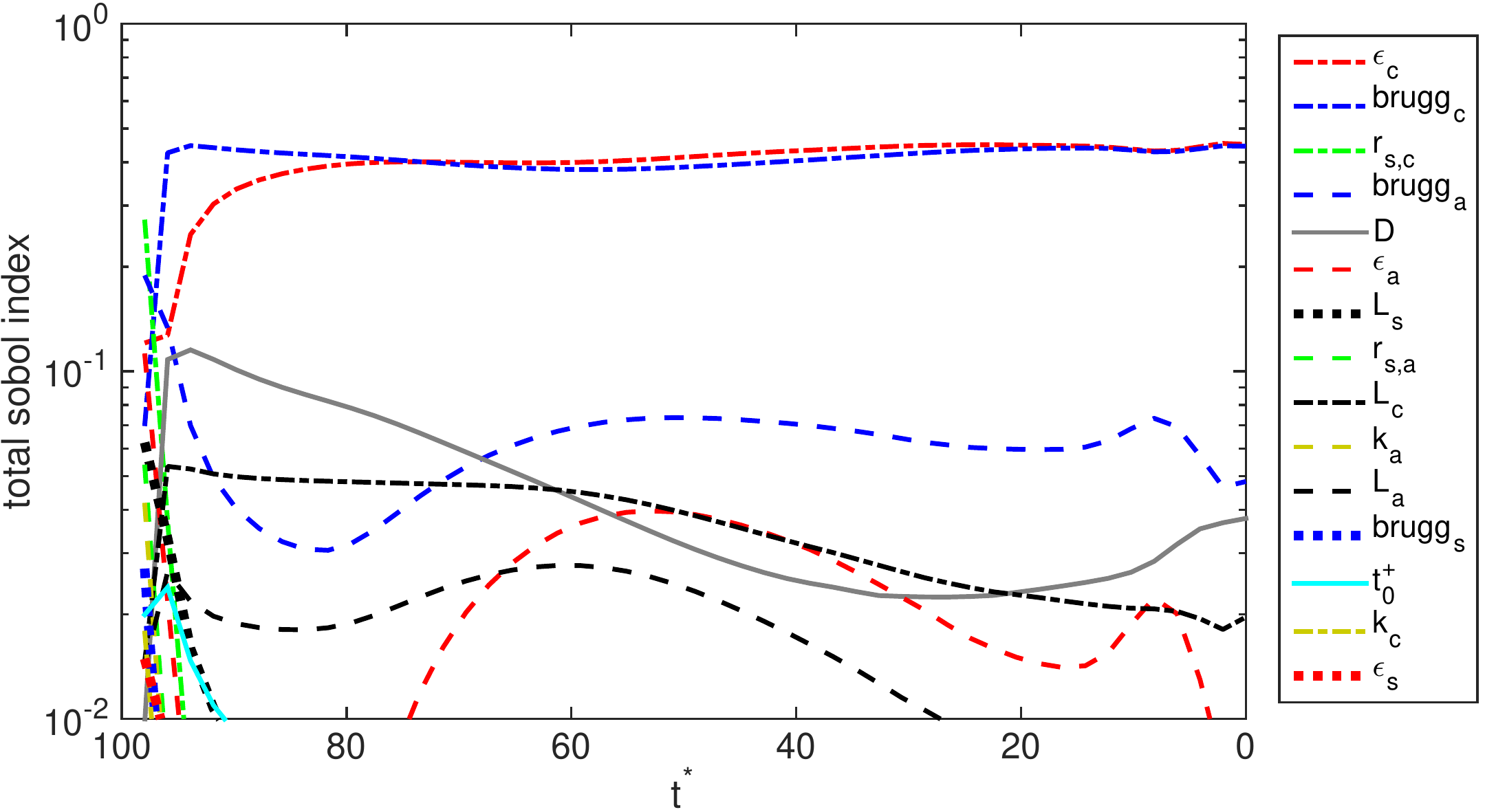}} 
\subfloat[]{\includegraphics[width = 3.2in]{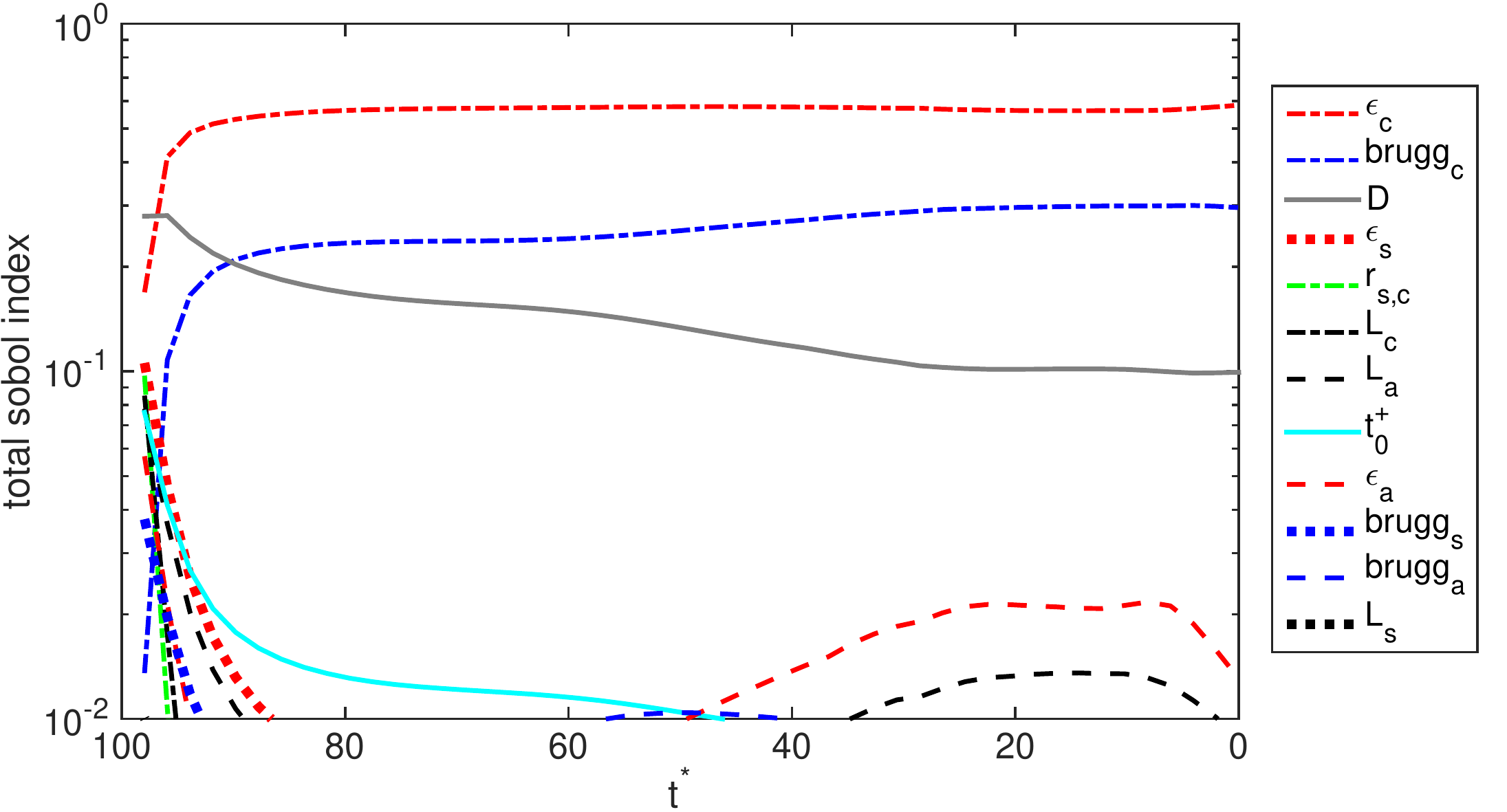}} \par 
\hspace{-1.0in}
\subfloat[]{\includegraphics[width = 3.2in]{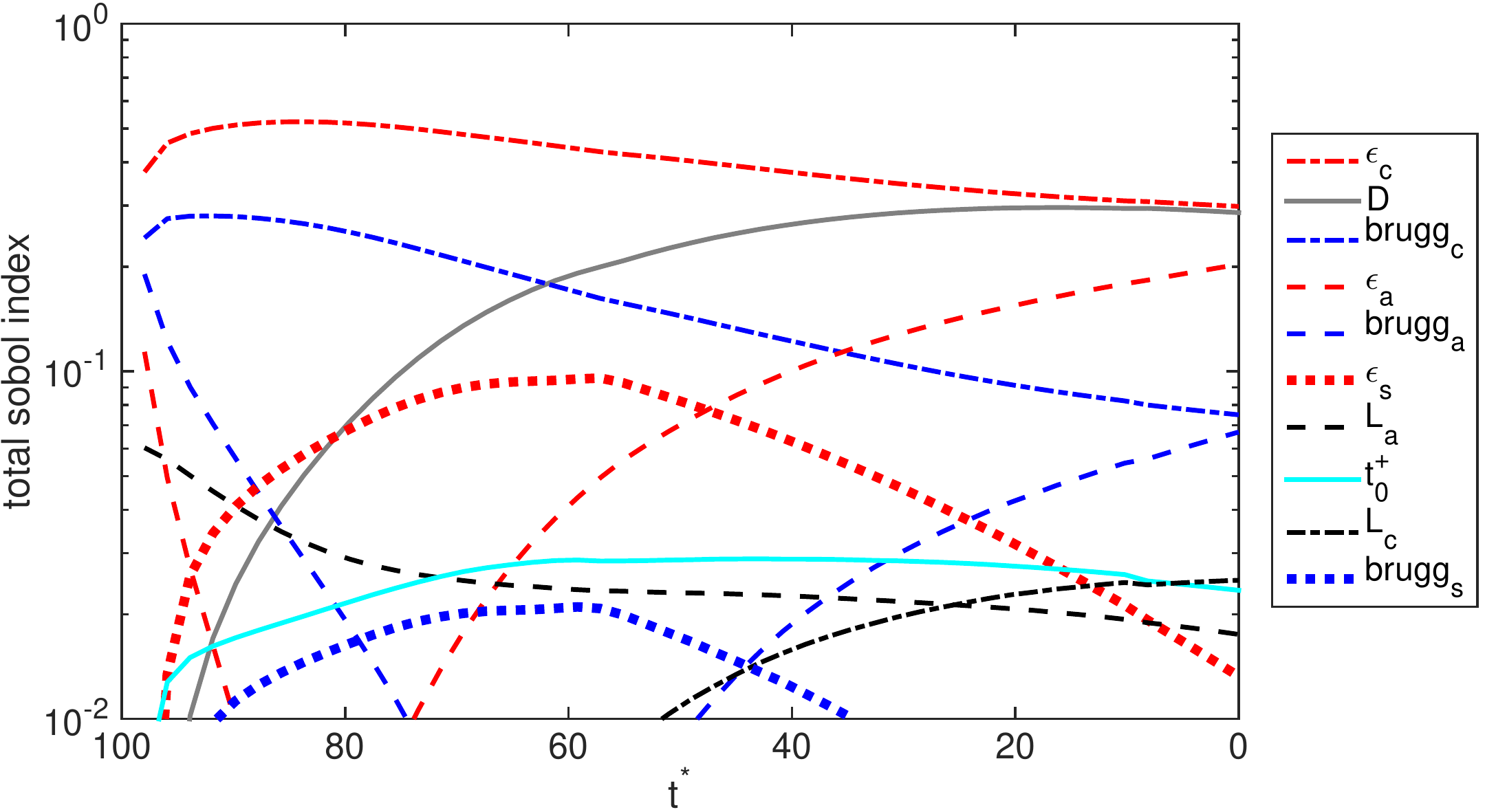}} 
\subfloat[]{\includegraphics[width = 3.2in]{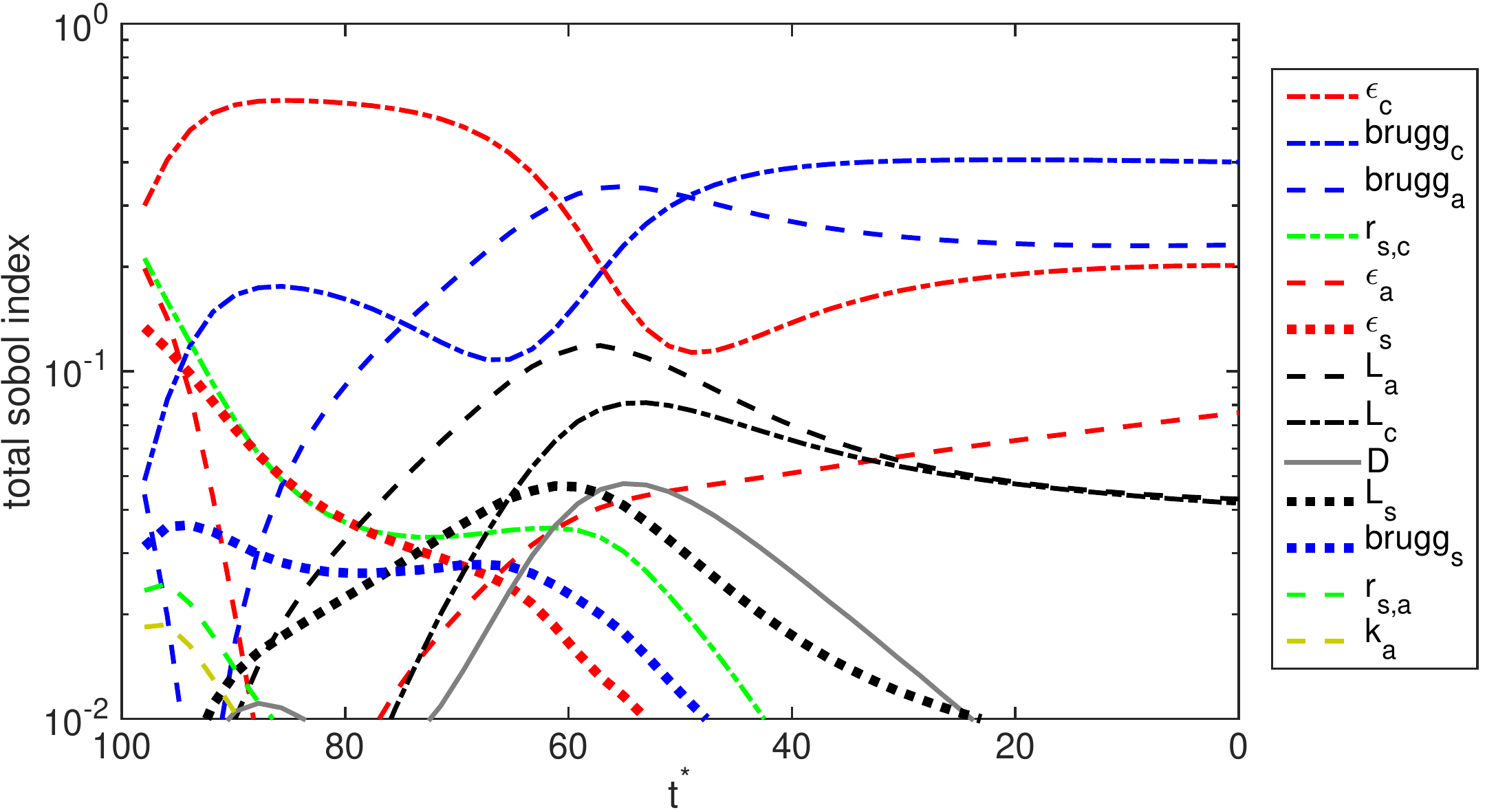}} 
\subfloat[]{\includegraphics[width = 3.2in]{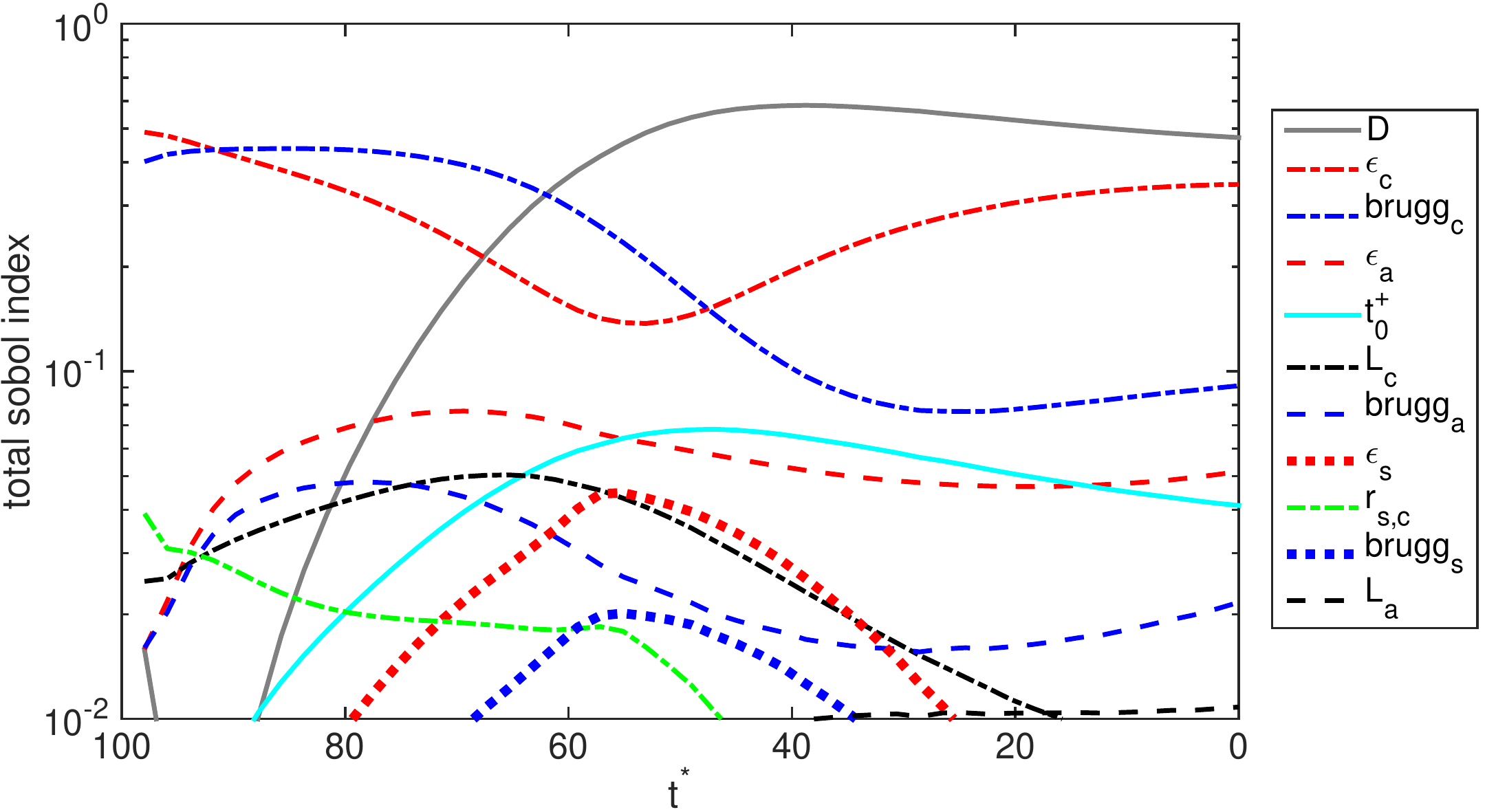}} 
\caption{Global sensitivity analysis of the liquid phase concentration for $I=$ 0.25C, 1C and 4C rates of discharge. (a) $c$ in the middle of anode with $I=$ 0.25C; (b) $c$ in the middle of separator with $I=$ 0.25C; (c) $c$ in the middle of cathode with $I=$ 0.25C; (d) $c$ in the middle of anode with $I=$ 1C; (e) $c$ in the middle of separator with $I=$ 1C; (f) $c$ in the middle of cathode with $I=$ 1C; (g) $c$ in the middle of anode with $I=$ 4C; (h) $c$ in the middle of separator with $I=$ 4C; (i) $c$ in the middle of cathode with $I=$ 4C.}
\label{fig:c_SOC_sobol}
\end{sidewaysfigure}

Figs. \ref{fig:c_SOC_mean_sd_er}(a)-(c) show the mean values of $c$ and their three standard deviation bounds at three locations for 0.25C, 1C and 4C rates of discharge. As the LIB is discharged at higher rates, larger concentration gradients are developed across the cell to balance the migration of anions \citep{Fuller94}, as a result of which the liquid concentration in the cathode approaches to zero near the end of discharge for 4C discharge rate (Fig. \ref{fig:c_SOC_mean_sd_er}(c)). This may lead to zero concentrations in a region at the back of the cathode electrode, which means the active materials may not be utilized further.  {\color{black}This limits the capability of the LIB to be discharged at higher rates}. On the other hand, high concentrations in the anode electrode may be problematic when {\color{black} the lithium salt/solvent system used has a solubility limit. For example, such a limit is 2100 mol/m$^3$ for Lithium Perchlorate in Propylene Carbonate at room temperature \citep{Fuller94}.} Consequently, analysis of {\color{black}the} variability of liquid phase concentration near the end of discharge may provide a more accurate understanding of the rate limiting mechanisms associated with diffusion of lithium ions. As can be observed from this figure, our definition of $t^*$ does not impose any constraints on $c$; hence, unlike the cell voltage, variability of $c$ is not equal to zero at $t^*$ = 0.} 

As mentioned in Section \ref{sec:rv_LIB}, it has been shown that $\sigma$ and $D_s$ depend on the solid phase concentration. Taking into account the variability in the solid phase concentration due to input variations may help to assess such dependencies more accurately. A similar analysis is performed for the solid phase concentration at the surface of the solid particle $c_s^{\mathrm{surf}}$. {\color{black}The mean} values of $c_s^{\mathrm{surf}}$ bounded by three standard deviations in the middle of electrodes for $I=$ 0.25C, 1C and 4C rates of discharge are also presented in Figs. \ref{fig:c_SOC_mean_sd_er}(d)-(f). It can be observed that the variability of $c_s^{\mathrm{surf}}$ in the middle of {\color{black}the} anode does not grow significantly as the battery is discharged at low and medium rates, while for {\color{black}the} high discharge rate, standard deviation of  $c_s^{\mathrm{surf}}$ in both electrodes is increased monotonically over the course of discharge. 

Fig. \ref{fig:c_SOC_sobol} shows the corresponding total Sobol' indices of $c$ in three regions for $I=$ 0.25C, 1C and 4C rates of discharge. {\color{black}We observe} that for low to medium discharge rates, the influential random inputs on the variability of $c$ vary considerably from one region to another. As an instance, for $I=$ 0.25C, variations in $c$ in the middle of cathode are mainly affected by uncertainty in $\epsilon_c$, $\mathrm{brugg}_c$, and $D$, while in the anode, theses variations are affected by the {\color{black}uncertainty} in $D$, $\epsilon_c$, $\mathrm{brugg}_c$, $\epsilon_a$, $\mathrm{brugg}_a$, $\epsilon_s$, $t_+^0$, $L_c$, and $\mathrm{brugg}_s$. In other words, for low discharge rates, variability of $c$ in the anode is affected by the random inputs associated with the other two regions, i.e., separator and cathode, while this is not true for the variation of $c$ in the cathode. {\color{black}Additionally, we note that:}

\begin{itemize}
  \item Similar to the cell capacity and voltage, uncertainties in $\sigma_a$, $\sigma_c$, $D_{s,a}$, and $D_{s,c}$ {\color{black}have} no significant effects on the variability of $c$. {\color{black}Moreover, although reaction rate constants $k_a$ and $k_c$ appear in Figs. \ref{fig:c_SOC_sobol}(d), (e), and (h),} their impact on the variability of $c$ is {\color{black}small}. 
  \item Despite the discharge rate in all three regions, {\color{black}uncertainty} in $D$ and $t_+^0$ has more pronounced effects on the variation of $c$ in both electrodes in comparison to separator.        
  \item $S^T_{D}$ is increased in {\color{black}the} electrodes and decreased in {\color{black}the} separator as the LIB is discharged at higher rates. 
  \item Unlike the cell capacity and voltage, {\color{black}the} effect of uncertainties in the length of {\color{black}the} electrodes $L_a$ and $L_c$ on the variations of $c$ increases as the rate of discharge increases. 
\end{itemize}

A global SA of $c_s^{\mathrm{surf}}$ in both electrodes is also presented in Fig. \ref{fig:cs_SOC_sobol}. We {\color{black}see} that uncertainty in the solid phase diffusion coefficient $D_s$ {\color{black}has} no significant effects on the variability of cell capacity, voltage and liquid phase concentration. Surprisingly, $D_s$ may be labeled as an insignificant random input even on the variations of $c_s^{\mathrm{surf}}$ in both electrodes for all three discharge rates. Moreover, the {\color{black}uncertainty} in $\sigma_a$ and $\sigma_c$ also has no impacts on the variability of $c_s^{\mathrm{surf}}$ {\color{black}independent of} the discharge rate. As the LIB experiences higher discharge rates, the total Sobol' index of $r_{s,a}$ decreases, while for $r_{s,c}$, an increase in the total Sobol' index is observed. 

In summary, we {\color{black}see} that in our stochastic LIB example, the {\color{black}uncertainty} in $\sigma$ and $D_s$ {\color{black}has} no significant effects on the variability of the cell capacity, voltage and concentrations, hence, may be treated as deterministic inputs. {\color{black}Because of this,} tight quality control measures {\color{black}are not needed for these parameters}. On the other hand, $\epsilon$ and $\mathrm{brugg}$ are the most important random inputs {\color{black}independent of} the discharge rate{\color{black},} and require tight quality control measures to reduce the LIB cell-to-cell variations. Moreover, we {\color{black}observe} that {\color{black}the relative contribution of uncertainty in the model parameters in the overall performance variability} is highly affected by the battery discharge rate.

{\color{black}We emphasize that the results presented in this section {\color{black}apply only to} the LiC$_6$/LiCoO$_2$ cell we consider in this study subjected to the input variabilities presented in Table \ref{table:random_parameters_battery}. Any changes in these uncertainties or cell chemistry/configuration may result in different observations. This is also the case if a more accurate LIB model, which {\color{black}accounts for additional physical and chemical phenomena}, is employed. }

\begin{figure}
\vspace{-0.5in}
\hspace{-0.5in}
\subfloat[]{\includegraphics[width = 3.2in]{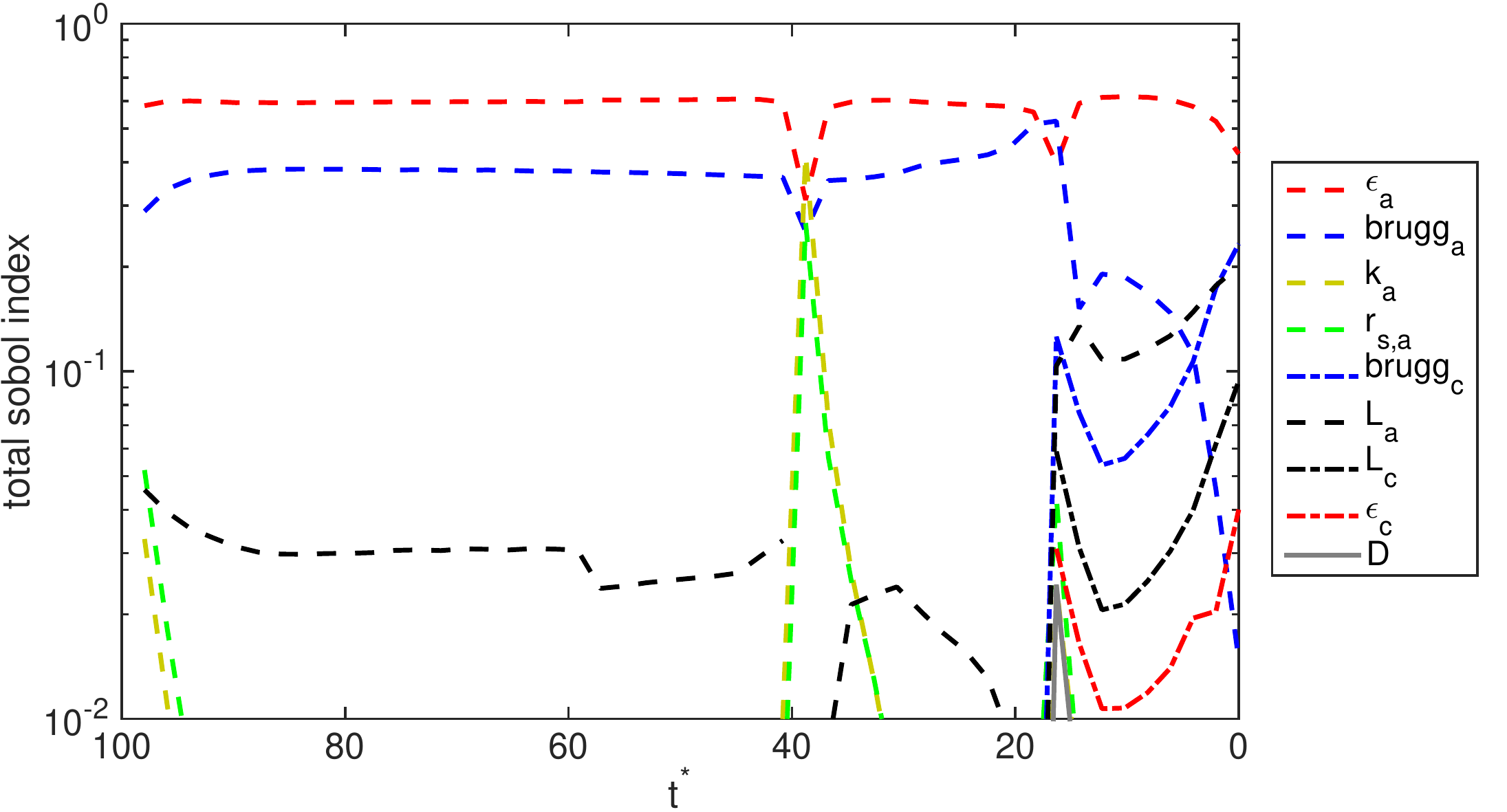}} 
\subfloat[]{\includegraphics[width = 3.2in]{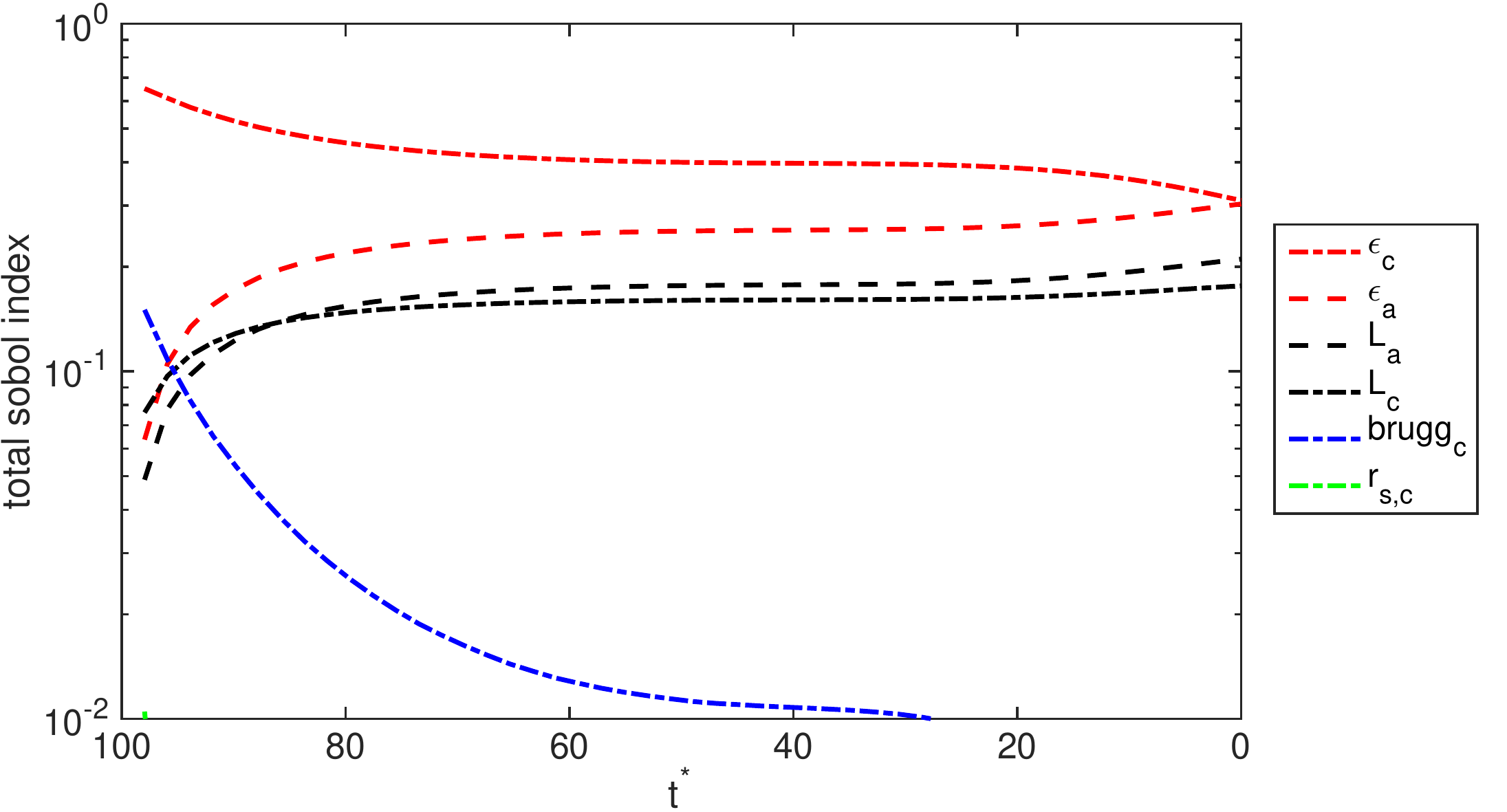}} \par 
\hspace{-0.5in}
\subfloat[]{\includegraphics[width = 3.2in]{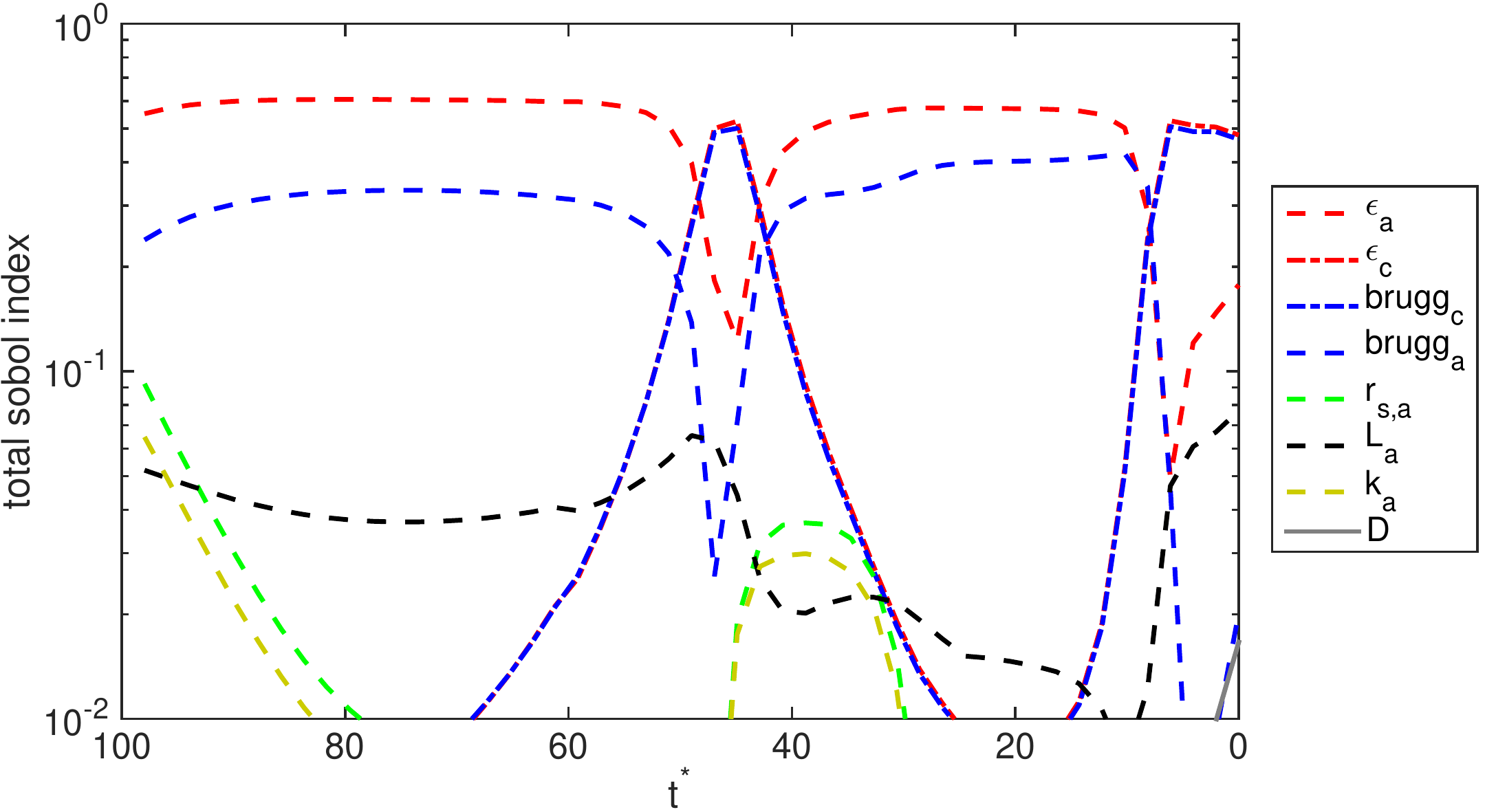}} 
\subfloat[]{\includegraphics[width = 3.2in]{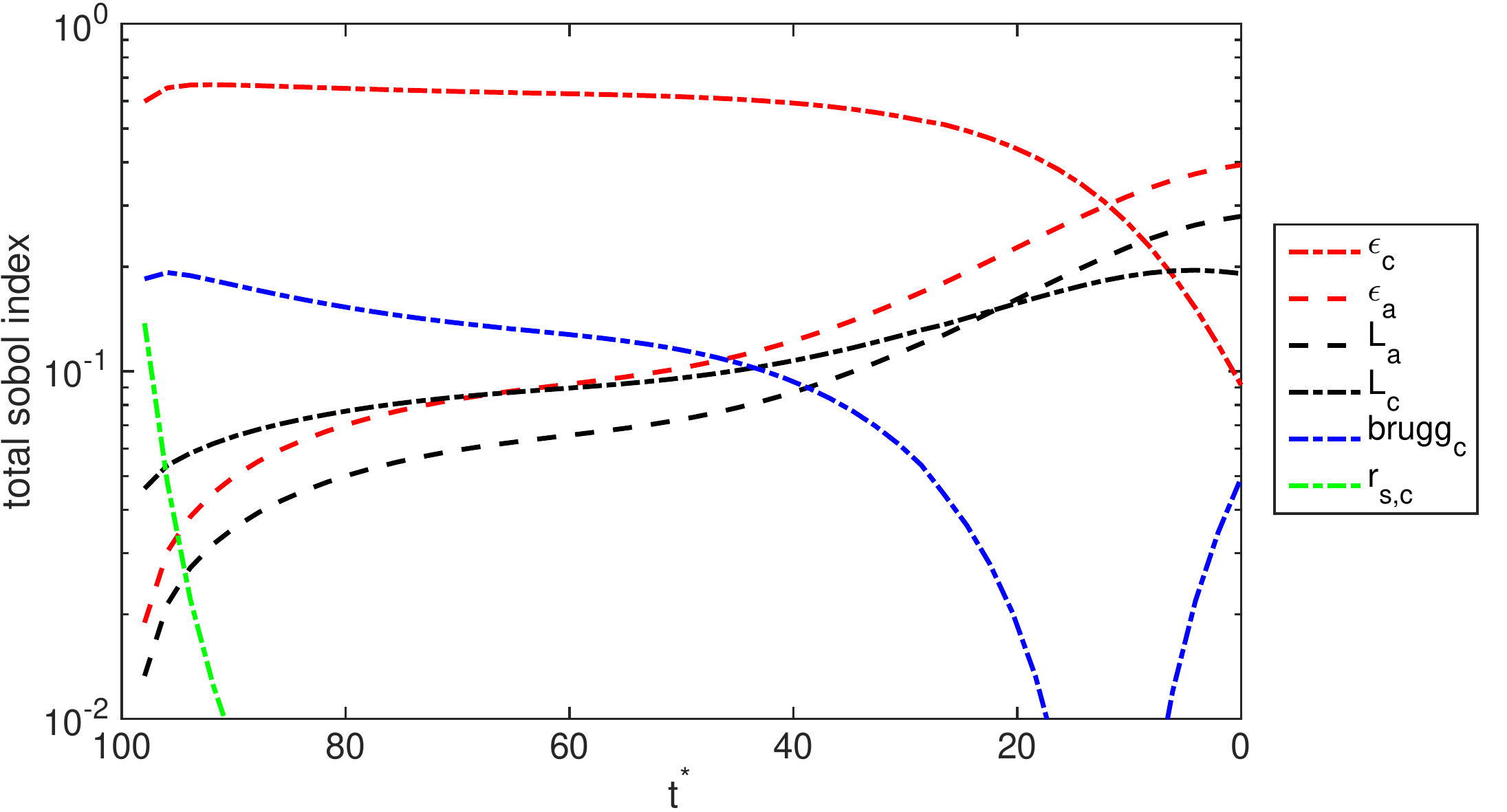}} \par 
\hspace{-0.5in}
\subfloat[]{\includegraphics[width = 3.2in]{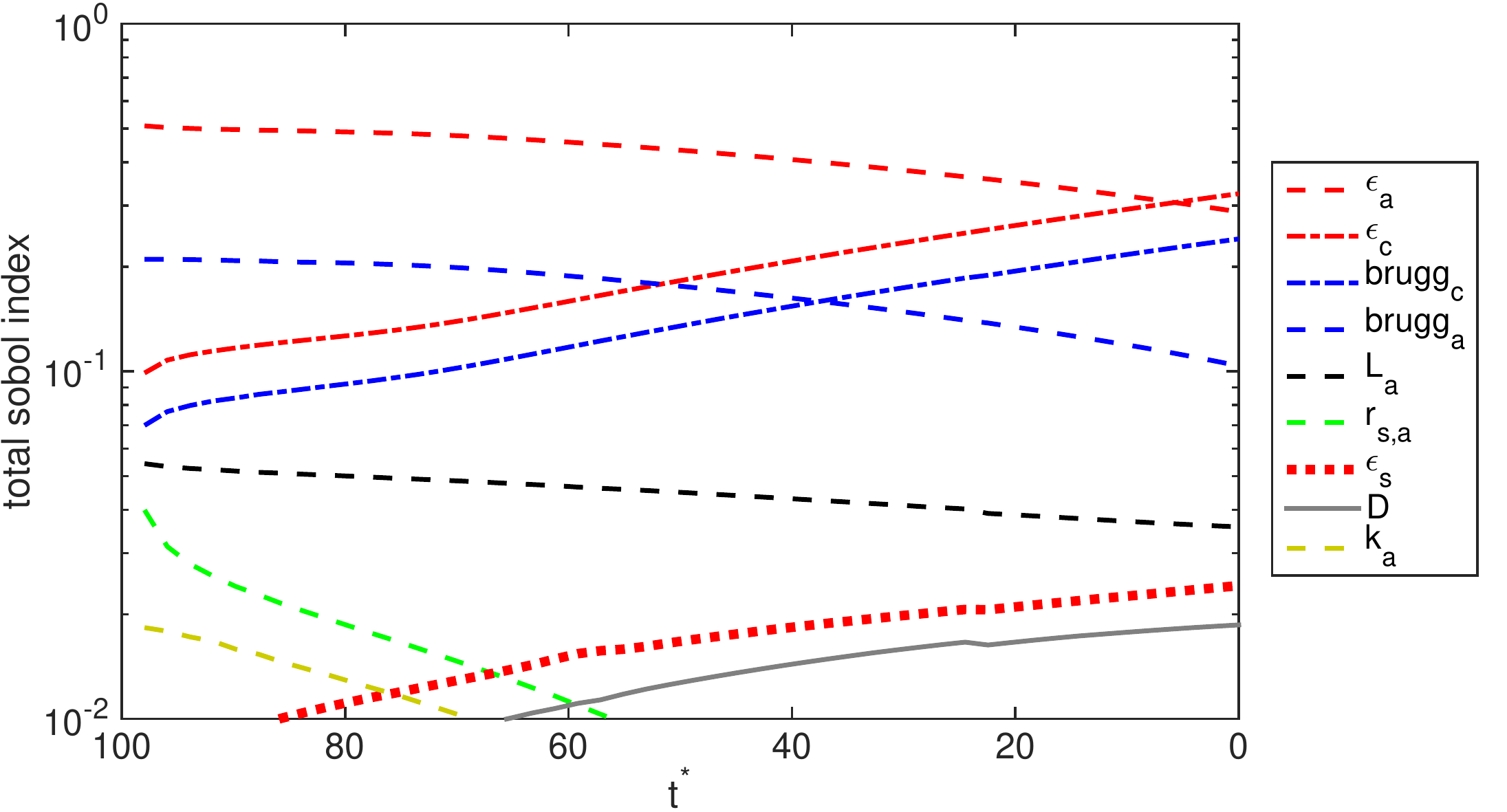}} 
\subfloat[]{\includegraphics[width = 3.2in]{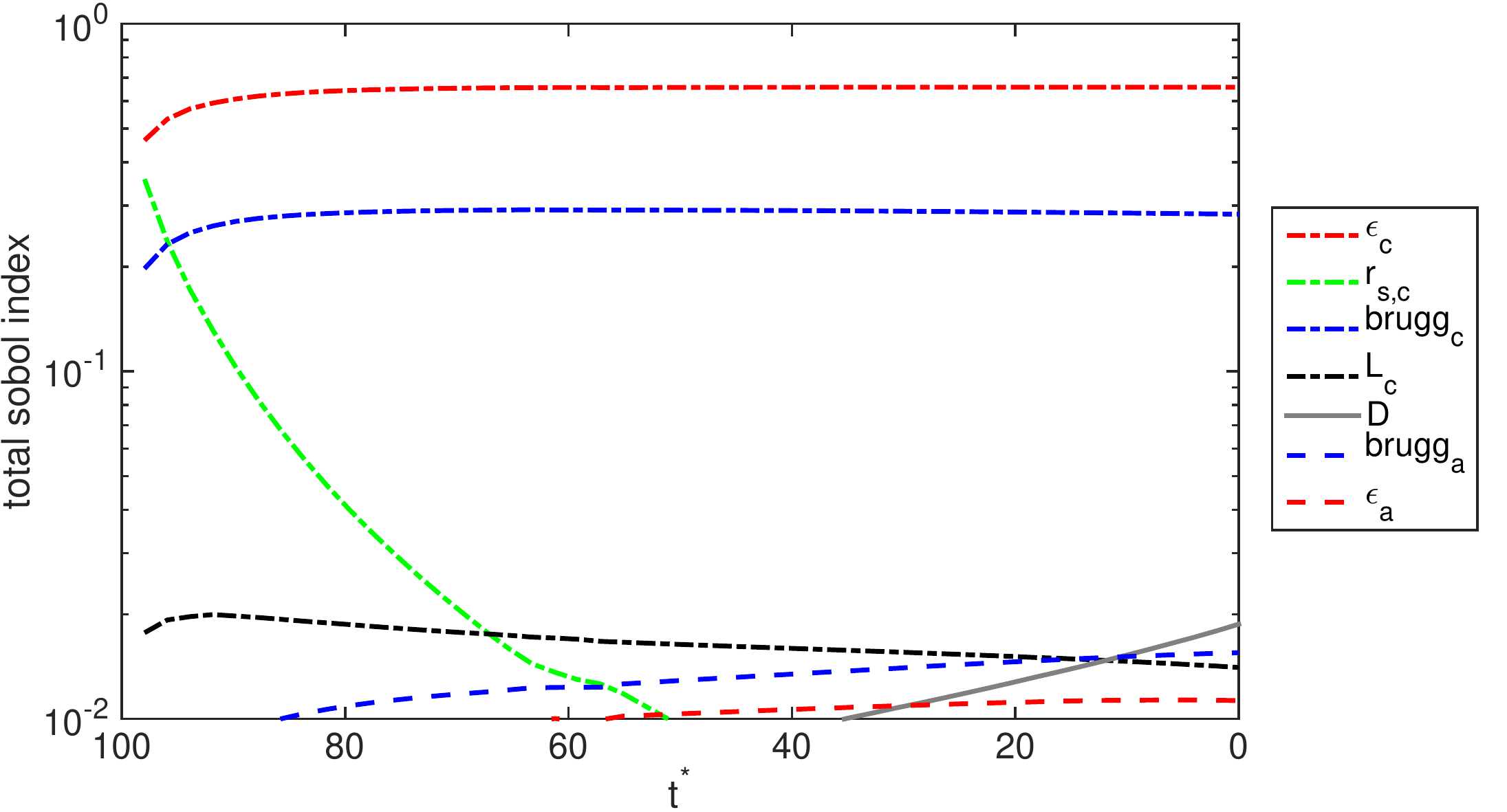}} 
\caption{Sensitivity analysis of the solid phase concentration for $I=$ 0.25C, 1C and 4C rates of discharge. (a) $c_{s}^{\mathrm{surf}}$  in the middle of anode with $I=$ 0.25C; (b) $c_{s}^{\mathrm{surf}}$  in the middle of cathode with $I=$ 0.25C; (c) $c_{s}^{\mathrm{surf}}$  in the middle of anode with $I=$ 1C; (d) $c_{s}^{\mathrm{surf}}$  in the middle of cathode with $I=$ 1C; (e) $c_{s}^{\mathrm{surf}}$  in the middle of anode with $I=$ 4C; (f) $c_{s}^{\mathrm{surf}}$  in the middle of cathode with $I=$ 4C.}
\label{fig:cs_SOC_sobol}
\end{figure}

\section{Conclusion}
\label{sec:conclusion}

A {\color{black}sampling-based} UQ approach was introduced to study the effects of input uncertainties on the performance of the LIBs. The proposed UQ approach is based on polynomial chaos expansion framework and hinges on a sparse approximation technique to achieve an accurate estimation of solution statistics with a small number of LIB simulations. Additionally, the proposed method enables one to identify the most important random inputs for any QoI such as capacity, voltage, and concentrations by performing a global {\color{black}sensitivity analysis via} computing the total Sobol' indices. Such an analysis is helpful in designing more efficient and targeted quality control measures{\color{black},} from material selection to cell assembly, {\color{black}and} reducing the manufacturing cost of the LIBs. 

Performance of the proposed UQ approach was explored through its application to an LiC$_6$/LiCoO$_2$ LIB discharged at three different rates. It was shown that the proposed UQ method can accurately compute the variability in the output QoIs such as the cell capacity, voltage, and concentrations with a small number of battery simulations, 1000 in this example. The global sensitivity analysis results corresponding to these QoIs showed that the identification of the most important uncertain inputs is highly affected by the battery discharge rate. For all three discharge rates, we found that the porosity and Bruggeman coefficient are among the most significant uncertain parameters in the performance variability of the examined LIB.  

We acknowledge that {\color{black}the} LIB model we considered in this study suffers from {\color{black}many} unmodeled phenomena such as side reactions, stresses associated with volume changes, cell degradation, and temperature dependence, which may affect the stochastic behavior of the cell. {\color{black}Since the proposed stochastic LIB model is based on sampling, in order to include these phenomena in the model, {\color{black}more accurate LIB models can be incorporated without changing the overall UQ framework}. This may, however, increase the number of required battery simulations and, hence, the overall computational cost.} {\color{black}Additionally, the accuracy of the uncertainty models we adopted for the LIB input parameters is limited by the available experimental data. For some of these inputs, data are highly sparse or non-existent. We, therefore, resorted to assumptions in describing the uncertainty. When data becomes available, these uncertainty models may be improved accordingly.

{\color{black}
The proposed UQ framework in this study was developed for forward propagation of the uncertainties in the LIB simulations. Interesting future research directions include using this forward UQ framework along with actual experimental data to infer battery model parameters as well as quantitative validation of the model itself. }}

\section*{Acknowledgements}

MH is immensely grateful to Dr. Jan Reimers for his fruitful comments on coding the decoupled formulation of LIBs and also Dr. Sehee Lee for his comments on the experimental techniques. 

MH {\color{black}and KM's} work was supported by the National Science Foundation grant CMMI-1201207. This material is based upon work of AD supported by the U.S. Department of Energy Office of Science, Office of Advanced Scientific Computing Research, under Award Number DE-SC0006402 and NSF grant CMMI-1454601.

This work utilized the Janus supercomputer, which is supported by the National Science Foundation (award number CNS-0821794) and the University of Colorado Boulder. The Janus supercomputer is a joint effort of the University of Colorado Boulder, the University of Colorado Denver and the National Center for Atmospheric Research.

\pagebreak

%
\appendix
\addappheadtotoc
\section{A decoupled formulation for LIBs}
\label{sec:LIB_decoupled}

The decoupling procedure starts with combining the potential equations for the solid phase (\ref{eqn:phi_s}), liquid phase (\ref{eqn:phi_e}) and the Butler-Volmer (BV) equation (\ref{eqn:j_vol}) in order to first reduce the number of equations to be solved. This is motivated by the fact that in each electrode, there are only three tightly coupled quantities, i.e., $c$, $c_s$ and the over-potential $\eta = \phi_{s} - \phi_{e} - V$, which need to be solved simultaneously \citep{Reimers13}. Consequently, the following non-linear equation for the over-potential $\eta$ can be obtained

\begin{eqnarray} 
\label{eqn:eta}
&& \nabla ( \sigma^{h} \nabla \eta) = a i_{ex} \Big[ \exp \Big( \frac{0.5 F \eta}
{RT} \Big) - \exp \Big( -\frac{0.5 F \eta}{RT} \Big)\Big] \nonumber \\
&& \qquad \qquad \qquad -\nabla \Big[ \frac{I}{\sigma^{\mathrm{eff}}} + \frac{\kappa_D^{\mathrm{eff}} }{\kappa^{\mathrm{eff}} } \nabla \ln c(\eta) + \nabla V(\eta) \Big],
\end{eqnarray}
subjected to the boundary conditions
\begin{eqnarray} 
\label{eqn:eta_bcs}
&& \nabla \eta |_{x=0,L} = -\Big[ \frac{I}{\sigma^{\mathrm{eff}}} + \nabla V(\eta) \Big]_{x=0,L}, \\
&& \nabla \eta |_{x=L_a,L_a+L_s} = \Big[ \frac{I}{\kappa^{\mathrm{eff}}} - \frac{\kappa_D^{\mathrm{eff}}}{\kappa^{\mathrm{eff}}} \nabla \ln c(\eta) - \nabla V(\eta) \Big]_{x=L_a,L_a+L_s}.
\end{eqnarray}

Here, $\sigma^{h}$ is the harmonic mean conductivity defined by $\frac{1}{\sigma^h} = \frac{1}{\sigma^{\mathrm{eff}}} + \frac{1}{\kappa^{\mathrm{eff}}}$. So far, instead of solving the coupled system of equations in Table \ref{table:coupled_GE}, one needs to solve a system of equations including Eqs. (\ref{eqn:c}), (\ref{eqn:c_s}) and (\ref{eqn:eta}) simultaneously. We note that all of the nonlinearities are isolated in Eq. (\ref{eqn:eta}) for the over-potential. 

The next step of deriving the reformulation is decoupling these three equations via the concept of particular and homogeneous solutions to ODEs. Application of the implicit finite time differencing to the solid phase diffusion equation (\ref{eqn:c_s}) results in the following non-homogeneous ODE with non-homogeneous BCs 

\begin{eqnarray} 
\label{eqn:cs_finite_time}
&& c_s(r,t+\Delta t) - \frac{\Delta t}{r^2} \frac{\partial}{ \partial r} \Big( D_s r^2 \frac{\partial}{ \partial r} c_s(r,t+\Delta t) \Big) = c_s(r,t), \\
&& \nabla c_s|_{r=0} = 0, \qquad \nabla c_s|_{r=r_s} = -\frac{j_{vol}}{a F D_s}. 
\end{eqnarray}

One may write the general solution to Eq. (\ref{eqn:cs_finite_time}) as

\begin{equation} 
\label{eqn:cs_general_sol}
c_s(r,t+\Delta t) = c_s^o(r,t+\Delta t) + \frac{j_{vol}}{aF} c_s^j(r,t+\Delta t),
\end{equation} 
{\color{black}where $c_s^o$ is the solution to the following non-homogeneous ODE with homogeneous BCs
\begin{eqnarray} 
\label{eqn:c_s^o}
&& c_s^o(r,t+\Delta t) - \frac{\Delta t}{r^2} \frac{\partial}{ \partial r} \Big( D_s r^2 \frac{\partial}{ \partial r} c_s^o(r,t+\Delta t) \Big) = c_s(r,t),  \\
&& \nabla c_s^o|_{r=0} = 0, \qquad \nabla c_s^o|_{r=r_s} = 0, 
\end{eqnarray}
and $c_s^j$ is the solution to the following the homogeneous ODE with non-homogeneous BCs
\begin{eqnarray} 
\label{eqn:c_s^j}
&& c_s^j(r,t+\Delta t) - \frac{\Delta t}{r^2} \frac{\partial}{ \partial r} \Big( D_s r^2 
\frac{\partial}{ \partial r} c_s^j(r,t+\Delta t) \Big) = 0,  \\
&& \nabla c_s^j|_{r=0} = 0, \qquad \nabla c_s^j|_{r=r_s} = -\frac{1}{D_s}. 
\end{eqnarray}
}

A similar approach may be employed {\color{black}for} decoupling the solution of the electrolyte phase diffusion equation (\ref{eqn:c}) from the over-potential solution $\eta$ and the volumetric pore wall flux $j_{vol}$. The only difference here is that $j_{vol}$ in Eq. (\ref{eqn:c}) is a position dependent source term while in Eq. (\ref{eqn:c_s}) it appears on the boundary conditions. Consequently, the general solution to the discretized version of Eq. (\ref{eqn:c}) in the time domain can be obtained via 

\begin{equation} 
\label{eqn:c_general_sol}
c(x,t+\Delta t) = c^o(x,t+\Delta t) + \frac{1-t_+}{F} \int j_{vol}(x_0) c^j(x,x_0,t+\Delta t) dx_0.
\end{equation} 

Here in Eq. (\ref{eqn:c_general_sol}), $c^o$ and $c^j$ are the solutions to the following ODEs both subjected to zero flux boundary conditions 

\begin{equation} 
\label{eqn:c^o} 
\frac{\epsilon [c^o(x,t+\Delta t) - c(x,t)]}{\Delta t} = \nabla [\epsilon D^{\mathrm{eff}} \nabla c^o(x,t+\Delta t)],
\end{equation}

\begin{equation} 
\label{eqn:c^j} 
\frac{\epsilon c^j(x,x_0,t+\Delta t)}{\Delta t} = \nabla [\epsilon D^{\mathrm{eff}} \nabla c^j(x,x_0,t+\Delta t)]+\delta(x-x_0),
\end{equation}
with $\delta(x-x_0)$ being the Dirac delta function. 

At this point, solutions to the solid and electrolyte phase diffusion equations are decoupled from the over-potential and pore wall flux since in solving Eqs. (\ref{eqn:c_s^o}), (\ref{eqn:c_s^j}), (\ref{eqn:c^o}) and (\ref{eqn:c^j}), $\eta$ and $j_{vol}$ are not needed. Moreover, for the cases when $D_s$, $D^{\mathrm{eff}}$ and $\epsilon$ are constant, analytic solutions are available for Eqs. (\ref{eqn:c_s^j}) and (\ref{eqn:c^j}) \citep{Reimers13}. $c_s^j$ and $c^j$ can also be used to obtain a linearized form for the terms including $V(\eta)$ and $\ln c(\eta)$ on the RHS of Eq. (\ref{eqn:eta}), respectively. In conclusion, one needs to solve the decoupled Eqs. (\ref{eqn:c_s^o}), (\ref{eqn:c_s^j}), (\ref{eqn:c^o}), (\ref{eqn:c^j}), and (\ref{eqn:eta}) instead of solving the coupled system of non-linear equations in Table \ref{table:coupled_GE} for LIB modeling. In addition, if a constant time step is used, Eqs. (\ref{eqn:c_s^j}) and (\ref{eqn:c^j}) need to be solved once at the beginning of the simulation. The only major approximation in deriving this decoupled formulation is the finite time differencing which is inevitable in numerical simulations. {\color{black}An} extended version of this decoupled formulation{\color{black},} which includes SEI resistance and double layer capacitive effects coupled with a thermal model{\color{black}, is given in} \citep{Reimers14a, Reimers14b}. Interested reader {\color{black}is referred} to \citep{Reimers13} for more details on linearizing the over-potential equation (\ref{eqn:eta}), grid generation and the time evolution strategy.

\pagebreak

\section*{References}

\bibliographystyle{unsrt}
{ \bibliography{Bib_V1}}

\pagebreak
\begin{multicols}{2}
\nomenclature[]{$\Omega$}{sample set}
\nomenclature[]{$\bm{\xi}$}{vector of input random variables}
\nomenclature[]{$d$}{number of random inputs}
\nomenclature[]{$\psi_{\bm{i}}$}{multivariate PC basis functions}
\nomenclature[]{$\alpha_{\bm{i}}$}{PC coefficients}
\nomenclature[]{$p$}{total order of the PC expansion}
\nomenclature[]{$\rho(\xi)$}{probability measure of random variable $\xi$}
\nomenclature[]{$\mathscr{I}_{d,p}$}{set of $p\mathrm{th}$ order multi-indices in dimension $d$}
\nomenclature[]{$P$}{number of PC basis functions of $p\mathrm{th}$ total order in dimension $d$}
\nomenclature[]{$\delta$}{Kronecker delta or Dirac delta function}
\nomenclature[]{$\gamma$}{the truncation error tolerance of BPDN problem}
\nomenclature[]{$N$}{number of samples}
\nomenclature[]{$c$}{salt concentration in liquid phase [$\mathrm{mol} \cdot \mathrm{m^{-3}}$]}
\nomenclature[]{$c_s$}{lithium concentration in solid phase [$\mathrm{mol} \cdot \mathrm{m^{-3}}$]}
\nomenclature[]{$c_s^{\mathrm{surf}}$}{lithium concentration in solid phase at $r=r_s$ [$\mathrm{mol} \cdot \mathrm{m^{-3}}$]}
\nomenclature[]{$\phi_{e}$}{Li$^+$ ion potential in liquid phase [$\mathrm{V}$]}
\nomenclature[]{$\phi_{s}$}{electron potential in the solid phase [$\mathrm{V}$]}
\nomenclature[]{$\eta$}{over-potential in electrodes [$\mathrm{V}$]}
\nomenclature[]{$L$}{width [m]}
\nomenclature[]{$x$}{distance from anode [m]}
\nomenclature[]{$r$}{micro-scale distance from the center of solid particle [m]}
\nomenclature[]{$F$}{Faraday's constant = 97484 [$\mathrm{C} \cdot \mathrm{mol}^{-1}$]}
\nomenclature[]{$t$}{time [s]}
\nomenclature[]{$a$}{active particle surface area per unit volume of electrode [$\mathrm{m}^{2} \cdot \mathrm{m^{-3}}$]}
\nomenclature[]{$\epsilon$}{porosity of electrodes and stack}
\nomenclature[]{$j_{vol}$}{volumetric reaction flux in the pore walls [$\mathrm{amp} \cdot \mathrm{m^{-3}}$]}
\nomenclature[]{$I$}{total current density across the stack [$\mathrm{amp} \cdot \mathrm{m^{-2}}$]}
\nomenclature[]{$i_{ex}$}{exchange current density of an electrode reaction [$\mathrm{amp} \cdot \mathrm{m^{-2}}$]}
\nomenclature[]{$T$}{temperature [K]}
\nomenclature[]{$r_s$}{solid particle size [m]}
\nomenclature[]{$\mathrm{brugg}$}{Bruggeman coefficient}
\nomenclature[]{$t_+^0$}{Li$^+$ transference number}
\nomenclature[]{$\tau$}{tortuosity}
\nomenclature[]{$D_s$}{diffusion coefficient of the solid phase [$\mathrm{m^{-2}} \cdot \mathrm{s^{-1}}$]}
\nomenclature[]{$D$}{diffusion coefficient of the liquid phase [$\mathrm{m^{-2}} \cdot \mathrm{s^{-1}}$]}
\nomenclature[]{$\sigma$}{electronic conductivity of the solid phase [$\mathrm{S} \cdot \mathrm{m^{-1}}$]}
\nomenclature[]{$\kappa$}{electronic conductivity of the liquid phase [$\mathrm{S} \cdot \mathrm{m^{-1}}$]}
\nomenclature[]{$\kappa_D$}{liquid phase diffusional conductivity [$\mathrm{S} \cdot \mathrm{m^{-1}}$]}
\nomenclature[]{$k$}{reaction rate constant [$\mathrm{m}^4 \cdot \mathrm{mol} \cdot \mathrm{s}$]}
\nomenclature[]{$V$}{open circuit potential of the active material [$\mathrm{V}$]}
\nomenclature[]{$\phi_{cell}$}{cell voltage [$\mathrm{V}$]}
\nomenclature[]{$S_k^T$ }{total Sobol' index of $k\mathrm{th}$ random input}
\nomenclature[]{$S_k$ }{first order Sobol' index of $k\mathrm{th}$ random input}
\nomenclature[U]{$a$}{anode}
\nomenclature[U]{$s$}{separator}
\nomenclature[U]{$c$}{cathode}
\nomenclature[U]{$\mathrm{max}$}{maximum}
\nomenclature[U]{$\bm{i}$}{multi-index}
\nomenclature[S]{$\mathrm{eff}$}{effective value}
\nomenclature[S]{$o$}{internal mixing portion}
\nomenclature[S]{$j$}{unit flux portion}
\nomenclature[S]{$h$}{harmonic mean}
{
\printnomenclature}
\end{multicols}



\end{document}